\documentclass[11pt,a4paper]{article}
\usepackage{epsfig}
\usepackage[T1]{fontenc}    
\usepackage{graphics}
\usepackage{graphicx}
\usepackage{pstricks,pst-coil,pst-fill,pst-plot}
\usepackage[fleqn]{amsmath}    
\usepackage{amssymb}    
\usepackage{amsfonts}   
\usepackage{verbatim}   
\usepackage{mathrsfs}   
\usepackage{dsfont}
\usepackage{euscript}
\usepackage{yfonts}
\usepackage{enumerate}     
\usepackage{amsthm}         
\usepackage{txfonts}
\usepackage{marvosym}
\usepackage{stmaryrd}
\usepackage{vmargin}        
\usepackage{wasysym}		

\setmarginsrb{1.8cm}{2cm}{1.8cm}{2cm}{1cm}{1cm}{1cm}{1.6cm}
 \makeatletter
 \@addtoreset{equation}{section}
 \makeatother

\providecommand{\bysame}{\leavevmode\hbox to3em{\hrulefill}\thinspace}
\providecommand{\MR}{\relax\ifhmode\unskip\space\fi MR }

\providecommand{\href}[2]{#2}

\newcommand{\bra}[1]{\big< \,#1\,|}
\newcommand{\ket}[1]{|\,#1\, \big>}

\let\ua=\uparrow
\let\da=\downarrow
\let\tend=\rightarrow

\long\def\symbolfootnote[#1]#2{\begingroup%
\def\thefootnote{\fnsymbol{footnote}}\footnote[#1]{#2}\endgroup}

\newtheorem{theorem}{Theorem}[section]
\newtheorem{prop}[theorem]{Proposition}
\newtheorem*{theorem*}{Theorem}

\newtheorem{conj}[theorem]{Conjecture}

\newtheorem{lemme}[theorem]{Lemma}

\def\Proof{\medskip\noindent {\it Proof --- \ }}

\def\qed{\hfill\rule{2mm}{2mm}}

\newcommand\beq{\begin{equation}}
\newcommand\enq{\end{equation}}
\newcommand\bem{\begin{multline}}
\newcommand\enm{\end{multline}}

\def\beqa{\begin{eqnarray}}
\def\eeqa{\end{eqnarray}}
\def\ba{\begin{array}}
\def\ea{\end{array}}

\newcommand{\f}[2]{{\ensuremath{%
    \mathchoice%
    {\dfrac{#1}{#2}}
    {\dfrac{#1}{#2}}
    {\frac{#1}{#2}}
    {\frac{#1}{#2}}
}}}
\newcommand{\tf}[2]{\ensuremath{#1/#2}}
\def\a{\alpha}

\def\ga{\gamma}
\def\Ga{\Gamma}

\def\de{\delta}

\def\De{\Delta}
\def\eps{\epsilon}
\def\veps{\varepsilon}
\def\la{\lambda}

\def\sg{\sigma}
\def\vsg{\varsigma}

\def\Ups{\Upsilon}
\def\ups{\upsilon}

\def\vth{\vartheta}

\def\Om{\Omega}
\def\om{\omega}
\def\vp{\varphi}

\newcommand{\mc}[1]{\ensuremath{\mathcal{#1}}}
\newcommand{\mf}[1]{\ensuremath{\mathfrak{#1}}}
\newcommand{\msc}[1]{\ensuremath{\mathscr{#1}}}

\newcommand{\bs}[1]{\ensuremath{\boldsymbol{#1}}}

\DeclareFontFamily{OT1}{pzc}{}
\DeclareFontShape{OT1}{pzc}{m}{it}{<-> s * [1.10] pzcmi7t}{}
\DeclareMathAlphabet{\mathpzc}{OT1}{pzc}{m}{it}

\def \i{ \mathrm i}

\newcommand{\wt}[1]{\ensuremath{\widetilde{#1}}}
\newcommand{\wh}[1]{\ensuremath{\widehat{#1}}}

\newcommand{\Int}[2]{\ensuremath{\int\limits_{#1}^{#2}}}

\newcommand{\sul}[2]{\ensuremath{\sum\limits_{#1}^{#2}}}
\newcommand{\pl}[2]{\ensuremath{\prod\limits_{#1}^{#2}}}

\newcommand{\R}{\ensuremath{\mathbb{R}}}
\newcommand{\Cx}{\ensuremath{\mathbb{C}}}

\newcommand{\Dp}[1]{\ensuremath{\partial_{#1}}}

\newcommand{\ex}[1]{\ensuremath{\e{e}^{#1}}}

\def\Res{\operatorname{Res}}

\newcommand{\op}[1]{ \boldsymbol{ \texttt{#1} } }

\newcommand{\dd}{\mathrm{d}}
\newcommand{\e}[1]{\ensuremath{\mathrm{#1}}}

\newcommand{\intff}[2]{\ensuremath{ [  #1 \,; #2 ] }}

\newcommand{\intof}[2]{\ensuremath{ ]  #1 \,; #2 ] }}
\newcommand{\intoo}[2]{\ensuremath{ ]  #1 \,; #2 [ }}

\newcommand{\intn}[2]{\ensuremath{[\![ \, #1 \,;\, #2 \,]\!]}}

\begin{document}

\begin{center}
\begin{LARGE}
{\bf Long-distance and large-time asymptotic behaviour of dynamic correlation functions in the massless regime of the XXZ spin-1/2 chain}
\end{LARGE}

\vspace{1cm}

\vspace{4mm}
{\large Karol K. Kozlowski \footnote{e-mail: karol.kozlowski@ens-lyon.fr}}%
\\[1ex]
Univ Lyon, ENS de Lyon, Univ Claude Bernard Lyon 1, CNRS, Laboratoire de Physique, F-69342 Lyon, France \\[2.5ex]

\par 

\end{center}

\vspace{40pt}

\centerline{\bf Abstract} \vspace{1cm}
\parbox{12cm}{\small}

Starting from the massless form factor expansion for the two-point dynamical correlation functions obtained recently, I 
extract the long-distance and large-time asymptotics of these correlators. 
 The analysis yields the critical exponents and associated amplitudes characterising the asymptotics. 
 The results are obtained on the basis of  exact and first principle based considerations: they  \textit{do not rely, at any stage}, on 
some hypothetical correspondence with a field theory  or the use of any other phenomenological approach. Being based on form factor
expansion, the method allows one to clearly identify which contributions to the asymptotics issues from which
class of excited states.  
All this permits to settle the long-standing question of the
contribution of bound states to the asymptotics of two-point functions. For instance, when considering the long-distance $m$ behaviour of equal-time correlators,
the analysis shows that while,  \textit{in fine}, the bound states only produce contributions that are exponentially small in $m$,
they also play a key role in cancelling out certain power-law contributions which, should they be present, would 
break explicitly the universality structure of the long-distance behaviour.

\vspace{40pt}
\tableofcontents

\section{Introduction}

One dimensional quantum critical Hamiltonians, \textit{i.e.} those having no gap above their ground state in the thermodynamic limit, are expected to exhibit various universal 
features. In particular, the long-distance asymptotic behaviour of their ground state correlation functions is expected to be grasped by a two-dimensional conformal 
field theory. This belief stems from certain heuristics relative to the connection which exists between the structure of the large-volume behaviour of the model's 
low-lying excited states energies and the structure of the excited states' energies in a conformal field theory \cite{AffleckCFTPreForLargeSizeCorrPartitionFctonAndLowTBehavior,BloteCardyNightingalePredictionL-1correctionsEnergyAscentralcharge}.
Such considerations allowed to deduce the critical exponents from the $1/L$ corrections to the low-lying excited states and this led to numerous 
explicit predictions for the critical exponents of many  quantum integrable models 
\cite{BogoluibovIzerginReshetikhinCriticalExponentsforXXZ,HaldaneCritExponentsAndSpectralPropXXZ,KlumperBatchelorNLIEApproachFiniteSizeCorSpin1XXZIntroMethod,
KlumperBatchelorPearceCentralChargesfor6And19VertexModelsNLIE,LutherPeschelCriticalExponentsXXZZeroFieldLuttLiquid}. 
The  case of dynamical correlation functions is more involved and appears to go beyond the CFT picture.  
For some models,  it seems that for large enough values of the ration $m/t$, the CFT picture still holds. This allowed
to predict the form (critical exponents and associated amplitudes) of the first few terms in the long-distance and large-time 
asymptotics of two-point correlation functions in the XXZ spin-$1/2$ chain at zero magnetic field \cite{LukyanovTerrasSpinSpinAmplitudesBetterTreatementXXZ}. 
The genuine behaviour of dynamical correlation functions was more recently argued to be grasped by the non-linear Luttinger liquid approach
\cite{GlazmanImambekovGeneralRelationEdgeExpForGeneralModel,GlazmanImambekovDvPMTCompletTheoryNNLL,GlazmanImambekovSchmidtReviewOnNLLuttingerTheory}. 
This approach incorporates certain non-linearities present in the model's dispersion curves. 
In fact, the approach takes 
the non-linear dispersion curves of a model as raw data to input into the parameters arising in the non-linear Luttinger model. 
These data can be accessed from the large-volume behaviour of the model's spectrum, what could have been done for several quantum integrable models
\cite{AffleckPereiraWhiteSpectralFunctionsfor1DLatticeFermionsBoundStatesContributions,CampbellGangardtMobileImpurityIntegrableModels,GlazmanImambekovComputationEdgeExpExact1DBose}
and provides then explicit predictions for the edge or critical exponents and also allows one to conjecture the expression for the associated amplitudes \cite{CauxGlazmanImambekovShasiNonUniversalPrefactorsFromFormFactors}. 

The rigorous characterisation of dynamical two-point functions in quantum integrable models was first possible for the 
free fermion equivalent models by using the important combinatorial simplifications that arise then in the description of the space of states.
This specific structure allowed to represent the dynamical two-point functions in terms of certain Painlevé transcendents 
and hence to study the regime of the long-distance or the long-distance and large-time asymptotics by obtaining connection formulae
for the Painlevé transcendent \cite{JimboMiwaMoriSatoSineKernelPVForBoseGaz,AbrahamBarouchMcCoy4ptcomputationtoGetlongTimeforTwo,McCoyPerkShrockSpinTimeSpaceAutoCorrAsModSineKernel,McCoyPerkShrockSpinTimeAutoCorrAsModSineKernel,
MullerShrockDynamicCorrFnctsTIandXXAsymptTimeAndFourier,PerkAuYangTimeDpdtTransverseIsing}. More recently,
the connection with Painlevé transcendents was elucidated within 
the form factor summation method \cite{KorepinSlavnovTimeDepCorrImpBoseGas} which allows one to represent directly the two-point dynamical correlation functions of free fermion equivalent models
as Fredholm determinants of integrable integral operators \cite{ColomoIzerginKorepinTognettiXX2ptFctsatGeneralTimeAndSpaceAndTempe,KorepinSlavnovTimeDepCorrImpBoseGas}. This, along with 
the development of the Riemann--Hilbert problems approach \cite{ItsIzerginKorepinSlavnovDifferentialeqnsforCorrelationfunctions} to such operators allowed  to access various asymptotic regimes of 
the correlators in free fermion equivalent models, see \textit{e.g.}
\cite{CheianovZvonarevZeroTempforFreeFermAndPureSine,ItsIzerginKorepinTemperatureLongDistAsympBoseGas,ItsIzerginKorepinVarguzinTimeSpaceAsymptImpBoseGaz}.

The obtention of explicit representations for the correlation  functions of quantum integrable models away from the free Fermi point was made possible by the development of the algebraic Bethe Ansatz \cite{FaddeevSklyaninTakhtajanSineGordonFieldModel}
and of the vertex operator approaches \cite{DavisFodaJimboMiwaNakayashikiDiagonalizationXXZinfiniteDelta>1}. 
First results consisted in  highly combinatorial expressions for the dynamical correlation functions 
in certain models \cite{IzerginKorepinQISMApproachToCorrFns2SiteModel,IzerginKorepinQISMApproachToCorrNextDiscussion,IzerginKorepinReshetikhinCommRelationFieldOpAndMonoMatrix}. 
After a long series of developments relative to obtaining manageable closed expressions for the static correlation functions
\cite{BoosJimboMiwaSmironovTakeyamaAlgebraicRepresentationofCorrelationFunctions=SepofIntegrals,JimboMikiMiwaNakayashikiElementaryBlocksXXZperiodicDelta>1,
JimboMiwaElementaryBlocksXXZperiodicMassless,KozKitMailSlaTer6VertexRMatrixMasterEquation,KitanineMailletSlavnovTerrasOriginalSeries,KitanineMailletTerrasElementaryBlocksPeriodicXXZ},
some representations for the dynamical two-point functions were obtained \cite{JimboMiwaFormFactorsInMassiveXXZ,KitanineMailletSlavnovTerrasDynamicalCorrelationFunctions}. 
The progress achieved in simplifying the representations for the correlation functions opened the possibility to devise techniques of asymptotic analysis 
that would allow to extract, from such representations, the asymptotic behaviour of the correlators. 
The first exact analysis of the long-distance asymptotics of two-point static functions in an interacting model
was achieved in \cite{KozKitMailSlaTerXXZsgZsgZAsymptotics} and this work paved the road to the systematic study of the asymptotic
behaviour of correlation functions in massless quantum integrable models. In particular, the long-time large-distance asymptotics of the dynamical 
two-point functions in the non-linear Schrödinger model at zero temperature was obtained in  the series of works \cite{KozTimeDepGSKandNatteSeries,KozReducedDensityMatrixAsymptNLSE,KozTerNatteSeriesNLSECurrentCurrent}
by means of the multidimensional deformation flow method. Those results demonstrated the incompleteness of the purely CFT-based characterisation of the asymptotics
in the dynamical correlation function case \cite{BerkovichMurthyWrongCFTBasedPredictionTimeMultiCorrNLSE}.

A strong conceptual simplification of the techniques of asymptotic analysis of correlation functions was provided by the form factor approach. 
The possibility to analyse the long-distance asymptotic behaviour on the level of the form factor series expansion of a correlation function
was first observed in  \cite{KozMailletSlaLowTLimitNLSE} and the method was subsequently developed in 
\cite{KozKitMailSlaTerRestrictedSums,KozKitMailSlaTerRestrictedSumsEdgeAndLongTime,KozKitMailTerMultiRestrictedSums,KozRagoucyAsymptoticsHigherRankModels}. 
The method relied on the large-volume behaviour of the model's form factors \cite{KozKitMailSlaTerEffectiveFormFactorsForXXZ,KozKitMailSlaTerThermoLimPartHoleFormFactorsForXXZ,SlavnovFormFactorsNLSE}. 
In particular, the work \cite{KozKitMailSlaTerRestrictedSumsEdgeAndLongTime} extracted the long-time and large-distance
asymptotics of two-point functions in the non-linear Schr\"{o}dinger model. However, while remaining close to the microscopic model, that method of analysis 
was still rather heuristics and based on several non-trivial assumptions. 
To make it exact, one first needed to provide a proper way of giving sense to a  form factor series expansions for massless models directly in the 
thermodynamic limit. This was achieved in \cite{KozMasslessFFSeriesXXZ} where a scheme for dealing with the apparent infrared divergencies of the series was devised. This work provided a series of multiple
integral representation for the dynamical two-point functions in the zero temperature XXZ chain. 
Recently, I have demonstrated \cite{KozSingularitiesSpectralFctsXXZ} that this series representation allows one to investigate thoroughly the singularity structure of dynamic response functions 
in the XXZ chain. In the present work, I will show that this series also allows one to grasp the 
long-time and large-distance asymptotics of the two-point dynamical functions in the XXZ chain at finite magnetic field.

The paper is organised as follows. Section \ref{Section notations resultats presents et passes} reviews various properties of the XXZ chain and presents the main results obtained in this work. 
Subsection \ref{Sous Section chaine XXZ} describes elementary facts about the XXZ Hamiltonian, Subsection \ref {Sous Section spectre XXZ limite thermo} recalls the description of
the XXZ chain's spectrum in the thermodynamic limit while Subsection \ref{Sous Section resultats principaux} contains the \textit{resumé} of the 
long-distance and large-time asymptotics of dynamical two-point functions obtained in the core of the paper. Then, Subsection \ref{Sous section serie de facteurs de formes}
describes the massless form factor series representation obtained in \cite{KozMasslessFFSeriesXXZ}. Finally, 
Subsection \ref{SousSection FF gd L} outlines the properties of the series building blocks, the so-called form factor densities, which are 
crucial for the setting up of the asymptotic analysis. Section \ref{Section Auxiliary rewriting of the FF series} contains some preliminary considerations that are necessary for setting in the steepest descent. 
Subsection \ref{Sous Section preprietes pour locus saddel pts} discusses the properties of the phase functions which determine the \textit{loci}
of the saddle-points and the form of the steepest descent paths. Then, Subsection \ref{Sous Section reecriture de la series de FF} rewrites the massless form 
factor series in a form more adapted for carrying out contour deformations. 
Section \ref{Section deformation des contours dans integrales auxiliaires}
establishes certain reorganisation properties of a family of auxiliary integrals under contour deformations. Subsection \ref{Sous Section integrales modeles pour deformation de contours} 
presents the class of auxiliary integrals of interest. 
Subsection \ref{Sous Section secteur a 2 particules} deals with integrals subordinate to a two-hole excitation sector while Subsection
\ref{Sous Section secteur a 3 particules} deals with integrals subordinate to a three-hole excitation sector. 
Then, Subsection \ref{Sous Section conjecture generale sur forme contours deformes} details a conjecture on transformations under contour deformations 
for the general $n$-hole excitation sector. Finally, Section \ref{Section steepest descent analysis} carries out the steepest descent asymptotic analysis of the massless form factor series. 
Sub-section \ref {Sous section partie algebriquement dominante de Sn} identifies all the sub-constituents of the massless form factor expansion 
which can generate a power-law behaviour in $(m,t)$. Subsection \ref{Sous section regime conforme} determines the asymptotic expansion in the 
conformal regime while Subsection \ref{Sous section regime profond asymptotiques} determines the asymptotics in the other region of the space of parameters. 
Section \ref{Section conclusion} is  the conclusion. The paper contain an appendix  which gathers several facts about the description of the spectrum of the 
XXZ chain in the thermodynamic limit. Sub-Appendix \ref{Appendix Sols Line Int Eqns} reviews the various solutions to linear integral equations which appear in the description of the spectrum. 
Sub-appendix  \ref{Appendix Sous-section cordes} discusses the domains of existence of $r$-string solutions for $r=2,\dots,8$. 
 Finally, Sub-appendix \ref{Appendix Section phase oscillante dpdte de la vitesse} provides a detailed study of the velocity of the various excitations in the model.

\section{The massless form factor expansion}
\label{Section notations resultats presents et passes}
\subsection{The XXZ chain }
\label{Sous Section chaine XXZ}

The XXZ spin-$\tf{1}{2}$ periodic chain  Hamiltonian  
\beq
\op{H} \, = \, J \sum_{a=1}^{L} \Big\{ \sigma^x_a \,\sigma^x_{a+1} +
  \sigma^y_a\,\sigma^y_{a+1} + \cos(\zeta)  \,\sigma^z_a\,\sigma^z_{a+1}\Big\} \, - \, \f{h}{2} \sul{a=1}{L} \sg_{a}^{z} \; , \qquad \e{with} \qquad \sg_{a+L}^{\ga}=\sg_{a}^{\ga} \;,  
\label{ecriture hamiltonien XXZ}
\enq
 is an operator on the Hilbert space $\mf{h}_{XXZ}=\otimes_{a=1}^{L}\mf{h}_a$ with $\mf{h}_a \simeq \Cx^2$. 
The matrices $\sg^{\ga}$, $\ga=x,y,z$ are the Pauli matrices and the operator $\sg_a^{\ga}$  is defined as 
\beq
\sg_a^{\ga} \; = \; \underbrace{ \e{id}\otimes \cdots \otimes \e{id} }_{ a-1 \; \e{times} } \otimes \; \sg^{\ga} \otimes \underbrace{ \e{id}\otimes \cdots \otimes \e{id} }_{ L-a \; \e{times} }  \;. 
\enq
I shall focus on the antiferromagnetic regime, namely the case of  positive exchange interaction  $J>0$, and assume that the model is in its genuine XXZ finite positive magnetisation massless phase. 
Thus, the anisotropy $\cos(\zeta)$ and the overall external magnetic field $h$ are restricted to the domains:  
\beq
 \zeta \in \intoo{0}{\pi} \;, \;\; i.e. \;\; -1< \cos(\zeta) <1 \; ,  \qquad \e{and} \qquad 0 < h  < h_{\e{c}}=8J \cos^2(\tf{\zeta}{2}). 
\enq
The case of the XXX chain at finite magnetic field can also be taken care of  by the present analysis, although the explicit treatment would demand to introduce different notations, so that 
I shall not develop further on this special point. Also, to avoid unnecessary technicalities that would obscure the main features of the analysis,  
 I shall assume that $\tf{\zeta}{\pi}$ is irrational.

The translational invariance of the model allows one to represent the Heisenberg picture time evolved
spin operator $\sg^{(\ga)}_{m+1}$ as 
\beq
\sg^{\ga}_{m+1}(t)\, = \, \ex{\i m \op{P} + \i \op{H}  t} \cdot \sg^{\ga}_{1} \cdot \ex{ -\i t \op{H} -\i m \op{P}} \;, 
\label{ecriture evolution temporelle op local}
\enq
where $\op{P}$ is the momentum operator and hence, $\ex{\i  \op{P} } $ is  the translation operator by one-site. Note also that the relative sign 
in the $t$ and  $m$ dependent terms in \eqref{ecriture evolution temporelle op local} issues from the choice of the sign of the Hamiltonian and of the anisotropy $\cos(\zeta)$ in \eqref{ecriture hamiltonien XXZ}

The zero temperature dynamical two-point functions are defined as 
\beq
\big< \sg_1^{\ga^{\prime}}\!(t)\, \sg_{m+1}^{\ga}(0) \big> \; = \; \lim_{L \tend + \infty} \Big\{ \bra{\Om}\sg_1^{\ga^{\prime}}(t) \sg_{m+1}^{\ga}(0) \ket{\Om}  \Big\} \;, 
\label{ecriture definition fct correlation zero temperature}
\enq
where $\dagger$ stands for the Hermitian conjugation and where $\ket{\Om}$ corresponds to the ground state of the model in finite volume $L$.  

 The  main purpose of the work  \cite{KozMasslessFFSeriesXXZ} was to obtain a well-defined, \textit{viz}. free of any divergencies, form factor series expansion
for the dynamical two-point functions in the massless regime of the XXZ spin $1/2$ chain. The purpose of this work is to show the effectiveness of that series 
relatively to the analysis of the long-distance and large-time asymptotics of the two-point functions $\big< \sg_1^{\ga^{\prime}}\!(t)\, \sg_{m+1}^{\ga} \big> $. 
Prior to discussing the very form of the series and presenting the details of the analysis, I shall describe the main result obtained in this work. 
For that, however, I need to review the structure of the model's spectrum in the thermodynamic limit.

\subsection{The spectrum of the XXZ chain in the thermodynamic limit}
\label{Sous Section spectre XXZ limite thermo}

In the thermodynamic limit, a given excited state is built up from various species of particles:  the holes, 
the particles (also called $1$-strings) and the bound states (also called $r$-strings). The various bound states species are 
labelled by integers belonging to the set $\mf{N}_{\e{st}} \subset \big\{ \mathbb{N} \setminus \{0,1\} \big\}$.  The set $\mf{N}_{\e{st}}$ may be finite or infinite, 
depending\symbolfootnote[2]{When $\zeta$ is a rational multiple of $\pi$ the set is finite and it is expected to be 
infinite otherwise, \textit{c.f.} \cite{TakahashiThermodynamics1DSolvModels}, although the last property has not been proven yet.}  on the value of $\zeta$.
It is sometimes convenient to parameterise it as $\mf{N}_{\e{st}} =\{ r_2,\dots , r_{|\mf{N}|}\}$. I stress that the integers $r_a$ appearing in this parameterisation do depend on $\zeta$ and 
that this collection may or may not be infinite. It will also sometimes be useful to denote $\mf{N} =  \{1\} \cup \mf{N}_{\e{st}}=\{ r_1,\dots , r_{|\mf{N}|}\}$, where $r_1=1$. 

A given excited state consists of
\begin{itemize}
 
 \item $n_h \in \mathbb{N}$ hole excitations each carrying a rapidity $\mu_a \in \intff{-q}{q}$, $a=1,\dots, n_h$;

 \item $n_r \in \mathbb{N}$, $r\in \mf{N}$, $r$-string excitations each carrying a rapidity $ \nu_{a}^{(r)}$, $a=1,\dots, n_r$, such that 
\beq
\nu_{a}^{(1)} \in \Big\{ \R \setminus \intoo{-q}{q} \Big\} \cup \Big\{ \R + \i \tfrac{\pi}{2} \Big\} \qquad \e{while} \qquad \nu_{a}^{(r)} \in \R \, + \, \i \sg_{r} \f{\pi}{2}  \;,  
\label{ecriture domaine evolution des diverses rapidites particules et cordes}
\enq
 in which $\sg_{r} = 0$ or $1$ is called the string parity. An $r$-string with $r\geq 2$ has only one possible parity;   
\item left and right Umklapp excitations characterised by the deficiencies $\ell_{\pm} \in \mathbb{Z}$ which encode the 
discrepancy between the numbers of massless particle and hole excitations that build up the swarm of microscopic excitations lying directly on the two endpoints $\pm q$ 
of the Fermi zone $\intff{-q}{q}$.

\end{itemize}
  Finally, a given excited state $\Ups$ belongs to  a sector of relative spin $-\op{s}_{\Ups}$ in respect to the ground state. 

  \vspace{4mm}

 The integers appearing above can be gathered in a single vector with integer components 
\beq
\bs{n} = (\ell_{+},\ell_{-}; n_h,  n_{r_1},\dots, n_{ r_k}, \dots ) \, .
\label{definition vecteur entier pour nombres excitations}
\enq
The set of allowed $\bs{n}$s builds up the set 
\beq
\mf{S} \, = \, \bigg\{ \bs{n} = (\ell_{+},\ell_{-}; n_h,   n_{r_1},\dots, n_{ r_k}, \dots )\; : \; \ell_{\pm}\in \mathbb{Z}\; , \; n_h, n_r \in \mathbb{N} \; \quad \e{and} \quad 
\; n_h + \op{s}_{\Ups} = \sul{r \in \mf{N} }{} r n_r \, + \, \sul{\ups=\pm}{} \ell_{\ups} \bigg\}  \;. 
\label{ecriture range summation dans series 2pt dynamic}
\enq
Note that while, in principle, one has $|\mf{N}|=+\infty$ so that $\bs{n}$ has infinitely many components, the constraint in \eqref{ecriture range summation dans series 2pt dynamic} ensures that, for a given excited state, 
$n_r=0$ for any $r$ large enough so that $\bs{n}$ has only a finite number of non-zero components. The vector $\bs{n}$ thus makes sense as an inductive limit. 

It is convenient to gather all the rapidities arising in  a given excited state into  the set  
\beq
\mf{Y} \; = \; \Big\{ -\op{s}_{\Ups};  \big\{ \ell_{\ups}\big\} \; \big| \; \big\{ \mu_{a} \big\}_{1}^{ n_{h} } \, ; \, \big\{ \nu_{a}^{(r_1)} \big\}_{a=1}^{n_1}; \dots ;  \big\{ \nu_{a}^{(r_k)} \big\}_{a=1}^{n_{r_k}} ;  \dots  \Big\} \;. 
\label{definition rapidites massives reduites}
\enq
I remind that $r_k$ correspond to the sequence of integers such that $\mf{N}=\{r_1, r_2, \dots,\}$ and that $r_1=1$. 
Since for a given excited state it holds that  $n_r=0$ provided that $r$ is large enough, $\mf{Y}$ is built up from a collection of sets  where only finitely many of them are non-empty, 
so that, again, it makes sense as an inductive limit. 

Note that the use of sets to parameterise an excited state is unambiguous since, in the thermodynamic limit, all the physical observables become set functions of the excited state's 
rapidities. For instance, the thermodynamic limit of the excitation energy and momentum, relatively to the ground state,
of an excited state characterised by the set of rapidities  $\mf{Y}$ takes the form   
\beqa
 \mc{E}\big( \mf{Y}  \big) & = &  \sul{ r\in \mf{N}  }{ } \sul{a=1}{n_r } \veps_{r}\big( \nu_{a}^{(r)} \big)    \, - \, \sul{ a=1 }{ n_h  } \veps_1( \mu_{a}  )   					     
\label{ecriture forme leading energie ex}\\
 \mc{P}\big( \mf{Y} \big)  & = &   \sul{ r\in \mf{N}  }{ } \sul{a=1}{n_r } p_{r}\big( \nu_{a}^{(r)} \big)    \, - \, \sul{ a=1 }{ n_h  } p_1( \mu_{a} )    
					   \, + \, p_{F} \sul{\ups=\pm }{} \ups \ell_{\ups}   \; + \; \pi \op{s}_{\Ups}  \;. 
\label{ecriture forme leading impulsion ex}
\eeqa
in which $p_{F}=p_1(q)$ corresponds to the Fermi momentum.  The functions $\veps_{a}$, resp. $p_{a}$, are the dressed energies and momenta of the individual excitations. 
They are defined as solutions to linear integral equation, \textit{c.f.} equations \eqref{definition r energie habille} and \eqref{definition r moment habille} in 
Appendix \ref{Appendix Sols Line Int Eqns}. In particular, the dressed energies are such that 
\begin{itemize}
\item[$ \rm i)$]  for $r \in \mf{N}_{\e{st}}$ and $\sg_r$ as in \eqref{ecriture domaine evolution des diverses rapidites particules et cordes} there exists a constant $c_r>0$ such that 
it holds $\veps_{r} > c_r >0$ on $\R+\sg_{r}\i\tfrac{\pi}{2} $;
\item[$ \rm ii) $]  $\veps_{1}(\pm q)=0$ and 
$$ \veps_1 >0  \quad \e{on} \quad \Big\{ \R + \i\tfrac{ \pi }{ 2 } \Big\} \cup \Big\{ \R \setminus \intff{ -q }{ q } \Big\} \qquad  
\e{while} \quad  \veps_1<0 \quad \e{ on } \quad  \intoo{ - q }{ q } \; .$$
\end{itemize}

The properties i)-ii) above ensure that bound states only carry massive excitations while particle and hole may also exhibit massless excitations when their rapidities collapse on either
endpoint of the Fermi zone. I refer to \cite{KozProofOfAsymptoticsofFormFactorsXXZBoundStates} where all these facts are discussed at length.  
 Property ii) was rigorously established in \cite{KozDugaveGohmannThermoFunctionsZeroTXXZMassless}. See Appendix \ref{Appendix Sols Line Int Eqns} for more details on i).

For further purposes, it is convenient to introduce a compact notation for a combination of individual momenta and energies building up an excited state
characterised by the rapidities \eqref{definition rapidites massives reduites}
\beqa
\msc{U}\big( \mf{Y}, \op{v} \big) & = &   \mc{P}\big( \mf{Y} \big)\, - \, \f{ 1 }{ \op{v} }  \mc{E}\big( \mf{Y} \big) \; - \; \pi \op{s}_{\Ups} \\
                                 &  =  &   \sul{ r\in \mf{N} }{ } \sul{a=1}{ n_r } u_{r}\big( \nu_{a}^{(r)}, \op{v} \big)   \, - \, \sul{ a=1 }{ n_h  } u_1( \mu_{a}, \op{v} )    \, + \, p_{F} \sul{\ups=\pm }{} \ups \ell_{\ups} \;. 
\label{definition de energi impuslion reduite condensation massless}
\eeqa
Above $\op{v}$ refers to the ratio of the distance to time $\op{v}=m/t$ while 
\beq
u_{r}\big( \la, \op{v} \big)     \; = \; p_r(\la) - \f{ \veps_{r}(\la) }{ \op{v} } \;. 
\label{definition fct ur}
\enq

\subsection{Main result: the asymptotic expansion of two-point functions}
\label{Sous Section resultats principaux}

In this work, I focus on directional asymptotics of the two-point function  $\big< \sg_1^{\ga^{\prime}}\!(t)\, \sg_{m+1}^{\ga}(0) \big> $, namely when 
\beq
(m,t) \tend \infty \quad \e{so} \, \e{that} \, \e{the} \, \e{ratio} \quad  \op{v}=\f{m}{t}
\label{definition ratio distance au tps}
\enq
 remains fixed. This is the most interesting regime from the point of view of physics. For technical reasons, I shall also assume that 
\beq
\op{v} \not=  \pm \op{v}_F \;   \qquad \e{where} \qquad \op{v}_F \; = \; \f{ \veps^{\prime}_1(q) }{  p_1^{\prime}(q) }  
\enq
stands for the Fermi velocity, namely the velocity of the massless excitations in the model, \textit{viz}. those whose rapidities coincide with one of
the endpoints of the Fermi zone. The treatment of the regimes $\op{v}=\pm \op{v}_F$ would demand to develop completely different tools.

 The structure of the asymptotic expansion depends on the value of $\op{v}$ but also on the range of the velocity functions $\op{v}_r(\la)=\tf{ \veps_r^{\prime}(\la)}{ p_r^{\prime}(\la) }$.  
This range depends on the value of the endpoint of the Fermi zone $q$ and also on the anisotropy $\zeta$. 
The asymptotic expansion given below holds in the regime of the model's parameters $q$ and $\zeta$ where the function 
$u_r$ enjoy the "minimal structure property" and when $\op{v}_F<\op{v}_{\infty}$ in which 
\beq
\op{v}_{\infty}=\lim_{\la \tend + \infty} \Big\{  \f{ \veps^{\prime}_1(\la) }{  p_1^{\prime}(\la) }   \Big\} \; ,  
\enq
see \eqref{definition cste V infty} for its explicit expression. The  "minimal structure property"  corresponds to a setting where 
\begin{itemize}
 \item $u^{\prime}_r(\la,\op{v})$, $r \in \mf{N}\cup\{0\}$, does not vanish on $\R+\i \sg_r \tfrac{\pi}{2}$ for $|\op{v}|>\op{v}_{\infty}$;
 \item $u^{\prime}_r(\la,\op{v})$, $r \in \mf{N}_{\e{st}}$, admits a unique\symbolfootnote[2]{See Appendix \ref{Appendix Section phase oscillante dpdte de la vitesse} for more details.}
 zero $\om_r$ on   $\R+\i \sg_r \tfrac{\pi}{2}$ for $|\op{v}|<\op{v}_{\infty}$, \textit{i.e.}
\beq
p^{\prime}_{r}(\om_r) \; - \; \frac{ 1 }{ \op{v} } \veps^{\prime}_{r}(\om_r) = 0 \qquad \e{with} \quad \om_r  \in \R + \i \sg_{r} \f{\pi}{2} \;;  
\enq
\item For $r=1$,  and $|\op{v}|<\op{v}_{\infty}$ there are two saddle-points 
\beq
\om_1 \in \R + \i \tf{\pi}{2} \quad \e{and} \quad \om_0 \in \R \;, \quad \e{such} \; \e{that} \qquad 
p^{\prime}_{1}(\om_a) \; - \; \frac{ 1 }{ \op{v} } \veps^{\prime}_{1}(\om_a) = 0 \quad a\in \{0,1\} \;. 
\enq

\end{itemize}  

As shown in Appendix \ref{Appendix Section phase oscillante dpdte de la vitesse}, numerics and perturbative consideration 
show that the "minimal structure property" does hold for a certain range of parameters, in particular when $|\zeta-\tf{\pi}{2}|$ is small enough. 
However, numerical investigation shows that, in general, more complicated situations may also arise. 
The form of the asymptotic expansion in the most general case can be found in Subsection \ref{SousSection asymptotiques cas general}.

 The form of the asymptotic expansion depends on the value of $\op{v}$. In the case when the "minimal structure property" holds, 
one may basically single out two distinct regimes of the directional asymptotics. The distinction between the two regime takes its origin in a different type of universality grasping the asymptotic behaviour. 
In the general case, those two kinds of regimes persist, but the limiting velocity distinguishing between the two has a different interpretation. 
In the case when the "minimal structure property" holds, The border between the two regimes arises at the velocity scale $\op{v}_{\infty}$ whose explicit expression can be found in \eqref{definition cste V infty}. 
There second characteristic velocity of the problem, the Fermi velocity $\op{v}_F$ singles out two sub-regimes building up to the regime $|\op{v}| < \op{v}_{\infty}$. 
\begin{itemize}
 \item The  conformal regime corresponds to $|\op{v}|> \op{v}_{\infty}$. It governs  the regime of  moderate time asymptotics.
 In this regime, the leading structure of the asymptotics has a pure CFT-like interpretation in terms of the operator content of a free boson model, see \cite{KozMailletMicroOriginOfc=1CFTUniversality}
 where this identification was explicitly achieved. 
\item The genuine asymptotic regime  $|\op{v}| < \op{v}_{\infty}$, which governs the regime of large time asymptotics. In this regime, also 
the massive excitations do contribute explicitly to the power-law asymptotics. The genuine asymptotics sector splits in two sub-regimes: 

\begin{itemize}
 
 \item[i)] the space-like regime $\op{v}_F < |\op{v}| <  \op{v}_{\infty}$, in which case the hole excitations do not contribute to the power-law asymptotics; 
 
 \item[ii)] the time-like regime $\op{v}_F > |\op{v}| $, where the hole excitations contribute to the power-law asymptotics.  
 
\end{itemize}

\end{itemize}

 This predictions for the  asymptotics of dynamical correlation functions in the conformal regime were obtained in \cite{LukyanovTerrasSpinSpinAmplitudesBetterTreatementXXZ} by means 
of field theoretic arguments for the XXZ spin-$1/2$ chain at zero magnetic field, under the restriction that $|\op{v}|\gg \op{v}_F$. 
Also, the work \cite{AffleckPereiraWhiteEdgeSingInSpin1-2} predicted first few terms of the large-time asymptotics of a two-point autocorrelation function, again for the XXZ spin-$1/2$ chain at zero magnetic field.

\subsubsection*{ $\bullet$ The conformal regime:  $|\op{v}|> \op{v}_{\infty}$}

If $|\op{v}|> \op{v}_{\infty}$, then it holds 
\beq
\big< \big(\sg_1^{\ga} (t)\big)^{\dagger} \sg_{m+1}^{\ga}(0) \big>_{\e{c}} \; = \;   (-1)^{m \op{s}_{\ga} } \sul{ \ell \in \mathbb{Z} }{} 
\,  \f{  \mc{F}_{ \ell }^{(\ga)} \cdot \ex{2\i m  \ell p_{F} }  }{  \pl{\ups=\pm}{}  [ -\i (\ups m - \op{v}_F t)  ]^{  \De_{\ups,\ell}^2   } }
\cdot \Bigg\{ 1 \; + \; \e{O}\bigg( \,\f{1}{m^{1-0^+}}  \bigg)  \Bigg\} \;. 
\label{ecriture DA cas purement conforme}
\enq
The summation parameter $\ell$ corresponds to the Umklapp deficiency of the right Fermi boundary, so that the left boundary has  deficiency $-\ell$. 

The summation over $\ell$ corresponds to summing up over all possible Umklapp excitations in the model. 
In \eqref{ecriture DA cas purement conforme}, the contribution of a given Umklapp excitation with deficiency $\ell$  is weighted by an oscillatory pre-factor 
which oscillates with a wave vector $2 \ell p_{F}$. The latter corresponds to the macroscopic momentum carried by the given Umklapp excitation. 
The asymptotic expansion \eqref{ecriture DA cas purement conforme} has several other building blocks.

\begin{itemize}

 \item  The integer $\op{s}_{\ga}$ arising in the oscillatory pre-factor corresponds to the pseudo-spin of the operator $\sg_{1}^{\ga}$:
\beq
\op{s}_{z}=0 \quad \e{and}  \quad \op{s}_{\pm}=\mp 1 \;. 
\label{definition spin operateur}
\enq
In other words, $-\op{s}_{\ga}$ corresponds to the relative to the ground state, longitudinal spin of the excited states that are connected to the ground state by the operator $\sg_{1}^{\ga}$. 

 \item  The critical exponents 
\beq
\De_{\ups,\ell} \, = \,  \ell Z(q) - \frac{\ups \op{s}_{\ga} }{ 2 Z(q) } \;, \quad \ups=\pm \; , 
\label{ecriture explicite exposant Delta ups ell dans regime conforme}
\enq
are expressed only in terms of the value of the dressed charge Z at the right endpoint $q$ of the Fermi zone, \textit{c.f.} \eqref{definition dressed charge}.

\item The amplitudes $\mc{F}_{ \ell }^{(\ga)}$ are given by the properly normalised in the volume thermodynamic limit of the form factor squared of the operator $\sg_{1}^{\ga}$
taken between the lowest lying  Umklapp excited state associated with a macroscopic momentum $2\ell p_F$ and the model's ground state.

\end{itemize}

\subsubsection*{ $\bullet$ The genuine asymptotic regime: $\op{v}_{\infty} > |\op{v}|$ }

In order to gather the space-like and the time-like regimes in one formula, it is convenient to introduce the parameter
\beq
\varkappa_{\op{v}} \; = \; \left\{ \ba{cc}   1  &  \op{v}_{\infty} > |\op{v}| > \op{v}_F \\ 
					   - 1  &  \op{v}_{F} > |\op{v}|  \ea \right.  \;. 
\enq

These regimes contain new types of power-law behaviour in $m$ stemming from the appearance of Gaussian saddle-points in the spectrum of the massive excitations. 
The asymptotics take the form 
\beq
\big< \big(\sg_1^{\ga} (t)\big)^{\dagger} \sg_{m+1}^{\ga}(0) \big>_{\e{c}} \; = \;   (-1)^{m \op{s}_{\ga} } \sul{ \bs{n} \in \mf{S}_{\op{v}} }{} 
\, \op{C}_{\bs{n}} \cdot \mc{F}_{ \bs{n} }^{(\ga)} \cdot  \pl{\ups=\pm}{} \Bigg\{  \f{  \ex{ \i  \ups \ell_{\ups} m   p_{F} }  }{    [ -\i (\ups m - \op{v}_F t)  ]^{  \De_{\ups,\bs{n}}^2   } } \Bigg\}
\cdot \f{ \ex{\i m \vp_{\bs{n}}(\op{v}) } }{  m^{\De_{\e{sp};\bs{n}}}  }
\cdot \Bigg\{ 1 \; + \; \e{O}\bigg( \,\f{1}{  m^{1-0^+} }  \bigg)  \Bigg\}
\enq

\begin{itemize}

 \item $\op{C}_{\bs{n}}$ represents the universal part of the amplitude. It is expressed in terms of the Barnes $G$ function \cite{BarnesDoubleGaFctn1} as 
\beq
\op{C}_{\bs{n}} \; = \;   \f{G(1+n_0)}{ (2\pi)^{ \frac{n_0}{2} } } \cdot 
\bigg(  \f{  \i \varkappa_{\op{v}} }{ p^{\prime\prime}_1(\om_0) - \frac{1}{ \op{v} } \veps_1^{\prime\prime}(\om_0)  } \bigg)^{ \frac{1}{2}n_0^2 }    \cdot 
\pl{r \in \mf{N}  }{} \Bigg\{  
  \f{ G(1+n_r) \cdot \Big\{ \e{sgn}\big[ p^{\prime}_r(\om_r)\big]  \Big\}^{n_r}  }{ (2\pi)^{ \frac{n_r}{2} }\cdot  \Big(-\i \big[p^{\prime\prime}_r(\om_r) - \frac{1}{ \op{v} } \veps_r^{\prime\prime}(\om_r) \big]\Big)^{ \frac{1}{2}n_r^2 } }      \Bigg\} \;. 
\enq

 \item The summation runs through all possible choices of vectors of integers belonging to the set 
\beq
\mf{S}_{\op{v}} \; = \; \bigg\{ \bs{n}=(\ell_+, \ell_-; n_0, n_{r_1},n_{r_2}, \dots  ) \; \, : \,  \; n_r \in \mathbb{N}\; , \;  \ell_{\pm} \in \mathbb{Z}  \quad \e{and} \quad 
\op{s}_{\ga}= \sul{\ups=\pm}{} \ell_{\ups} \; + \; \varkappa_{\op{v}} n_0 + \sul{r \in \mf{N} }{} r n_r  \bigg\} \;. 
\enq

\item  $\ups \ell_{\ups}  p_{F}$ corresponds to the macroscopic momentum carried out by a Umklapp excitation on the $\ups q$ Fermi boundary with a deficiency $\ell_{\ups}$.
$\De_{\ups,\bs{n}}$ is the critical exponent carried by  this branch of excitations. It takes the explicit form 
\beq
\De_{\ups,\bs{n}}\; = \; - \, \ups \ell_{\ups} \, + \, \tfrac{ 1 }{ 2 } \op{s}_{\ga}  Z(  q ) \, - \varkappa_{\op{v}} n_0  \phi_1( \ups q , \om_{0} )   
\; - \; \sul{  r \in \mf{N}  }{  }n_r  \phi_{r}(\ups q , \om_r ) \; -\sul{ \ups^{\prime} \in \{ \pm \} }{}\ell_{ \ups^{\prime} }  \phi_1(\ups q , \ups^{\prime} q \, )  \;, 
\enq
where $Z$ is the dressed charge \eqref{definition dressed charge} and  $\phi_{r}$ is the dressed phase \eqref{definition dressed phase} associated with an $r$-string excitation.
The critical exponent $\De_{\ups,\bs{n}}$ depends also on the location of the massive saddle-points, and thus on the ratio $\op{v}$ of $m$ and $t$.

 \item $ \vp_{\bs{n}}(\op{v}) $ is the oscillatory phase which takes its origin in the contributions of the massive excitations saddle-points to  the asymptotics. It takes the explicit form 
\beq
\vp_{\bs{n}}(\op{v}) \; = \; \varkappa_{\op{v}} n_0 u_{1}(\om_0,\op{v}) \; + \;  \sul{ r \in \mf{N}  }{} n_r u_{r}(\om_r,\op{v}) \;. 
\enq
Finally, $\De_{\e{sp};\bs{n}}$ captures the contribution of the massive modes to the power-law decay. It reads
\beq
\De_{\e{sp};\bs{n}} \; = \; \f{1}{2} \sul{ r \in \mf{N}\cup \{0\} }{} n_r^2 \;.  
\enq

 \item Last but not least, the amplitude $ \mc{F}_{ \bs{n} }^{(\ga)}$ represents the non-universal part of the asymptotics, and corresponds to a properly normalised in the volume
 thermodynamic limit of the form factor squared of the $\sg_{1}^{\ga}$ operator 
taken between the model's ground state and the lowest lying  excited state having left/right Umklapp deficiencies $\ell_{-/+}$
and,  in the thermodynamic limit, giving rise to hole, particle and $r$-string excitations subordinate to the below distribution of rapidities
\beq
\mf{Y}^{(\e{sp})}_{\bs{n}} \; = \; \left\{ \ba{cc}  
\bigg\{ -\op{s}_{\ga} \, ;\,  \big\{ \ell_{\ups} \big\} \; \big| \; \emptyset \, ; \,   \Big\{ \big\{ \om_0 \big\}^{n_0} \cup   \big\{ \om_1 \big\}^{n_1}   \Big\}  ;  \big\{ \om_{r_2} \big\}^{n_{r_2}} ; \dots    \bigg\}  
&  \qquad  \op{v}_F < |\op{v}| < \op{v}_{\infty}   \vspace{3mm} \\
\bigg\{  -\op{s}_{\ga} \, ;\,   \big\{ \ell_{\ups} \big\}\; \big| \; \big\{ \om_0 \big\}^{n_0}  ;    \big\{ \om_1 \big\}^{n_1} ;  \big\{ \om_{r_2} \big\}^{n_{r_2}} ; \dots     \bigg\}   
&  \qquad     |\op{v}| < \op{v}_{F}   \ea \right.  \;. 
\label{ecriture saddle point configuration of roots}
\enq

\end{itemize}

Note that, in \eqref{ecriture saddle point configuration of roots}, the notation $\Big\{ \big\{ \om_0 \big\}^{n_0} \cup   \big\{ \om_1 \big\}^{n_1}   \Big\}$ associated with the regime $\op{v}_F < |\op{v}| < \op{v}_{\infty}$
refers to the set of particle rapidities built up from $n_0$ particles with rapidity $\om_0$ and $n_1$ particles with rapidity $\om_1$. There are no-hole excitations in that regime.

 \subsection{The massless form factor series}
 \label{Sous section serie de facteurs de formes}
 
The thermodynamic limit of the form factor series obtained in \cite{KozMasslessFFSeriesXXZ} takes the form: 
\beq
\big< \sg_1^{\ga^{\prime}}\!(t)\,  \sg_{m+1}^{\ga}(0) \big> \; = \;   (-1)^{m \op{s}_{\ga} }   \sul{  \bs{n}\in \mf{S}   }{}   \mc{S}_{\bs{n}}(m,t)\;. 
\label{ecriture serie FF massless sg ga sg ga dagger}
\enq
There, the summation runs through all the possible integers  $\bs{n}\in \mf{S}$, 
with $\mf{S}$ as defined in \eqref{ecriture range summation dans series 2pt dynamic}, associated with the various  excitations in the model. 
The integer $\op{s}_{\ga}$ arising in the oscillatory pre-factor is as introduced in \eqref{definition spin operateur}. 
Finally, $\mc{S}_{\bs{n}}(m,t)$ represents the contribution to the form factor expansion of the sector whose excitation integers are gathered in the coordinates of $\bs{n}$. 
It is expressed in terms of a multiple integral as 
\beq
\mc{S}_{\bs{n}}(m,t) \; = \;  \pl{ r \in  \mf{N} }{} \;\Bigg\{  \Int{ \big( \msc{C}_r \big)^{n_r} }{} \hspace{-3mm} \f{ \dd^{n_r}\nu^{(r)} }{ n_r! \cdot (2\pi)^{n_r}  }   \Bigg\} 
\cdot \hspace{-3mm} \Int{ \big( \msc{C}_{h} \big)^{n_h}   }{} \hspace{-3mm} \f{ \dd^{n_h}\mu  }{ n_h! \cdot (2\pi)^{n_h}  } \cdot
  \f{   \mc{F}^{(\ga)}\big( \mf{Y} \big) \cdot    \ex{\i     m \msc{U}(\mf{Y},\op{v})  }    }{  \pl{\ups=\pm}{}  \big[  -  \i (\ups m - \op{v}_F t)   \big]^{   \vth_{\ups}^2(\mf{Y}) }  }  
\cdot  \bigg( 1+ \mf{r}_{\de,m, t}\big( \mf{Y} \big) \bigg)  \;.   
%
%
%
\label{ecriture serie FF limite thermo pour fct 2 pts}
\enq
The above form factor expansion is valid everywhere on the $(m,t)$ plane with the exception of the rays 
\beq
\op{v}=\f{m}{t} \not= \pm \op{v}_F   
\label{definition parametre vitesse} 
\enq
and the point $(m,t)=(0,0)$. There are several building blocks to formula \eqref{ecriture serie FF limite thermo pour fct 2 pts}. 

\begin{figure}[ht]
\begin{center}

\includegraphics{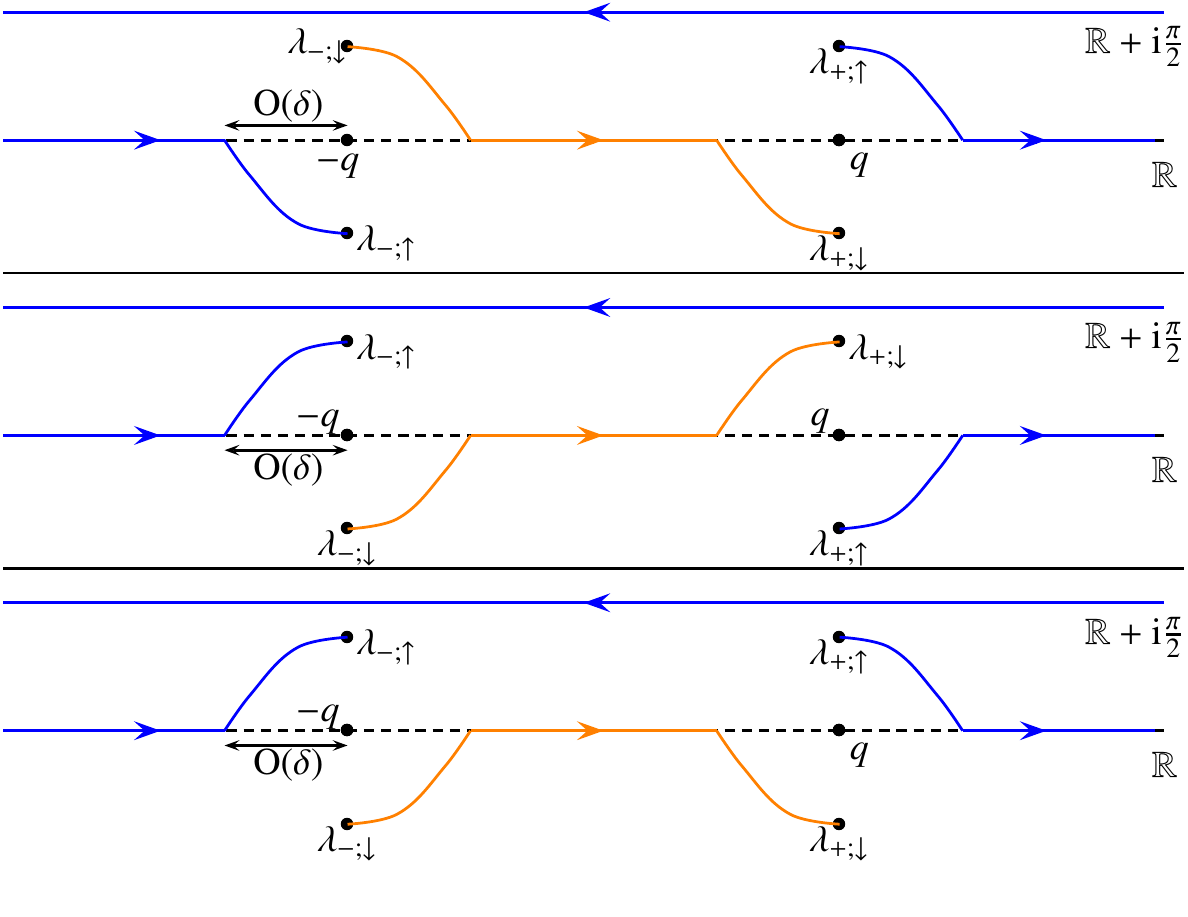}
\caption{  Particle $\msc{C}_{p}$ -in blue- and hole $\msc{C}_{h}$ -in orange- contours in the vicinity of $\R$. The contours are plotted for the three regimes of the velocity 
$\op{v}=\tf{m}{t}$ appearing from bottom to top $|\op{v}|>\op{v}_F$, $\op{v}_F>\op{v}>0$ and $0>\op{v}>-\op{v}_F$. The contour $\msc{C}_{p}$ and $\msc{C}_{h}$
start at the points $\la_{\pm;\ua/\da}=\pm q + \e{O}(\de)$ and then, over a distance of the order of $\de$,  joint with the real axis. 
The points $\la_{\pm;\ua/\da}$ are defined as the unique solutions in the neighbourhood of $\pm q$ of the equations $u_1(\la_{\pm;\ua},\op{v})=\i\de + u_1( \pm  q ,\op{v})$
and  $u_1(\la_{\pm;\da},\op{v})= - \i\de + u_1(\pm q ,\op{v})$.
\label{Figure contour des particules et trous delta deformes} }
\end{center}

\end{figure}

The integration runs through the curves $\msc{C}_{h}$ for the hole rapidities, and  $\msc{C}_r$ for the $r$-string rapidities. 
The hole $\msc{C}_{h}$ and the particle $\msc{C}_{1}$ contours are depicted in Figure \ref{Figure contour des particules et trous delta deformes}.
For $r \geq 2$, the $r$-string integration curves $\msc{C}_r$ coincide with 
$\varrho_{r}\R+\i \sg_{r} \tfrac{\pi}{2}$ where 
\beq
\varrho_{r}=\e{sgn}[{p^{\prime}_r}_{\mid \R+\i \sg_{r} \tfrac{\pi}{2} } ]\, = \, (-1)^{\sg_r} \mf{s}_r \,  \qquad \mf{s}_r=\e{sgn}\big[\sin(r\zeta) \big] \,,
\enq
so that it is oriented
in the direction of increasing momenta\symbolfootnote[2]{Here and in the following, I agree that given an oriented curve $\msc{C}$, $\sg \msc{C}$ with $\sg\in \{\pm 1\}$
stands for the curve $\msc{C}$ is $\sg=+1$ and the curve $\msc{C}$ endowed with the opposite orientation if $\sg=-1$.}, \textit{c.f.} Figure \ref{Figure contour integration r cordes} and Section  \ref{Appendix Sous-section cordes}
of the Appendix. Also, $\sg_r$ is as introduced in \eqref{ecriture domaine evolution des diverses rapidites particules et cordes}.

There appear two functions in the integrand.  $\mc{F}^{(\ga)}\big( \mf{Y} \big)$, corresponds to the form factor density squared of the operator $\sg_1^{\ga}$ taken between the ground state and an excited state 
whose massive excitations are parameterised by $\mf{Y}$.  It will be discussed in Subsection \ref{SousSection FF gd L} to come. 
The exponent $\vth_{\ups}^2(\mf{Y})$ corresponds to the local contribution to the full critical exponent of the Fermi boundary $\ups q$.   It is defined as 
\beq
\vth_{\ups}(\mf{Y}) \, = \, \vth( \ups q\mid \mf{Y}) \, - \, \ups \ell_{\ups}  \;,  
\label{definition exponsant critique ell shifte}
\enq
where the function $\vth ( \om \mid \mf{Y} )$ is the opposite of the shift function associated with the massive excitation $\mf{Y}$ and  
takes the explicit form 
\beq
\vth ( \om \mid \, \mf{Y} ) \, = \,  \sul{ a=1 }{ n_h } \phi_1( \om, \mu_{a} )   
\, + \, \tfrac{ 1 }{ 2 } \op{s}_{\ga}  Z( \om ) \; - \; \sul{  r \in \mf{N}  }{  } \sul{ a=1 }{ n_r } \phi_{r}(\om,\nu_a^{(r)} )
\; - \hspace{-1mm} \sul{ \ups^{\prime} \in \{ \pm \} }{}\ell_{ \ups^{\prime} }  \phi_1(\om, \ups^{\prime} q \, )  \;. 
\label{definition phase habilee totale excitation}
\enq
$\vth$ is expressed in terms of the dressed charge $Z$, \textit{c.f.}  \eqref{definition dressed charge},
and of the dressed phase  $\phi_r$ of an $r$-string, \textit{c.f.}  \eqref{definition dressed phase}.

It is important to stress that $\mc{S}_{\bs{n}}(m,t)$ depends on an auxiliary parameter $\de>0$ which is arbitrary and can be taken as small as necessary. 
The parameter $\de$ does not appear in the model and solely plays the role of an \textit{ad hoc} regulator. 
Physically,  this parameter plays the role of a separating scale between the massless and massive particle-hole excitations. 
The $\de$ dependence manifests itself on the level of the remainder $ \mf{r}_{\de,m, t}\big( \mf{Y} \big)$  and in the way the integration curves  $\msc{C}_{1}$ and 
$\msc{C}_{h}$ are deformed in the vicinity of the endpoints $ \pm q$ of the Fermi zone (\textit{c.f.} Figure \ref{Figure contour des particules et trous delta deformes}). While the two-point function, as a whole, does not depend on this parameter, 
taking the $\de \tend 0^+$ limit on the level of each multiple integral is delicate due to the presence of numerous singularities (see \eqref{ecriture comportement local FF smooth a rapidite coincidantes}-\eqref{ecriture fonction singuliere D}). 
However, the freedom of choosing $\de$ small enough -even in a $m$ and $t$ dependent way- is sufficient for  
carrying out the analysis of the large $(m,t)$ behaviour of the series. 
The remainder is controlled as 
\beq
\mf{r}_{\de,m, t} \big( \mf{Y} \big) \; = \; \e{O}\bigg( \de |\ln \de|\,  \, + \, \sul{\ups=\pm }{} \big\{ \de^{2} |m_{\ups}| \,+ \, \de  \big| \ln |m_{\ups}| \big| \, + \, \ex{-\de |m_{\ups}|} \big\}  \bigg) \;, 
\quad \e{where} \quad m_{\ups}=\ups m -\op{v}_F t \;. 
\label{ecriture du controle sur le reste}
\enq
\begin{figure}[ht]
\begin{center}

\includegraphics{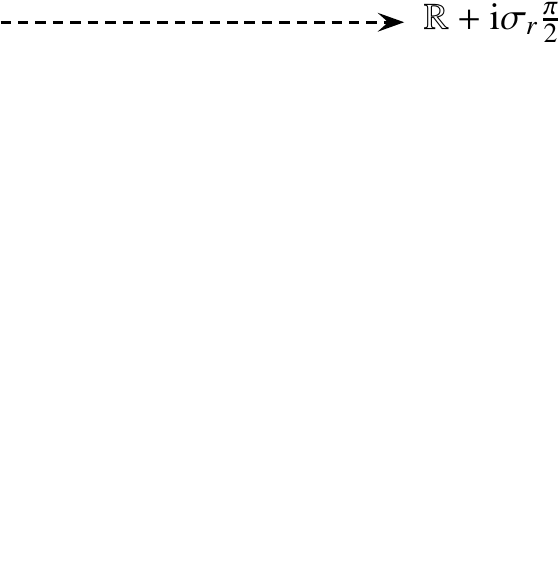}

\caption{ Contour of integration for the $r$-string rapidities: $\msc{C}_{r}= \varrho_r \R+\i\sg_r \tfrac{\pi}{2}$. Here $\sg_r = 0$ or $1$ is the string parity and depends on $r$. 
Moreover, the curve  $\msc{C}_{r}$ is oriented in the direction of increasing momenta of the $r$-strings, \textit{viz}.  $\varrho_r=\e{sgn}\big[p_r^{\prime} \big]_{ \mid \R+\i\sg_r \tf{\pi}{2}  }$.  \label{Figure contour integration r cordes} }
\end{center}

\end{figure}

\subsection{The form factors density}
\label{SousSection FF gd L}

The explicit expression for the form factor density $ \mc{F}^{(\ga)}\big( \mf{Y} \big)$ can be found in \cite{KozProofOfAsymptoticsofFormFactorsXXZBoundStates}. 
Since it is rather bulky, I do not reproduce it here. 
In fact, only a reduced number of properties of   $ \mc{F}^{(\ga)}\big( \mf{Y} \big)$ matters for the $(m,t) \tend \infty$ analysis of the form factor series:

\begin{enumerate}[i)]
  
  \item   $ \mc{F}^{(\ga)}\big( \mf{Y} \big)$  is an analytic function of the rapidities $\mf{Y}$, as long as each rapidity stays in a small vicinity of its respective integration curve $\msc{C}_h$ or $\msc{C}_r$.

    \item  The form factor density decays as $\ex{ \mp 2 r \nu_a^{(r)}}$ when the real part of a rapidity $\nu^{(r)}_a$ goes to $\pm \infty$.
  
   \item  $ \mc{F}^{(\ga)}\big( \mf{Y} \big)$   admits cuts along the lines 
\beq
\nu_a^{(1)} \in \intoo{-\infty}{-q} + \i \mf{f}_{ \zeta } \quad \e{and} \quad 
\nu_a^{(r)} \in \intoo{-\infty}{-q} + \i \mf{f}_{ \f{r \pm 1}{2}\zeta } \quad \e{for} \quad r \in \mf{N}_{\e{st}} \quad \e{and} \; modulo \; \i\pi \;. 
\label{ecriture des domaines de saut pour FF density}
\enq
Above,  $\mf{f}_{ \eta }=\e{min}\big( \eta- \pi \lfloor \tfrac{\eta}{\pi} \rfloor, \pi - \eta +  \pi \lfloor \tfrac{\eta}{\pi} \rfloor \big)$.  The associated jump discontinuities
take the form 
\beq
\mc{F}^{(\ga)}\big( \mf{Y}_{\pm;(b,s)}^{\ua} \big)  \;  = \; \mc{F}^{(\ga)}\big( \mf{Y}_{\pm;(b,s)}^{\da} \big) \;, 
\label{ecriture conditions saut FF}
\enq
in which the two sets or rapidities take the explicit form 
\beqa
\mf{Y}_{\sg;(b,s)}^{\ua} & = & \Big\{ -\op{s}_{\ga} \, ;\,  \big\{ \ell_{\ups}\big\} \; \big| \; 
\big\{ \mu_{a} \big\}_{1}^{ n_{h} } \, ; \,     \big\{ \nu_{a}^{(r_1)} \big\}_{a=1}^{n_{r_1}}  \, ; \,\dots  \, ; \,  \big\{ \nu_{a}^{(r_k)} \big\}_{a=1}^{n_{r_k}}  ; \, \dots  \Big\}_{ \nu^{(s)}_b=x+\i\mf{f}_{ \f{s +\sg }{2}\zeta } + \i 0^+ } \nonumber \\
\mf{Y}_{\sg;(b,s)}^{\da} & = & \Big\{ -\op{s}_{\ga}; \big\{ \ell_{\ups}+\ups u_s^{\sg} \big\}  \; \big| \;
\big\{ \mu_{a} \big\}_{1}^{ n_{h} } \, ; \, \big\{ \nu_{a}^{(r_1)} \big\}_{a=1}^{n_{r_1}}  \, ; \,\dots  \, ; \,\big\{ \nu_{a}^{(r_k)} \big\}_{a=1}^{n_{r_k}}  ; \, \dots 
 \Big\}_{ \nu^{(s)}_b=x+\i\mf{f}_{ \f{s +\sg }{2}\zeta } - \i 0^+ } 
\label{ecriture variables pour saut FF}
\eeqa
and one has $x<-q$.  In this notation, the rapidities in $\mf{Y}_{\sg;(b,s)}^{\ua/\da}$ are assumed to be in generic positions with the exception of $\nu^{(s)}_b$ which ought to be specialised as stated. 
Above, $\sg \in \{ \pm 1\}$ if $s\geq 2$ and $\sg =1$ if $s=1$. Furthermore, one has 
\beq 
u_r^{\sg} \, = \,  -\e{sgn}\Big( \pi + 2\pi \lfloor \tfrac{r+ \sg}{2\pi}\zeta \rfloor - (r+ \sg)\zeta \Big)\cdot \big(1-\de_{\sg,-}\de_{r,1}\big) \;.
\label{definition parametre u r sigma}
\enq

  \item  The form factor density has at least a double zero when two rapidities of the same type (\textit{i.e.} corresponding to a particle, hole or $r$-string excitations) coincide.

    \item The form factor density exhibits  power-law singularities  at $\pm q$ in respect to the particle or hole rapidities $\nu_{a}^{(1)}$ and $\mu_a$.

  \item The form factor density admits dynamical poles at shifted rapidities which take the form 
\beq
\mc{F}^{(\ga)}\big( \mf{Y} \big)  \;  = \;   \pl{   s,r \in \mf{N}    }{ }    \underset{  (s,a)\not= (r,b) } { \pl{ a=1 }{ n_s }   \pl{ b=1 }{ n_r } }\Phi_{r,s}\big( \nu_a^{(s)} - \nu_b^{(r)}\big)   \cdot 
\mc{F}^{(\ga)}_{\e{hol}}\big( \mf{Y}  \big). 
\label{factorisation sur poles de la densite de FF}
\enq
The set function $\mc{F}^{(\ga)}_{\e{hol}}\big( \mf{Y}  \big)$ is holomorphic in the neighbourhood of rapidities going to $\infty$  and 
\beq
\Phi_{r,p}\big( \la \big) \, = \,   
\pl{\ell = \lfloor \tfrac{r-p+1}{2} \rfloor }{r-1} \Bigg\{ \f{  \sinh\Big(\la +\i\zeta\big[ \tfrac{p-r}{2}+\ell \big] \Big)    }{ \sinh\Big(  \la - \i\zeta \big[\tfrac{p-r}{2} +\ell+1\big] \Big)   }     \Bigg\}^{ w_{\ell}^{(r,p)}- w_{\ell+1}^{(r,p)}} 
\enq
where $w_p^{(r,s)}\;= \; \min(r,s+p)-\max(0,p)$. Again, since for any given excited state $n_r=0$ for $r$  large enough,  
the product in \eqref{factorisation sur poles de la densite de FF} runs, effectively, over a finite set.

\item The form factor density associated with an excited state containing various $r$-string excitations can be recovered from the form factors only involving particle and hole rapidities, upon computing an 
appropriate residue. Indeed, consider the sets  
\beqa
 \mf{Y} & = & \Bigg\{ -\op{s}_{\ga}\, ; \, \{\ell_{\ups} \} \; \big| \; \{\mu_a \}_1^{n_h} \;; \; 
 \bigg\{   \big\{ \nu_a^{(1)} \big\}_{a=1}^{n_1}    \cup    \Big\{  \;   \Big\{ \;  \big\{ \nu_{a;k}^{(r)} \big\}_{k=1}^{r} \, \Big\}_{a=1}^{n_r} \Big\}_{r \in \mf{N}_{\e{st}} } \bigg\} \, ; \, 
   \emptyset    \, ; \, \dots   \Bigg\}  \, ,  \nonumber \\
 \mf{Y}_{\e{red}} & = & \Bigg\{ -\op{s}_{\ga}\, ; \, \{\ell_{\ups} \} \; \big| \; \{\mu_a \}_1^{n_h} \; ; \;   \big\{ \nu_a^{(1)} \big\}_{a=1}^{n_1}   \, ; \,     
 \bigg\{      \Big\{ \;    \nu_{a;1}^{(r_2)}  - \i \zeta \tfrac{r_2-1}{2} \, \Big\}_{a=1}^{n_{r_2}} \bigg\}  \, ; \, \dots \Bigg\}  \;. 
\nonumber 
\eeqa
Then, it holds
\bem
\e{Res}\left( \rule{0cm}{0.7cm} \right. \mc{F}^{(\ga)}_{\e{tot}}\big( \mf{Y} \big) 
 \pl{r \in \mf{N}_{\e{st}} }{} \pl{ a = 1 }{ n_r } \pl{ k = 2 }{ r }   \dd \nu_{a;k}^{(r)} \;  , \;
\bigg\{ \Big\{ \nu_{a;k}^{(r)}=\nu_{a,k-1}^{(r)} \, - \, \i \zeta \, ; \, k=2,\dots, r  \Big\}_{a=1}^{ n_r } \bigg\}_{r \in \mf{N}_{\e{st}} }  \left. \rule{0cm}{0.7cm} \right)  \\
\; = \; \pl{ r \in \mf{N}_{\e{st}} }{} \Big\{  \big( -\i \big)^{ (r-1) n_r } \Big\}  \;  \cdot  \; 
 \mc{F}^{(\ga)}_{\e{tot}}\big( \mf{Y}_{\e{red}} \big)  \;, 
\label{ecriture reduction dans equation sur residus}
\end{multline}
where 
\beq
 \mc{F}^{(\ga)}_{\e{tot}}\big( \mf{Y} \big) \, = \, 
 \f{ \mc{F}^{(\ga)}\big( \mf{Y} \big) \cdot  \ex{\i m \msc{U}\big( \mf{Y} , \op{v} \big) }  }{   \pl{\ups=\pm}{}  \big[  -  \i (\ups m - \op{v}_F t)   \big]^{   \vth_{\ups}^2(\mf{Y}) }   } 
\cdot \Big(1 \, + \, r_{\de,m,t}\big( \mf{Y} \big) \Big) \;. 
\enq
Note that  the individual, one-dimensional, residues in respect to the variables  $\nu_{a;k}^{(r)}, k=2,\dots, r$ can be taken in any order.

\end{enumerate}
The properties iv)-v) can be written up explicitly as
\beq
\mc{F}^{(\ga)}\big( \mf{Y} \big) \; = \;  \pl{r \in \mf{N}_{\e{st}} }{}   \pl{a\not= b}{n_r} \sinh\Big(\nu_{a}^{(r)} - \nu_{b}^{(r)} \Big)
\cdot  D_{n_1;n_h}\Big( \big\{  \nu_{a}^{(1)} \big\}_1^{n_1} \, ; \,  \big\{  \mu_{a} \big\}_1^{n_h}  \Big) \cdot  \mc{F}^{(\ga)}_{\e{reg}}\big( \mf{Y} \big) \;. 
\label{ecriture comportement local FF smooth a rapidite coincidantes}
\enq
In the above decomposition, the Vandermonde determinants in the variables $\nu^{(r)}$ catch the double zero vanishing relatively to the contributions of the $r$-string modes, $r \in \mf{N}_{\e{st}}$, 
while the prefactor $D_{n_1;n_h}$ given below gathers all the singularities and zeroes that exist relatively to the particle and hole rapidities:
\bem
 D_{n_1;n_h}\Big( \big\{  \nu_{a}^{(1)} \big\}_1^{n_1} \, ; \,  \big\{  \mu_{a} \big\}_1^{n_h}  \Big)  \; = \; 
 \pl{a=1}{n_1} \bigg( \f{ \sinh[\nu^{(1)}_{a} + q] }{  \sinh[\nu_{a}^{(1)} - q]   }  \bigg)^{2 \vth (\nu_a^{(1)} \mid \, \mf{Y} )  }  \cdot 
\pl{a=1}{n_h} \bigg( \f{ \sinh[\mu_{a} - q] }{   \sinh[\mu_{a} + q]   }  \bigg)^{2 \vth (\mu_a  \mid \,  \mf{Y})  }  \\
\times \pl{\ups = \pm }{} \Bigg\{  \f{ \pl{a=1}{n_1} \sinh[\nu_a^{(1)}-\ups q ]  }{   \pl{a=1}{n_h} \sinh[\mu_a-\ups q ] }  \Bigg\}^{ 2 \ell_{\ups} } \cdot   
\f{   \pl{a \not= b}{n_h} \sinh\big[ \mu_{a} - \mu_{b} \big] \cdot \pl{a \not= b}{n_1} \sinh\big[  \nu^{(1)}_{a} -  \nu^{(1)}_{b} \big]   }{ \pl{a=1}{n_1}\pl{b=1}{n_h} \sinh^2\big[  \nu^{(1)}_a-\mu_b \big] } \;. 
\label{ecriture fonction singuliere D}
\end{multline}

The decomposition  \eqref{ecriture comportement local FF smooth a rapidite coincidantes}  is such that, for a generic collection of parameters, the function $ \mc{F}^{(\ga)}_{\e{reg}}\big( \mf{Y} \big)$
does not vanish when some rapidities coincide or when some of the particle and hole rapidities approach the endpoints $\pm q$ of the Fermi zone. 
Note also that the singularities present at the Fermi endpoints $\pm q$ in \eqref{ecriture comportement local FF smooth a rapidite coincidantes} that appear in 
the $D_{n_1;n_h}$ factor are reminiscent of the infrared divergences that where absorbed in \cite{KozMasslessFFSeriesXXZ} by carefully taking the thermodynamic limit of the series. 
Since the integration contours are uniformly away from $\pm q$ with a minimal distance controlled by $\de$, these are not, \textit{per-se} singularities of the series. 
However, they start to play a role should one want to take the $\de \tend 0^+$ limit of the series.

\vspace{3mm}

Note that the system of cuts enjoyed by the form factor density, property iii)  appears as a fundamental content of the model which, in fact, plays an important role in the 
large $(m,t)$ behaviour of the model's correlation function. The jump conditions on the various cuts discussed in iii), are also shared by 
the other observables in the model: the left/right Fermi endpoint critical exponent functions $\vth_{\ups}(\mf{Y})$ introduced in \eqref{definition exponsant critique ell shifte} 
and the exponents of the combination of dressed momenta and energies of the excited state $\msc{U}(\mf{Y},\op{v})$, \textit{c.f.} \eqref{definition de energi impuslion reduite condensation massless}. 
Namely, it holds
\beq
\ex{\i   \msc{U}\big(\mf{Y}_{\pm,(b,s)}^{\ua} ,\op{v}\big) }  \, = \, \ex{\i  \msc{U}\big( \mf{Y}_{\pm,(b,s)}^{\da} ,\op{v}\big) }   \qquad \e{and} \qquad \vth_{\ups}\big( \mf{Y}_{\pm,(b,s)}^{\ua} \big)
\, = \, \vth_{\ups}\big( \mf{Y}_{\pm,(b,s)}^{\da} \big) \;. 
\label{ecriture condition saut coupure U et vartheta}
\enq
 Finally,  the way the remainder is constructed in \cite{KozMasslessFFSeriesXXZ} also indicates that 
\beq
\mf{r}_{\de,m, t}\big(\mf{Y}_{\pm,(b,s)}^{\ua} ,\op{v}\big) \; = \; \mf{r}_{\de,m, t}\big(\mf{Y}_{\pm,(b,s)}^{\da} ,\op{v}\big) \;. 
\label{ecriture condition saut coupure reste}
\enq

\section{Some preliminaries to the steepest descent analysis}
\label{Section Auxiliary rewriting of the FF series}

The two-point dynamical correlation function is expressed through the massless form factor series expansion \eqref{ecriture serie FF massless sg ga sg ga dagger}. 
The very structure of the building blocks $\mc{S}_{\bs{n}}(m,t)$ \eqref{ecriture serie FF limite thermo pour fct 2 pts} suggests that the large $(m,t)$ behaviour
can be extracted by means of a steepest descend analysis. However, setting the latter in place demands some preliminary results that will be established in the present section. 

 First of all, the contour deformations arising in the steepest descent analysis are dictated by the properties of the oscillating phase $\msc{U}(\mf{Y},\op{v})$
introduced in \eqref{definition de energi impuslion reduite condensation massless}. Since the latter is given as a sum over the contributions $u_r(\nu_{a}^{(r)},\op{v})$, \textit{c.f.} \eqref{definition fct ur},  of each individual rapidity
building up an excited state, the problem reduces, effectively, to   one dimensional considerations.  
More precisely, in order to set up the appropriate contour deformation, one needs to identify the domains in the strip of width $\tf{\pi}{2}$ around the real axis 
where $\Im\big[ u_r(\la,\op{v}) \big] \, \geq \, 0$. These will correspond to the regions where the saddle-point contour deformation should go. 
Also, one should localise the saddle-points, \textit{viz}. the solutions to $u_r^{\prime}(\la,\op{v})=0$.
All these pieces of information are gathered in Subsection \ref{Sous Section preprietes pour locus saddel pts}

Second, the integrand has poles in the complex plane, \textit{c.f.} \eqref{factorisation sur poles de la densite de FF}. These have to be taken into account 
when deforming the contours. The form factor series has thus to be put in a form that allows one for a convenient computation of the effect of such poles. 
This will be carried out in Subsection \ref{Sous Section reecriture de la series de FF}.

 \subsection{Properties of the oscillating phase-factor} 
\label{Sous Section preprietes pour locus saddel pts}

Sub-section \ref{Appendix Section phase oscillante dpdte de la vitesse} provides a detailed analysis of the properties of the function $u_r(\la,\op{v})$ depending on the value of 
the distance to time  ratio $\op{v} = \tf{ m }{ t }$. Below, we provide a summary, with an emphasis of the behaviour that influences the locuum of the steepest descend paths. 

\vspace{2mm}

$\blacklozenge $ The sign of  $  \Im\big[ u_{r}(\la ,\op{v}) \big]  $

\vspace{2mm}

The first observation is that, irrespectively of the value of $r$:
\beq
u^{\prime}_r(\la,\op{v}) \; = \; \e{O}\Big( \ex{-2 |\Re(\la)| } \Big) \quad \e{when} \quad  \Re(\la) \tend \pm \infty \; .
\enq
 A precise analysis of this asymptotic behaviour shows that, 
for $\la=x+\i y $ with $\pm x >A$,  $A$ being large enough and $r$-independent, it holds 
\beq
\ba{|c|c|c|c|}
\hline
							& |\op{v}| \, > \,  \op{v}_{\infty}              &    0 \, < \, \op{v}  \,  <  \, \op{v}_{\infty} &  -\op{v}_{\infty} \, <  \, \op{v}  \, < \,  0 \\ \hline
\e{sgn}\Big[ \Im\Big( u_{r}(\la ,\op{v}) \Big)\Big] \rule{0cm}{0.5cm}   & \e{sgn}\big[ \sin(r\zeta) \sin(2y) \big]	&   \mp \e{sgn}\big[ \sin(r\zeta) \sin(2y) \big]   & \pm \e{sgn}\big[ \sin(r\zeta) \sin(2y) \big] \\
\hline
\ea
\label{tableau du signe de ur dans le cas de corede generales}
\enq

The integration curves $\msc{C}_r$ go to infinity  along  $\R+\i\sg_r\tfrac{\pi}{2}$, in what concerns the genuine bound states, 
\textit{viz}. the $r$-strings with $r \geq 2$, and $\msc{C}_1$ goes to infinity along $\R$ and $\R+\i\tfrac{\pi}{2}$. 
Since the phase factors $u_{r}(\la,\op{v})$ are bounded at $\infty$, the neighbourhood of infinity could provide some contributions to 
the power-law behaviour in $(m,t)$ of the two-point function. Due to the opposite orientation on $\R$ and $\R+\i\tfrac{\pi}{2}$ of the particle 
contour $\msc{C}_1$ (\textit{c.f.} Figure \ref{Figure contour des particules et trous delta deformes}),
one may deform the part of $\msc{C}_1$ in the vicinity of $\infty$ so that, up to a contribution of the resiudes of the poles 
that are crossed in the procedure, the deformed contour reduces to some compact curve located in the strip $-\tfrac{\pi}{2} \leq \Im(\la) \leq  \tfrac{\pi}{2}$. 
 Note that the $\i\pi$-periodicity of the integrand allows one to switch some portion of the integration on $\R+\i\tfrac{\pi}{2}$
with one over $\R- \i \tfrac{\pi}{2}$, hence allowing for a deformation of $\msc{C}_1$ that would be located in the strip $-\tfrac{\pi}{2} \leq  \Im(\la) <0$. 

For $r=1$, the sign properties listed in \eqref{tableau du signe de ur dans le cas de corede generales} may be expressed in the more explicit array
\beq
\ba{|c|c|c|c|}
\hline
	\e{sgn}\Big[ \Im\Big( u_{1}(\la ,\op{v}) \Big)\Big]^{} >0	&    \Re(\la) \in \intoo{-\infty}{ -A }       &  x \in \intoo{ A }{  + \infty }      \\ \hline
|\op{v}| \, > \,  \op{v}_{\infty}                                &         0< \Im(\la) <\tfrac{\pi}{2}   &    0< \Im(\la) <\tfrac{\pi}{2}     \\ \hline 
   0 \, < \, \op{v}  \,  <  \, \op{v}_{\infty}                  &         0< \Im(\la) <\tfrac{\pi}{2}  &    -\tfrac{\pi}{2} < \Im(\la) <0      \\   \hline
 -\op{v}_{\infty} \, <  \, \op{v}  \, < \,  0                  &    -\tfrac{\pi}{2} < \Im(\la) <0  &         0< \Im(\la) <\tfrac{\pi}{2}    \\ \hline 
\ea
\label{tableau recapitulatif region positivite phase oscillante une particule}
\enq
Thus, in the regions of the complex plane characterised by a sufficiently large real part, the deformation of
the contour $\msc{C}_1$ should close in the regions described in array \eqref{tableau recapitulatif region positivite phase oscillante une particule}.

\vspace{2mm}

$\blacklozenge $  The saddle-points

\vspace{2mm}

It is shown in Sub-section \ref{Appendix Section phase oscillante dpdte de la vitesse} that, for fixed $r\in \mf{N}$, $u_r^{\prime}(\la,\op{v})$
\begin{itemize}
 
 \item will have an even number of zeroes on $\R+\i\tfrac{\pi}{2} \sg$, with $\sg\in \{0,1\}$, if $|\op{v}|>\op{v}_{\infty}$;
 
 \item will have an odd number of zeroes on $\R+\i\tfrac{\pi}{2} \sg$, with $\sg\in \{0,1\}$, if $-\op{v}_{\infty} < \op{v} <\op{v}_{\infty}$.

\end{itemize}
Clearly, these zeroes determine the steepest descent path that has to be chosen for the asymptotic analysis of the multiple integrals $\mc{S}_{\bs{n}}(m,t)$. 
It is thus important to have more information on these zeroes. It is shown in Lemma \ref{Lemme borne inf sur impuslion energie sur courbes excitations}
there exists
\begin{itemize}
 
 \item[i)] an upper threshold velocity $\op{v}_r^{(M)} \geq \op{v}_{\infty}$ such that   $u_r^{\prime}(\la,\op{v})$ does not vanish on $\big\{\R+\i\tfrac{\pi}{2} \big\} \cup \R $
for $|\op{v}|> \op{v}_{r}^{(M)}$.
 
 \item[ii)]  an lower threshold velocity $\op{v}_r^{(m)} \leq \op{v}_{\infty}$ such that $u_r^{\prime}(\la,\op{v})$ only vanishes once on $\big\{\R+\i\tfrac{\pi}{2} \big\} \cup \R $ 
for $|\op{v}|< \op{v}_{r}^{(m)}$.
 
\end{itemize}

Numerical analysis indicates that, in fact, for any $r\in \mf{N}_{\e{st}}$, 
one has $ \op{v}_r^{(m)} \, = \,  \op{v}_{\infty}$ independently of $\zeta$ while $\op{v}_r^{(M)}=\op{v}_{\infty}$ provided that $|\zeta-\tf{\pi}{2}|$ is small enough.
Furthermore, for $q$ small enough, it holds $\op{v}_F<\op{v}_{\infty}$ but, for general values of $\zeta$ and $q$ it may happen that $\op{v}_F>\op{v}_{\infty}$. Then, for this range of parameters, 
the following scenario appears to hold for $r \in \mf{N}$
\begin{itemize}
 
\item $u^{\prime}_r(\la,\op{v})$ \textit{does not vanish} on $\R+\i \sg \tfrac{\pi}{2}$ for $|\op{v}|>\op{v}_{\infty}$;
 
\item  $u^{\prime}_r(\la,\op{v})$ only vanishes \textit{once} on $\R+\i \sg \tfrac{\pi}{2}$ when $-\op{v}_{\infty} < \op{v} <\op{v}_{\infty}$;
\end{itemize}
where one should take $\sg=\sg_{r}$ if $r \not=1$, while $\sg=0$ or $1$ for $r=1$.

This will be called the "minimal structure property" in the following. It allows to slightly simplify the combinatorics arising in the asymptotic expansion 
as compared to the most general situation.

 \vspace{2mm}
  {\bf Minimal structure property}
 \vspace{2mm}
 
\noindent The minimal structure property holds if the parameters $q$ and $\zeta$ are tuned so that 

\vspace{1mm}

$\bullet$  $\op{v}_F < \op{v}_{\infty}$ ; 
\vspace{1mm}

$\bullet$ for any $\op{v}$ such that $|\op{v}|>\op{v}_{\infty}$ one has that $|u_r^{\prime}(\la,\op{v})| >0$ on $\msc{C}_r$;
 
\vspace{1mm}
$\bullet$  for $\op{v}$  such that $-\op{v}_{\infty} < \op{v} < \op{v}_{\infty}$:
\begin{itemize}
 
 \item[i)] given any $r \in \mf{N}_{\e{st}}$, there exists a unique $\om_r \in \msc{C}_r=  (-1)^{\sg_r}\mf{s}_r \R \, + \, \i \sg_r \tfrac{\pi}{2}$ such that $ u_r^{\prime}(\om_r,\op{v}) \, = \, 0$.  
 
 \item[ii)] when $r=1$, there exist a unique $\om_0$ in $\R$ and a unique $\om_1\in \R + \i \tfrac{\pi}{2}$ such that $ u_1^{\prime}(\om_0,\op{v}) \, = \,  u_1^{\prime}(\om_0,\op{v}) \, = \, 0$.

\end{itemize}

Since,    $u_r$ is holomorphic in an open neighbourhood of $\msc{C}_r$,  there exist $\eta>0$ and  biholomorphisms $h_r: \mc{D}_{0,2\eta} \mapsto \Cx$, $r\in \mf{N}\cup\{0\}$ such that 
\begin{itemize}
 \item $h_r(0)=0$ \;  and \;  $h^{\prime}_r(0) = \Big\{ \tfrac{1}{2} \cdot |u_r^{\prime\prime}(\om_r,\op{v})|   \Big\}^{ \tfrac{1}{2} } $ ; 
 \item for $r \in \mf{N}_{\e{st}}$, 
\beq
u_r(\la,\op{v})\, = \,  u_r(\om_r,\op{v}) \, + \, \veps_r \cdot \big[ h_r(\la-\om_r)\big]^2 \qquad \e{with} \qquad \veps_r=\e{sgn}\big[ u_r^{\prime\prime}(\om_r,\op{v}) \big]
\label{definition ur et vepsr r corde generale}
\enq
for $\la \in \om_r+ \mc{D}_{0,2\eta}$;
\item for $r=1$
\beq
u_1(\la,\op{v})\, = \,  u_1(\om_a,\op{v}) \, + \, \veps_a \cdot \big[ h_a(\la-\om_a)\big]^2 \qquad \e{with} \qquad \veps_a=\e{sgn}\big[ u_1^{\prime\prime}(\om_a,\op{v}) \big]\; ,  \quad a \in \{0,1\}
\label{definition u1 et veps10 particule}
\enq
this for $\la \in \om_a+ \mc{D}_{0,2\eta}$. 
 
\end{itemize}

However, if $|\zeta-\tf{\pi}{2}|$ is large enough, the structure of zeroes of $u^{\prime}(\la;\op{v})$ appears slightly more involved. 

 \vspace{2mm}
  {\bf Most general structure of the saddle-points}
 \vspace{2mm}

There exits a certain subset $\mf{N}_{\e{sp}}\subset  \mf{N}_{\e{st}}$ such that $\op{v}_{r}^{(M)}\not= \op{v}_{\infty}$ for $r \in \mf{N}_{\e{sp}}$. 
Then, for $r \in \mf{N}_{\e{st}}$, one has 
\begin{itemize}
 
\item $u^{\prime}_r(\la,\op{v})$ \textit{does not vanish} on $\R+\i \sg_r \tfrac{\pi}{2}$ for $|\op{v}|>\op{v}_{r}^{(M)}$;

\item $u^{\prime}_r(\la,\op{v})$ \textit{vanishes}  $\vsg_r$ times on $\R+\i \sg_r \tfrac{\pi}{2}$ for $ |\op{v}|<\op{v}_{r}^{(M)}$.

\end{itemize}
As for  $u_1$,  it holds that
\begin{itemize}
\item $u^{\prime}_1(\la,\op{v})$ \textit{does not vanish} on $\R \cup \{ \R+\i  \tfrac{\pi}{2} \}$ for $|\op{v}|>\op{v}_{1}^{(M)}$;
\item $u^{\prime}_1(\la,\op{v})$ \textit{vanishes} twice on $\R\cup \{ \R+\i  \tfrac{\pi}{2} \}$ for $   | \op{v} | \, < \, \op{v}_{1}^{(M)} $. The zeroes may be both situated on the same line or belong to different lines;

\item $u^{\prime}_1(\la,\op{v})$ has one zero $\om_0 \in \intoo{-q}{q}$ and one in $\R\setminus \intff{-q}{q} \cup \big\{ \R+\i  \tfrac{\pi}{2}\big\}$ for $|\op{v}|<\op{v}_{F}$. 
\end{itemize}

The integer $\vsg_r$ \textit{depends} on $\op{v}$. It corresponds to the total number of zeroes, counting multiplicities. 
In the generic case, the zeroes will all be simple, but for well tuned parameters of the problem (in particular when $\op{v}=\op{v}_r^{(M)}$)
the zero may be of higher order\symbolfootnote[2]{The case of higher order zeroes will not be considered further, although taking these into account does not pose any special problem}. 

Numerical investigations seem to indicate that, in fact, 
\begin{itemize}
 
\item $u^{\prime}_r(\la,\op{v})$ \textit{does not vanish} on $\R+\i \sg_r \tfrac{\pi}{2}$ for $|\op{v}|>\op{v}_{r}^{(M)}$;

\item $u^{\prime}_r(\la,\op{v})$ \textit{vanishes twice}  on $\R+\i \sg_r \tfrac{\pi}{2}$ for $ \op{v}_{\infty} <|\op{v}|<\op{v}_{r}^{(M)}$;
 
\item  $u^{\prime}_r(\la,\op{v})$ only vanishes \textit{once} on $\R+\i \sg_r \tfrac{\pi}{2}$ when $-\op{v}_{\infty} < \op{v} <\op{v}_{\infty}$. 
\end{itemize}
There, the second condition is to be omitted if $\op{v}_r^{(M)}=\op{v}_{\infty}$.

 In the most general case, one also concludes that for a given velocity $\op{v}$, $u^{\prime}_r(\la,\op{v})$ has zeroes $\om_r^{(1)},\cdots, \om_r^{(\vsg_r)}$ and that 
 there exist $\eta>0$ and  biholomorphisms $h_r^{(a)}: \mc{D}_{0,2\eta} \mapsto \Cx$, $r\in \mf{N}$ such that 
\begin{itemize}
 \item $h_r^{(a)}(0)=0$ \;  and \;  $(h^{(a)}_r)^{\prime}(0) = \Big\{ \tfrac{1}{2} \cdot |u_r^{\prime\prime}(\om_r^{(a)},\op{v})|   \Big\}^{ \tfrac{1}{2} } $ ; 
 \item for $r \in \mf{N}_{\e{st}}$, $a=1,\dots, \vsg_r$
\beq
u_r(\la,\op{v})\, = \,  u_r(\om_r ^{(a)},\op{v}) \, + \, \veps_r ^{(a)} \cdot \big[ h_r ^{(a)}(\la-\om_r)\big]^2 \qquad \e{with} \qquad \veps_r ^{(a)}=\e{sgn}\big[ u_r^{\prime\prime}(\om_r ^{(a)},\op{v}) \big]
\label{definition ur et vepsr r corde case le plus general de point selle}
\enq
for $\la \in \om_r ^{(a)} + \mc{D}_{0,2\eta}$.
\end{itemize}

\subsection{A rewriting of the form factor series}
\label{Sous Section reecriture de la series de FF}

For the purposes of contour deformations as suggested by the results gathered in Subsection \ref{Sous Section preprietes pour locus saddel pts}, it is convenient to reorganise the massless form factor series.
One changes the summation variables as $\ell_+=p$ and 
$\ell_-=\ell-p$ in \eqref{ecriture serie FF massless sg ga sg ga dagger} so that
\beq
\big< \sg_1^{\ga^{\prime}}\!(t)\,  \sg_{m+1}^{\ga}(0) \big> \; = \; (-1)^{m \op{s}_{\ga} } \sul{   n_h \geq 0   }{} \sul{  \substack{ \ell \in \mathbb{Z} : \\  n_h-\ell +\op{s}_{\ga} \geq 0} }{}   \sul{ \bs{n}_{\e{ex}} \in \mf{S}_{n_h,\ell} }{} \; 
 \Int{ (\msc{C}_{h})^{n_h}   }{} \hspace{-3mm} \f{ \dd^{n_h}\mu  }{ n_h! \cdot (2\pi)^{n_h}  } 
 \cdot \mc{I}_{\bs{n}_{\e{ex}} }\big(  \{ \mu_a \}_{1}^{n_h} \big) 
\label{ecriture serie massless FF reorganisee}
\enq
where 
\beq
\mf{S}_{n_h,\ell} \; = \; \bigg\{ \bs{n}_{\e{ex}}=(n_{r_1},n_{r_2},\dots) \; : \; n_r \in \mathbb{N}\; \; , r\in \mf{N}  \quad  \e{and} \quad  \sul{r\in \mf{N} }{} r n_r \, = \, n_h - \ell + \op{s}_{\ga} \bigg\}
\enq
and 
\beq
\mc{I}_{\bs{n}_{\e{ex}} }\big(  \{ \mu_a \}_{1}^{n_h} \big)  \; = \;   
\pl{ r \in  \mf{N} }{} \;\Bigg\{  \Int{ \big( \msc{C}_r \big)^{n_r} }{} \hspace{-3mm} \f{ \dd^{n_r}\nu^{(r)} }{ n_r! \cdot (2\pi)^{n_r}  }   \Bigg\} 
\cdot 
\msc{F}^{(\ga)}_{\e{red}}\big( \mf{Y}_{\e{red}} \big) \;. 
\label{ecriture integrande I n ex}
\enq
Here, we agree upon  
\beq
\msc{F}^{(\ga)}_{\e{red}}\big( \mf{Y}_{\e{red}} \big) \; = \;  \sul{ p \in \mathbb{Z} }{} 
\f{   \mc{F}^{(\ga)}\big( \mf{Y}_{\e{red};p} \big) \cdot    \ex{\i     m \msc{U}(\mf{Y}_{\e{red};p},\op{v})  }    }{  \pl{\ups=\pm}{}  \big[  -  \i (\ups m - \op{v}_F t)   \big]^{   \vth_{\ups}^2(\mf{Y}_{\e{red};p}) }  }  
\cdot  \bigg( 1+ \mf{r}_{\de,m, t}\big( \mf{Y}_{\e{red};p} \big) \bigg)
\label{definition super form factor} 
\enq
where the variables are defined as
\beq
\mf{Y}_{\e{red}} \; = \;  \bigg\{ \ell;  \big\{ \mu_{a} \big\}_{1}^{ n_{h} }  \; \big| \;  \big\{ \nu_{a}^{(1)} \big\}_{a=1}^{n_1}; \dots ;  \big\{ \nu_{a}^{(r_k)} \big\}_{a=1}^{n_{r_k}};\dots  \bigg\}
\enq
and
\beq
\mf{Y}_{\e{red};p} \; = \; \bigg\{ -\op{s}_{\ga};  \big\{ p, \ell-p \big\} \; \big| \; \big\{ \mu_{a} \big\}_{1}^{ n_{h} } \, ; \, \big\{ \nu_{a}^{(1)} \big\}_{a=1}^{n_1}; \dots ;  \big\{ \nu_{a}^{(r_k)} \big\}_{a=1}^{n_{r_k}}; \dots  \bigg\} \;. 
\enq
The function $\msc{F}^{(\ga)}_{\e{red}}\big( \mf{Y}_{\e{red};p} \big) $ is analytic in each rapidity $\nu_{a}^{(r)}$ belonging to a small vicinity of the curves $\msc{C}_{r}$ and in the region $|\Re(\nu_{a}^{(r)})| \geq A/2$, $A$-large, 
with the exception of the lines 
\beq
\intof{-\infty}{q}+\i \mf{f}_{\tfrac{r\pm 1}{2}\zeta} \quad \e{for} \;  r \geq 2 \quad  \e{and} \quad  \intof{-\infty}{q}+\i \mf{f}_{ \zeta} \quad \e{when} \; r=1\; ,
\enq
\textit{c.f.} 
\eqref{ecriture conditions saut FF}-\eqref{ecriture variables pour saut FF}
and \eqref{ecriture condition saut coupure U et vartheta}-\eqref{ecriture condition saut coupure reste}. However, one may readily check that upon defining
\beq
\mf{Y}_{\e{red}}^{\ua/\da} \; = \; \bigg\{   \ell \, ;  \,  \big\{ \mu_{a} \big\}_{1}^{ n_{h} } \; \big| \; \big\{ \nu_{a}^{(1)} \big\}_{a=1}^{n_1}; \dots ;  \big\{ \nu_{a}^{(r_k)} \big\}_{a=1}^{n_{r_k}}, \dots 
\bigg\}_{ \nu_b^{(s)} = x \, + \, \i \mf{f}_{ \frac{r+\sg}{2}\zeta } +/- \i 0^+ }
\enq
it holds 
\beq
\msc{F}^{(\ga)}_{\e{red}}\big( \mf{Y}_{\e{red}}^{\ua}  \big) \; = \; \msc{F}^{(\ga)}\big( \mf{Y}_{\e{red}}^{\da}  \big) \;, 
\enq
so that $\msc{F}^{(\ga)}_{\e{red}}$ has already no cuts. 
This can be seen by making the shift in the summation variable $p \hookrightarrow p-u_{s}^{\sg}$ in the sum \eqref{definition super form factor}. 
Thus, $\msc{F}^{(\ga)}_{\e{red}}$ is analytic in any variable $\nu_{a}^{(s)}$ taken singly in the domains $ \big| \Re\big( \nu_{a}^{(s)} \big) \big| > A/2$, for some $A>0$ large enough.
Finally, it is clear that the function $\msc{F}^{(\ga)}_{\e{red}}$ enjoys the same reduction properties at the dynamical poles  as described in  \eqref{ecriture reduction dans equation sur residus}. 

\vspace{2mm}

The behaviour of the integrals $\mc{I}_{\bs{n}_{\e{ex}} }\big(  \{ \mu_a \}_{1}^{n_h} \big)$ under contour deformations will be analysed in the next section.

 \section{Contour deformations in auxiliary integrals}
\label{Section deformation des contours dans integrales auxiliaires}
 
This section is devoted to establishing two auxiliary results that are needed for the asymptotic analysis of the form factor series expansion \eqref{ecriture serie FF massless sg ga sg ga dagger}. 
More precisely, in this section, I compute explicitly the effect of deforming the particle-hole contours
 from the original contours $\msc{C}_{p}$ to the steepest descent ones in the building blocks of the form factor series expansion \eqref{ecriture serie FF massless sg ga sg ga dagger} corresponding to 
 the $2$ and  $3$ hole excitation sectors. While there would be no major obstacle to pursue this analysis to higher number hole excitation sectors, the calculations become 
 tedious. A conjecture on the form taken by the integrals at general $n$ under contour deformations, is given in the last subsection. 
 
 The presentation becomes slightly more convenient by introducing formal linear combinations of curves in $\Cx$. 
 Given a collection of curves $\ga_1,\dots, \ga_p$ in $\Cx$ and complex numbers $c_1,\dots, c_p$
one defines the formal curve $\ga=\sul{k=1}{p}c_k \ga_k$. The integral of any function $f$  on $\ga_1\cup \dots \cup \ga_n$ over the curve $\ga$ is to be understood as
\beq
\Int{ \ga }{} f(s) \cdot \dd s \; \equiv \; \sul{k=1}{p} c_k \Int{ \ga_k }{} f(s) \cdot \dd s  \;. 
\enq

Furthermore, we agree that given $a,b \in \Cx$, $\intff{a}{b}$ stands for the oriented segment  run from $a$ to $b$. Also, given an 
oriented curve $\ga$ in $\Cx$, given $\sg \in \{\pm 1\}$, the curve $\sg \ga$ stands for the curve $\ga$ whose orientation 
has been changed by $\sg$ in respect to the original orientation. Namely, if $\sg=+$ the orientation is the same, while if $\sg=-$,
it has opposite orientation to $\ga$. 
Finally, given $z\in \Cx$, and a segment $\intff{a}{b}$, $z+\intff{a}{b}$ denotes the segment $\intff{z+a}{z+b}$.

Throughout the section, I shall  employ the shorthand notation:
\beq
\mf{s}_{k} \, = \, \e{sgn}\big[ \sin(k \zeta) \big] \;. 
\enq
Taken   array \eqref{tableau recapitulatif region positivite phase oscillante une particule} which determines the regions where one should deform the curve $\msc{C}_1$,
it appears useful to introduce two parameters $\tau_{\op{v};\a}$, $\a\in \{L,R\}$,  depending on the velocity parameter $\op{v}=\tf{m}{t}$, 
such that 
\beq
\Im\Big( u_1(\la,\op{v}) \Big) \, > \, 0 \qquad \e{for} \qquad   0 \, < \,  \tau_{\op{v};\a} \Im(\la) \, <  \, \tfrac{ \pi }{ 2 } \qquad \e{when} \qquad  \Re(\la) \in \mf{I}_{\a}
\enq
with $\mf{I}_{L}=\intff{-\infty}{-A}$ and $\mf{I}_{R}=\intoo{A}{+\infty}$. The inspection of \eqref{tableau recapitulatif region positivite phase oscillante une particule} yields 
\beq
\tau_{\op{v};R} = 				\left\{ \ba{ccc} 1 & \e{if} &|\op{v}|>\op{v}_{\infty}   \\
											     -1 & \e{if}& 0<\op{v}<\op{v}_{\infty} \\ 
							      		 1 & \e{if} & -\op{v}_{\infty}< \op{v} <0   
\ea \right.  \qquad \e{and} \qquad
\tau_{\op{v};L} = 				\left\{ \ba{ccc} 1 & \e{if}& |\op{v}|>\op{v}_{\infty}   \\
											     1 & \e{if} & 0<\op{v}<\op{v}_{\infty} \\ 
							      		 -1 & \e{if} & -\op{v}_{\infty}< \op{v} <0 \ea \right.   \;. 
\label{definition parametres tau de v et alpha}
\enq

 Finally, for some $\eta>0$ small enough and $A>0$ large enough, I introduce the set 
\beq
\mc{D}_{\eta, A} \; = \; \Big\{ z \in \Cx \; : \; |\Im(z)| \leq \eta  \Big\} \cup  \Big\{ z \in \Cx \; : \; \big| \Im\big( z\pm \i \tfrac{\pi}{2} \big) \big| \leq \eta  \Big\}  \cup
 \Big\{ z \in \Cx \; : \; |\Re(z)| \geq \f{A}{2}  \Big\}  \;.
\label{definition domaine D eta A}
\enq

\subsection{The model integral for the $n$-excitation sector}
\label{Sous Section integrales modeles pour deformation de contours}

I introduce below a class of model integrals that contain, as a special case, the integrals $\mc{I}_{\bs{n}_{\e{ex}}}\big( \{ \mu_a \}_1^{n_h} \big)$ \textit{c.f.} \eqref{ecriture integrande I n ex}
which appear in the massless form factor series expansion obtained in the last section. 

Let $n\in \mathbb{N}$ be fixed. Recall that the set $\mf{N}$ built up from the allowed string lengths at fixed $\zeta$
can be represented as $\mf{N}=\{r_1,r_2,\dots\}$ where $r_1=1$, $r_2=2$ but $r_k\not=k$ in general. Let $m$ be such that $r_m$ is 
the largest element of $\mf{N}$ satisfying $r_m \leq n$. Then, introduce a subset $\mf{N}_m$ of $\mf{N}$
\beq
\mf{N}_m \; = \; \big\{ r_1,\dots, r_m \big\} \;. 
\enq

Let $\bs{n}_0=(n,0,\dots,0)\in \mathbb{N}^m$. Define
\beq
J_{\bs{n}_0}\big( \nu_1,\dots, \nu_n) \; = \; \pl{a \not= b}{n} \bigg\{ \f{ \sinh[\nu_{ab}] }{  \sinh[\nu_{ab}-\i\zeta ] } \bigg\} \cdot \wt{J}_{\bs{n}_0}\big( \nu_1,\dots, \nu_n) \qquad \e{with} \qquad \nu_{ab} \equiv \nu_a - \nu_b  \;. 
\enq
The function  $\wt{J}_{\bs{n}_0}$ appearing above is assumed 
\begin{itemize}
 
 \item to be holomorphic  on $(\Cx\setminus \ga)^n$ where $\ga$ is some compact curve in $\Cx$, and in particular on  $\mc{D}_{\eta,A}^n$; 
   \item to be  symmetric in $\nu_1,\dots, \nu_n$ and $\i\pi$-periodic in respect to each variable
$\nu_a$ taken singly;

\item  to have an exponential decay at infinity
\beq
\wt{J}_{\bs{n}_0}\big( \nu_1,\dots, \nu_n)  \; \leq \; C \pl{a=1}{n} \ex{-2|\Re(\nu_a)|}
\enq
for some constant $C>0$.

 \end{itemize}

By taking appropriate residues of $J_{\bs{n}_0}$, one defines a tower of subordinate functions $J_{\bs{n}}\big( \bs{\nu}_{\bs{n}})$.  
There, I agree upon 
\beq
\bs{n} = \big( n_{r_1},\dots, n_{r_m} \big) \qquad \e{with} \qquad \sul{ r \in \mf{N}_m }{} r \, n_{r} \; = \; n 
\label{ecriture multiindice n}
\enq
and 
\beq
\bs{\nu}_{\bs{n}} \; = \; \Big(\bs{\nu}^{(r_1)},\dots,  \bs{\nu}^{(r_m)} \Big) \qquad \e{with} \qquad \bs{\nu}^{(r)} \, = \, \Big( \nu^{(r)}_1,\dots, \nu^{(r)}_{n_r} \Big) \;.ç 
\label{ecriture multivecteur nu associe au multiindice n} 
\enq
In order to make the definition explicit, introduce
\beq
\bs{\la}^{(r)}_a \, = \, \Big( \la_{a;1}^{(r)}, \dots,  \la_{a;r}^{(r)} \Big) \; , \qquad 
\bs{\la}^{(r)}  \, = \, \Big( \bs{\la}^{(r)}_{1} , \dots,  \bs{\la}^{(r)}_{n_r}  \Big) \; , \qquad 
\bs{\la}  \, = \, \Big( \bs{\la}^{(r_1)} , \dots,  \bs{\la}^{(r_m)}   \Big)
\enq
and, finally, let $\bs{\nu}_{\bs{n}}$ be as defined in \eqref{ecriture multivecteur nu associe au multiindice n} with 
\beq
 \bs{\nu}^{(r)} \; = \; \Big( \la_{1;1}^{(r)}- \i \tfrac{r-1}{2}\zeta, \dots,  \la_{n_r;1}^{(r)}- \i \tfrac{r-1}{2}\zeta \Big) \;. 
\enq
Then, one sets 
\beq
J_{\bs{n}}\big( \bs{\nu}_{\bs{n}}) \; = \; \pl{r \in \mf{N}_m }{ } \Big\{ \i^{(r-1)n_{r}} \Big\} \cdot 
\e{Res}\Bigg( J_{\bs{n}_0}\big( \bs{\la} ) \pl{ r \in \mf{N}_m  }{  } \pl{ a=1 }{ n_{r} } \pl{ s=2 }{ r } \dd \la_{a;s}^{(r)} \; \mid  \;  
\Big\{ \la_{a;s}^{(r)} \, = \, \la_{a;s-1}^{(r)} \, - \, \i \zeta  \Big\}_{s=2}^{r}  \;   \ba{c} a=1,\dots, n_r  \\  r \in \mf{N}_m \ea \Bigg) \;. 
\enq
The symmetry of $\wt{J}_{\bs{n}_0}$ entails that the residues can be computed in any order. The functions $J_{\bs{n}}$
 give rise to a collection of integrals labelled by $\bs{n}$:
\beq
\mc{I}_{\bs{n}} \; = \; \pl{r \in \mf{N}_m }{ } \bigg\{ \Int{ \big( \msc{C}_{r} \big)^{n_{r}} }{}  \f{ \dd^{n_r}\nu^{(r)} }{ (2\pi)^{n_r} \cdot n_r! } \bigg\} \cdot J_{\bs{n}}\big( \bs{\nu}_{\bs{n}}) \;. 
\enq
Finally, one introduces the below combination of integrals
\beq
\mc{I}_{\e{tot}}^{(n)}\; = \; \sul{ \bs{n} \in \mf{S}^{(n)} }{} \mc{I}_{\bs{n}}  \qquad \e{where} 
\qquad \mf{S}^{(n)} \; = \; \Big\{ \bs{n}=(n_{r_1},\dots, n_{r_m}) \in \mathbb{N}^m \; : \; \sul{r \in \mf{N}_m }{}r n_r = n  \Big\}\;. 
\label{definition I tot n}
\enq
One may then carry out contour deformations in each of the integrals $\mc{I}_{\bs{n}} $. 
It turns out that all contributions around $\infty$  of the deformed  curves $\msc{C}^{(r)}$ 
cancel out between the various integrals building up  $\mc{I}_{\e{tot}}^{(n)}$. Thus, effectively, 
the contour deformations of $\msc{C}^{(r)}$, $r \in \mf{N}_m$ lead to an integration
over compact curves in $\Cx$. Furthermore, the contour deformation can be done in a way that is compatible
with the requirements of the steepest descent analysis to come. 
This mechanism is shown explicitly  for the $n=2$ and $n=3$ cases studied below. 
It is conjectured to hold for general $n$.

 \subsection{The $n=2$ sector}
 \label{Sous Section secteur a 2 particules}
 
 To start with, I particularise the general setting introduced above to the case $n=2$.
 Recall that $\mc{D}_{\eta, A}$ was introduced in \eqref{definition domaine D eta A}. 
 
 Let $J_{2,0}$ be the $\i\pi$ periodic, symmetric  function  of two variables  given as the product  of a ratio hyperbolic  polynomials by 
a holomorphic function $\wt{J}_{2,0}$ on $(\Cx\setminus \ga)^2 \supset \mc{D}_{\eta, A}^2$ by :
\beq
J_{2,0}(\nu_1,\nu_2) \, = \, \f{  \sinh^2[\nu_{12}] \cdot  \wt{J}_{2,0}(\nu_1,\nu_2)  }{   \sinh[\nu_{12}-\i\zeta] \cdot \sinh[\nu_{12} + \i\zeta]  }  
\quad \e{where} \quad  \nu_{12}=\nu_1-\nu_2 \;, 
\enq
 and
\beq
\Big| \wt{J}_{2,0}(\nu_1,\nu_2) \Big| \, \leq \, C \pl{a=1}{2}\ex{-2|\Re(\nu_a)|} \;, 
\enq
as $\Re(\nu_k )\tend \pm \infty$. 

 Define  an $\i\pi$-periodic function of one variable $J_{0,1}$ by
\beq
J_{0,1}\Big( \nu - \i \frac{\zeta}{2} \Big)  \, = \,    \i  \Res\bigg\{ J_{2,0}(\nu_1,\nu_2) \cdot \dd \nu_2, \nu_2=\nu_1-\i\zeta  \bigg\} \; = \; \f{ \sin^2(\zeta) }{ \sin(2\zeta) } \wt{J}_{2,0}\big(\nu, \nu-\i\zeta \big) \;. 
\enq
The symmetry properties of $J_{2,0}$ entail that 
\beq
J_{0,1}\Big( \nu  \pm  \i \frac{\zeta}{2} \Big)  \; = \; \f{ \sin^2(\zeta) }{ \sin(2\zeta) } \cdot \wt{J}_{2,0}\big(\nu, \nu \pm \i\zeta \big) \;. 
\enq
\begin{figure}[ht]
\begin{center}

\includegraphics{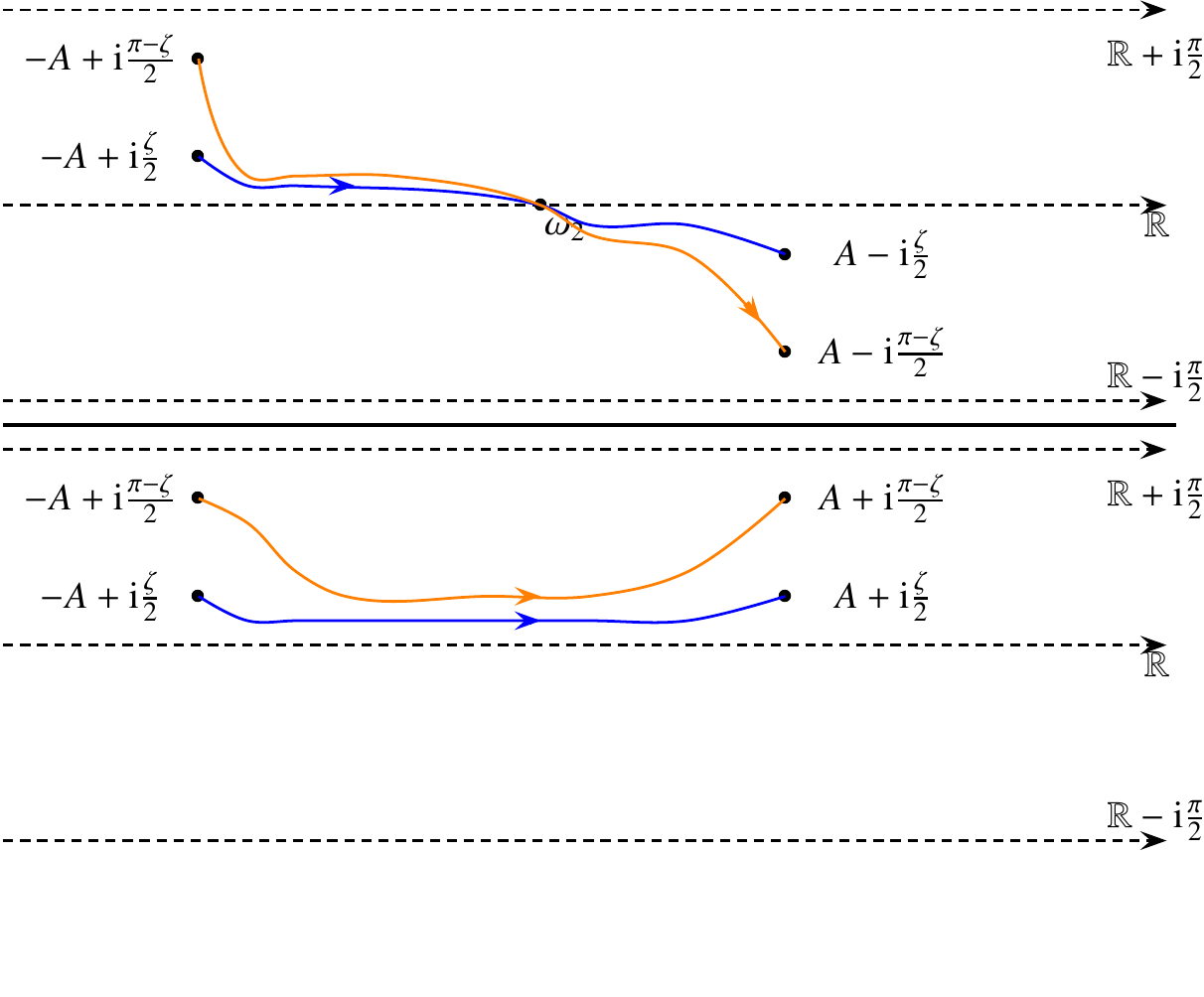}
\caption{  The curves building up  $\msc{C}_{2; A }$ in the regime $0 < \zeta  < \tfrac{\pi}{2} $. The lower graph corresponds to the regime $|\op{v}|> \op{v}_{\infty}$
 and the upper graph corresponds to the regime $0< \op{v} <  \op{v}_{\infty}$. The curve corresponding to the regime $-  \op{v}_{\infty} < \op{v} < 0$ is obtained from the one at 
 $0< \op{v} <  \op{v}_{\infty}$ through a left/right symmetry operation while keeping the same orientation of the curves. 
\label{Figure contour des 2-cordes apres reduction regime  0 zeta Pi sur 2} }
\end{center}

\end{figure}
\begin{figure}[ht]
\begin{center}
\includegraphics{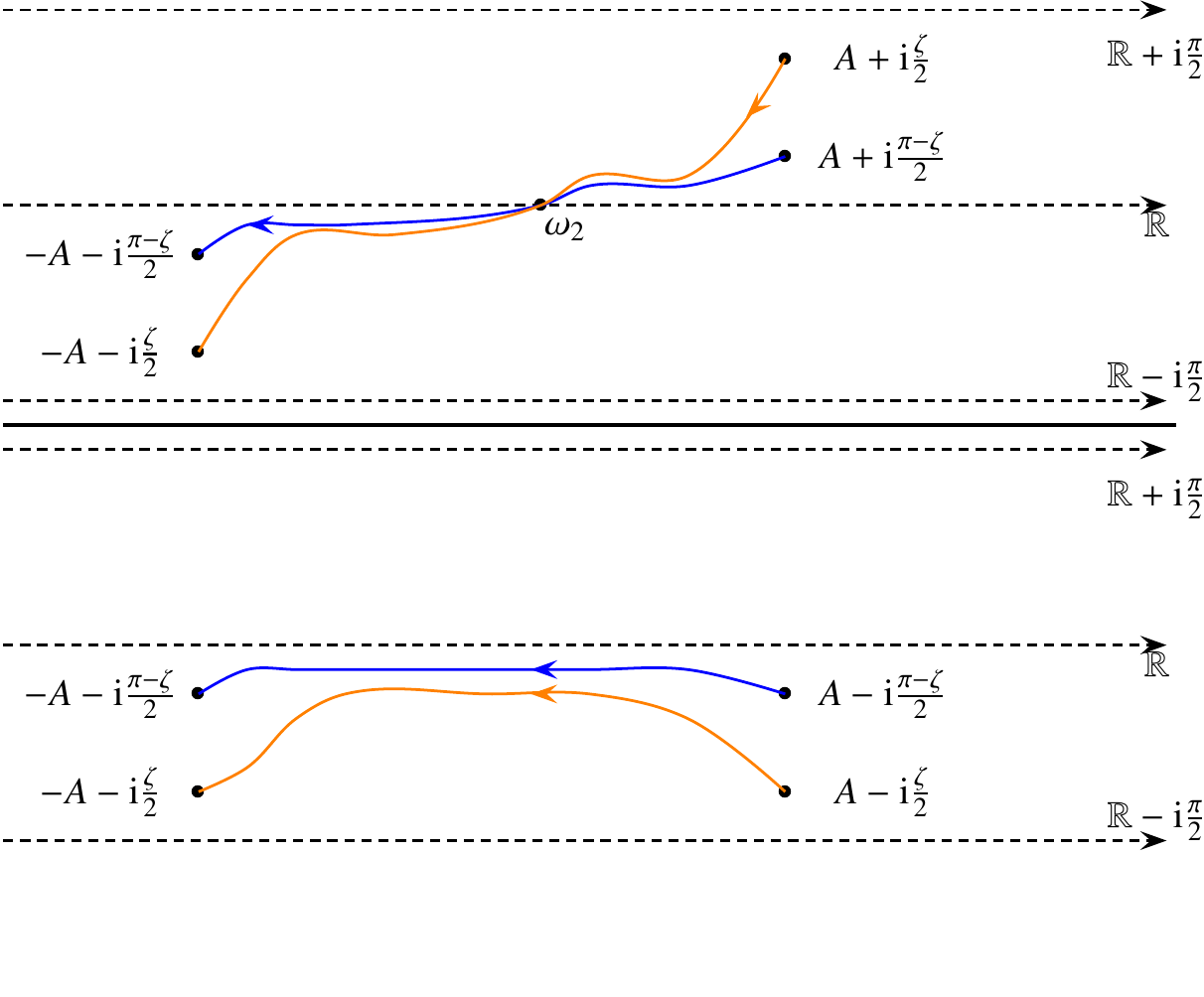}
\caption{  The curves building up  $\msc{C}_{2; A }$ in the regime $\tfrac{\pi}{2} < \zeta < \pi$. The lower graph corresponds to the regime $|\op{v}|> \op{v}_{\infty}$
 and the upper graph corresponds to the regime $0< \op{v} <  \op{v}_{\infty}$. The curve corresponding to the regime $-  \op{v}_{\infty} < \op{v} < 0$ is obtained from the one at 
 $0< \op{v} <  \op{v}_{\infty}$ through a left/right symmetry operation while keeping the same orientation of the curves. 
\label{Figure contour des 2-cordes apres reduction regime  Pi sur 2 zeta pi} }
\end{center}

\end{figure}
\begin{figure}[ht]
\begin{center}

\includegraphics{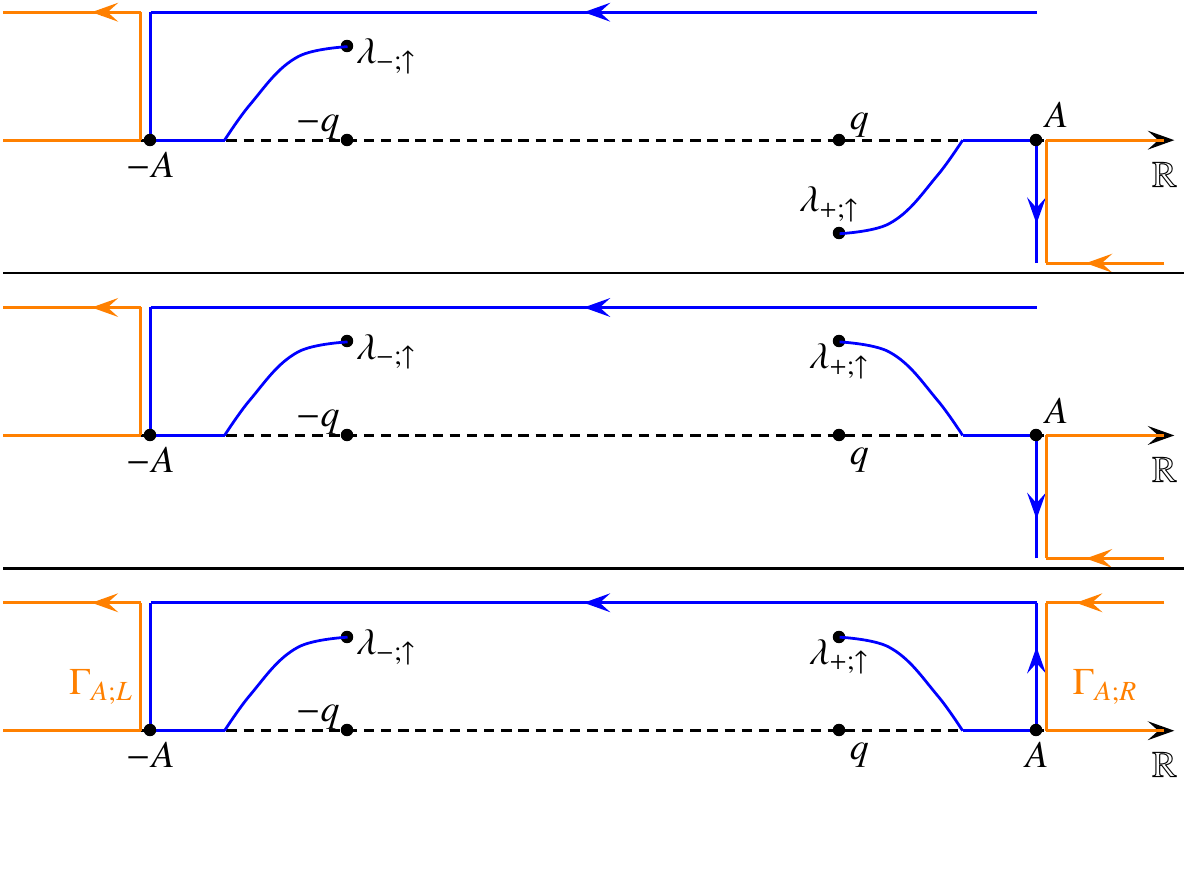}
\caption{  Compactified particle contours $\msc{C}_{1;A}$ -in blue- and the external contour $\Ga_{A}$ -in orange-. The contours are plotted for three regimes of the velocity 
$\op{v}=\tf{m}{t}$ appearing from bottom to top $\op{v}>\op{v}_{\infty}$, $\op{v}_{\infty}>\op{v}>\op{v}_F$ and $\op{v}_F> \op{v} >0$. 
The shape of the curves in the regimes associated with negative $\op{v}$s can be deduced by symmetry. 
The contour $\msc{C}_{1;A}$ 
starts at the points $\la_{\pm;\ua/\da}=\pm q + \e{O}(\de)$,
and then, over a distance of the order of $\de$, joins with the real axis. It closes with the upper part of the contour, going along $\R + \i\tfrac{\pi}{2}$,
along the lines $\Re(z)=\pm A$, for some $A>0$. The contours $\Ga_{A;R}$ appear to the right and $\Ga_{A;L}$ to the left. 
\label{Figure contour des particules apres compactification en var reelle A} }
\end{center}

\end{figure}

\begin{prop} 
Let $\mc{I}_{2,0}$ and $\mc{I}_{0,1}$ denote the integrals
\beq
\mc{I}_{2,0}=\Int{ \big( \msc{C}_{1} \big)^2 }{} \hspace{-2mm} \f{ \dd^2 \nu }{ 2 \cdot (2\pi)^2 } J_{2,0}(\nu_1,\nu_2)  \qquad  and \qquad 
\mc{I}_{0,1}=\Int{   \msc{C}_{2} }{}   \f{ \dd \nu }{ 2\pi} J_{0,1}(\nu) 
\;. 
\label{ecriture integrale 2 corde originale}
\enq
Here, I remind that $ \msc{C}_{2} = \mf{s}_2 \R $. Let $\msc{C}_{1; A } $ be the integration contour defined in Figure \ref{Figure contour des particules apres compactification en var reelle A}
and $\msc{C}_{2;A} $ the integration curve depicted in Figure \ref{Figure contour des 2-cordes apres reduction regime  0 zeta Pi sur 2} for $0<\zeta<\tf{\pi}{2}$ and Figure
\ref{Figure contour des 2-cordes apres reduction regime  Pi sur 2 zeta pi} for $\tf{\pi}{2} < \zeta < \pi $.

Then, it holds 
\beq
\mc{I}_{2,0} \, + \, \mc{I}_{0,1} \; =    \hspace{-4mm} \Int{   \msc{C}_{1; A } \supset \,  \msc{C}_{1; A }   }{}  \hspace{-4mm} J_{2,0}(\nu_1,\nu_2)   \f{ \dd^2 \nu }{ 2 \cdot (2\pi)^2 } \; + \; \f{1}{2} \Int{   \msc{C}_{2; A} }{}   \f{ \dd \nu }{ 2\pi} J_{0,1}(\nu) \;. 
\enq
Above, the notation $ \msc{C}_{1; A } \supset \,  \msc{C}_{1; A } $ refers to two contours $\msc{C}_{1; A }$ infinitesimally close to each other and such that 
the second lies totally inside the first one. 

\end{prop}

\Proof 

One first starts by splitting the contour $\msc{C}_1$ in $\mc{I}_{2,0}$ so as to be able to take into account the contribution of the poles at $\nu_1=\nu_2 \pm \i \zeta \; \e{mod}[\i\pi \mathbb{Z}]$. 
For this,  one decomposes the contours as in Figure \ref{Figure contour des particules apres compactification en var reelle A}, with $\msc{C}_{1;A}$ and  $\Ga_{A}$
being infinitesimally close to each other along the lines $\Re(z)=\pm A$:
\beq
\big( \msc{C}_{1} \big)^2 \;\; \hookrightarrow \; \;  \msc{C}_{1;\e{reg}}   
\times \Big\{ \msc{C}_{1;A} \times \Ga_{A} \Big\}
\enq
where $\msc{C}_{1;\e{reg}}$ is the regularised version of the contour $ \msc{C}_{1} $, where one has removed the points whose real part coincides with $\pm A$, \textit{viz}. 
\beq
\msc{C}_{1;\e{reg}} \, = \,  \msc{C}_{1} \setminus \Big\{\pm A, \pm A + \i\tfrac{\pi}{2} \Big\} \;.
\enq
The removal of the points $\pm A$ and $\pm A + \i\tf{\pi}{2}$ allows one to have a well-defined integral without apparent singularities. 
Then, 
\beq
\mc{I}_{2,0}=\Int{ \msc{C}_{1;\e{reg}} \times  \msc{C}_{1;A}   }{} \hspace{-4mm} \f{ \dd^2 \nu }{ 2 \cdot (2\pi)^2 } J_{2,0}(\nu_1,\nu_2)  \; + \; 
\Int{ \msc{C}_{1;\e{reg}} \times  \Ga_{A}   }{} \hspace{-4mm} \f{ \dd^2 \nu }{ 2 \cdot (2\pi)^2 } J_{2,0}(\nu_1,\nu_2)    \;. 
\enq
In the first integral, one may already shrink the contour $\msc{C}_{1;\e{reg}}$  to a curve that surrounds infinitesimally, from the exterior, the 
contour $ \msc{C}_{1;A}$. In the second integral, one computes the $\Ga_{A}$ contour integral in respect to the variable $\nu_2$
by computing the residues. In doing so, it appears useful to introduce the shorthand notations
\beq
\zeta_{\mf{p}}=\e{min}(\zeta, \pi-\zeta)  \;. 
\enq
Then, the integrand may be recast as 
\beq
J_{2,0}(\nu_1,\nu_2) \, = \, \f{  \sinh^2[\nu_{12}] \cdot  \wt{J}_{2,0}(\nu_1,\nu_2)  }{   \sinh[\nu_{12}-\i\tau_{\op{v};\a}\zeta_{\mf{p}}] \cdot \sinh[\nu_{12} + \i\tau_{\op{v};\a}\zeta_{\mf{p}}]  }  
\enq
with $\a\in \{L,R\}$ and $\tau_{\op{v};\a}$ as introduced in \eqref{definition parametres tau de v et alpha}. As shown in Figure \ref{Figure contour des particules apres compactification en var reelle A}, 
the contours $\Ga_{A;\a}$ are such that $0 < \tau_{\op{v};\a} \cdot \Im(\la) < \tf{\pi}{2}$ and they are endowed with the orientation $\tau_{\op{v};\a}$ in respect to the counterclockwise one. 
Thus, upon introducing $\mf{I}_L=\intoo{-\infty}{-A}$ and $\mf{I}_{R}=\intoo{A}{+\infty}$, one gets that $J_{2,0}(\nu_1,\nu_2)$ has poles at 
\beq
\ba{ccc}  \nu_2 \, = \, \nu_1 \, + \, \i\tau_{\op{v};\a}\zeta_{\mf{p}} &  \e{provided}\, \e{that} & \nu_1 \in \mf{I}_{\a} \vspace{2mm} \\
\nu_2 \, = \, \nu_1 \, - \, \i\tau_{\op{v};\a}\zeta_{\mf{p}} &  \e{provided}\, \e{that} & \nu_1 \in \mf{I}_{\a}  + \i \tau_{\op{v};\a} \tfrac{\pi}{2}  \ea 
\enq
This entails that 

\bem
\mc{I}_{2,0} \, =   \hspace{-3mm}  \Int{ \msc{C}_{1;A} \supset  \msc{C}_{1;A}   }{} \hspace{-4mm} \f{ \dd^2 \nu }{ 2 \cdot (2\pi)^2 } J_{2,0}(\nu_1,\nu_2)  \; + \; 
\sul{\a \in \{L,R\} }{}   \Int{ \msc{C}_{1;\e{reg}}  }{} \hspace{-2mm} \f{ \dd \nu_1 }{ 2 \cdot (2\pi)^2 }  \cdot 2\i\pi \tau_{\op{v};\a}\cdot 
\Bigg\{  \f{ \sinh^2[\i\tau_{\op{v};\a}\zeta_{\mf{p}} ] }{ \sinh[ 2 \i\tau_{\op{v};\a}\zeta_{\mf{p}} ] }  J_{2,0}(\nu_1,\nu_1+\i \tau_{\op{v};\a}\zeta_{\mf{p}} ) \bs{1}_{ \mf{I}_{\a} }(\nu_1) \\
 \; + \;  \f{ \sinh^2[-\i\tau_{\op{v};\a}\zeta_{\mf{p}} ] }{ \sinh[ -2 \i\tau_{\op{v};\a}\zeta_{\mf{p}}] } 
  J_{2,0}(\nu_1,\nu_1-\i \tau_{\op{v};\a}\zeta_{\mf{p}} ) \bs{1}_{ \mf{I}_{\a}  + \i\tau_{\op{v};\a} \frac{\pi}{2} }(\nu_1)
     \Bigg\}    \;.
\end{multline}
Thus, upon implementing the various simplifications and using that $\zeta_{\mf{p}}=\mf{s}_2 \zeta + \pi \bs{1}_{\intoo{\tf{\pi}{2}}{\pi}}(\zeta)$, one eventually gets 
\bem
\mc{I}_{2,0} \, =  \hspace{-3mm}   \Int{   \msc{C}_{1; A } \supset \msc{C}_{1; A }  }{} \hspace{-4mm} \f{ \dd^2 \nu }{ 2 \cdot (2\pi)^2 } J_{2,0}(\nu_1,\nu_2)  
 \, -  \, \sul{\a \in \{L,R\} }{} \mf{s}_2  \Int{ \mf{I}_{\a}  }{}   \f{ \dd \nu   }{  4\pi } \bigg\{   J_{0,1}\Big(\nu + \i \mf{s}_2 \tau_{\op{v};\a} \tfrac{\zeta}{2} \Big)  
\, + \, J_{0,1}\Big(\nu + \i \mf{s}_2 \tau_{\op{v};\a} \tfrac{\pi-\zeta}{2} \Big)  \bigg\} \;.
\label{ecriture integrale 2 corde forme reduite}
\end{multline}
The oriented intervals
\beq
\ba{ccc}  
\msc{J}_{A;\op{v}}^{(1)} & = & \Big] -\infty + \i \mf{s}_2\tau_{\op{v};L} \tfrac{ \zeta}{2} \, ; \,   -A + \i \mf{s}_2\tau_{\op{v};L} \tfrac{  \zeta}{2}\Big]
		  \cup \Big\{  \Big[ A + \i \mf{s}_2\tau_{\op{v};R} \tfrac{ \zeta}{2} \, ; \, +\infty + \i \mf{s}_2 \tau_{\op{v};R} \tfrac{ \zeta}{2}  \Big[ \Big\} \vspace{2mm}\\
\msc{J}_{A;\op{v}}^{(2)} & = &  \Big]-A + \i \mf{s}_2\tau_{\op{v};L}  \tfrac{\ (\pi-\zeta)}{2}  \, ; \,   -\infty + \i \mf{s}_2\tau_{\op{v};L}  \tfrac{  (\pi-\zeta)}{2}  \Big]
						\cup \Big\{ \Big[ A + \i \mf{s}_2\tau_{\op{v};R} \tfrac{  (\pi-\zeta)}{2}  \, ; \, +\infty + \i \mf{s}_2\tau_{\op{v};R} \tfrac{ (\pi-\zeta)}{2} \Big] \Big\}
\ea  \;, 
\enq
allow one to recast $\mc{I}_{2,0}$ in the compact form  
\beq
\mc{I}_{2,0} \, = \,  \Int{   \msc{C}_{1; A } \supset \msc{C}_{1; A }  }{} \hspace{-3mm}  J_{2,0}(\nu_1,\nu_2)   \f{ \dd^2 \nu }{ 2 \cdot (2\pi)^2 } \; - \;
\f{\mf{s}_{2}}{2}  \hspace{-3mm} \Int{ \msc{J}_{A;\op{v}}^{(1)} \cup \msc{J}_{A;\op{v}}^{(2)}}{}  \hspace{-4mm} J_{0,1}(\nu)   \f{ \dd \nu }{ 2\pi} \;. 
\enq
It is then enough to add the obtained contribution to the one issuing from $\mc{I}_{0,1}$ so as to get the claim upon deforming the integration contours
in $\mc{I}_{0,1}$ what, all-in-all, cancels out the contributions appearing above. \qed

 \subsection{The $n=3$ sector}
 \label{Sous Section secteur a 3 particules}

 Let $J_{3,0,0}$ be an $\i\pi$ periodic, symmetric, function  of three variables given as the ratio of a rational 
trigonometric function and a holomorphic function $\wt{J}_{3,0,0}$ on $(\Cx\setminus \ga)^3\supset \mc{D}_{\eta, A}^3$, where $\mc{D}_{\eta, A}$ is as it was introduced in \eqref{definition domaine D eta A}:
\beq
J_{3,0,0}(\nu_1,\nu_2,\nu_3) \, = \, \pl{a \not= b}{3} \bigg\{ \f{  \sinh[\nu_{ab}]   }{   \sinh[\nu_{ab}-\i\zeta]  }  \bigg\} \cdot \wt{J}_{3,0,0}(\nu_1,\nu_2,\nu_3)  
\enq
where I remind that $\nu_{ab} = \nu_a - \nu_b$ and that it is assumed 
\beq
 \big| \wt{J}_{3,0,0}(\nu_1,\nu_2,\nu_3)   \big| \; \leq \; C \pl{a=1}{3} \ex{-2|\Re(\nu_a)|}
\enq
as $\Re(\nu_a) \tend \pm \infty$.

 Then, define an $\i\pi$-periodic function of two variables $J_{1,1,0}$ and an $\i\pi$-periodic function of one variable $J_{0,0,1}$, as 
\beqa
J_{1,1,0}\Big( \nu_1, \nu_2 - \i \frac{\zeta}{2} \Big)  & = &   \i  \Res\bigg\{ J_{3,0,0}(\nu_1,\nu_2,\nu_3) \cdot \dd \nu_3, \nu_3=\nu_2-\i\zeta  \bigg\}  \; ,  \\
J_{0,0,1,}\Big( \nu_1 - \i \zeta \Big)  & = &   (\i)^2  \Res\bigg\{ J_{3,0,0}(\nu_1,\nu_2,\nu_3) \cdot \dd \nu_2 \, \dd \nu_3, \nu_3=\nu_2-\i\zeta, \nu_2=\nu_1-\i\zeta   \bigg\}   \;. 
\eeqa
Those two functions can be explicitly expressed as 
\beqa
J_{1,1,0}\Big( \nu_1, \nu_2 \mp \i \frac{\zeta}{2} \Big)  & = &     \f{ \sin^2(\zeta) }{ \sin(2\zeta) }  \cdot 
\f{ \sinh(\nu_{12}) \sinh(\nu_{12} \pm \i\zeta)  }{ \sinh(\nu_{12} \mp \i\zeta) \sinh(\nu_{12} \pm  2\i \zeta) } \cdot \wt{J}_{3,0,0}(\nu_1,\nu_2,\nu_2 \mp \i\zeta )   \label{ecriture relation J110 et J300}\\
J_{0,0,1,}\Big( \nu  \mp \i \zeta \Big)  & = &    \f{ \sin^3(\zeta) }{ \sin(3\zeta) }  \cdot \wt{J}_{3,0,0}(\nu ,  \nu  \mp \i\zeta, \nu \mp 2\i\zeta )     \;. 
\label{ecriture relation J001 et J300}
\eeqa
Finally, consider the three integral
\beq
\mc{I}_{3,0,0}\, = \, \Int{ \big( \msc{C}_{1} \big)^3 }{} \hspace{-2mm} \f{ \dd^3 \nu }{ 6 \cdot (2\pi)^3 } J_{3,0,0}(\nu_1,\nu_2,\nu_3) \quad , 
 \qquad 
\mc{I}_{1,1,0} \, = \, \Int{  \msc{C}_{1} }{} \f{ \dd \nu_1 }{ 2\pi } \Int{ \msc{C}_2 }{}  \f{ \dd \nu_2 }{ 2\pi }   J_{1,1,0}(\nu_1,\nu_2) \;, 
\label{ecriture integrale 3 corde originale trois particules et un particule et une deux corde}  
\enq
where $\msc{C}_2=\mf{s}_2 \R$ and 
\beq
\mc{I}_{0,0,1}\, = \, \Int{  \msc{C}_{3}  }{}   \f{ \dd \nu }{ 2\pi } J_{0,0,1}(\nu)  \qquad \e{with} \qquad 
\msc{C}_3 \; = \; \mf{s}_2 \mf{s}_3 \R + \i \f{\pi}{2} \bs{1}_{\intoo{ \tfrac{\pi}{2} }{ \pi } }(\zeta) \;. 
\label{ecriture integrale 3 corde originale une trois corde}
\enq
\begin{figure}[ht]
\begin{center}

\includegraphics{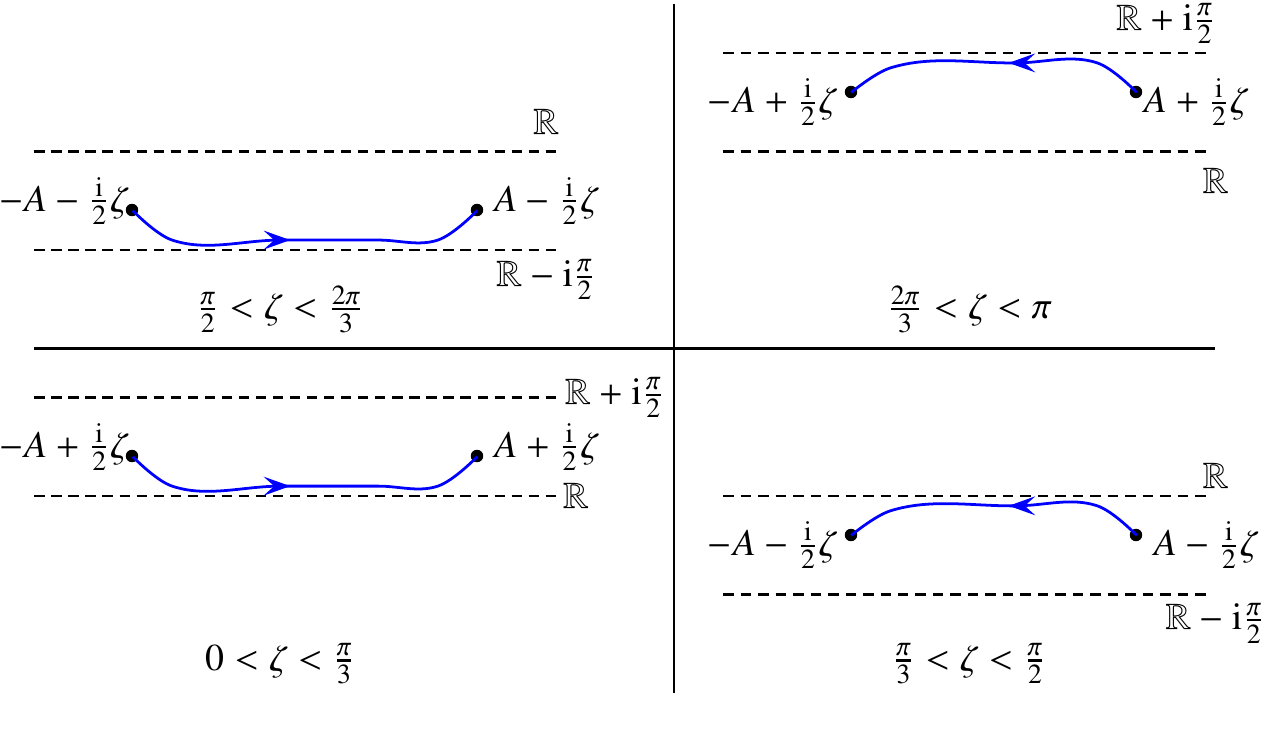}
\caption{  The contours $\msc{C}_{3;A}$, for various values of $\zeta$ and throughout the regime  $\op{v}>\op{v}_{\infty}$. 
\label{Figure contour C3A pour v plus grand que v infty} }
\end{center}

\end{figure}
\begin{figure}[ht]
\begin{center}
\includegraphics{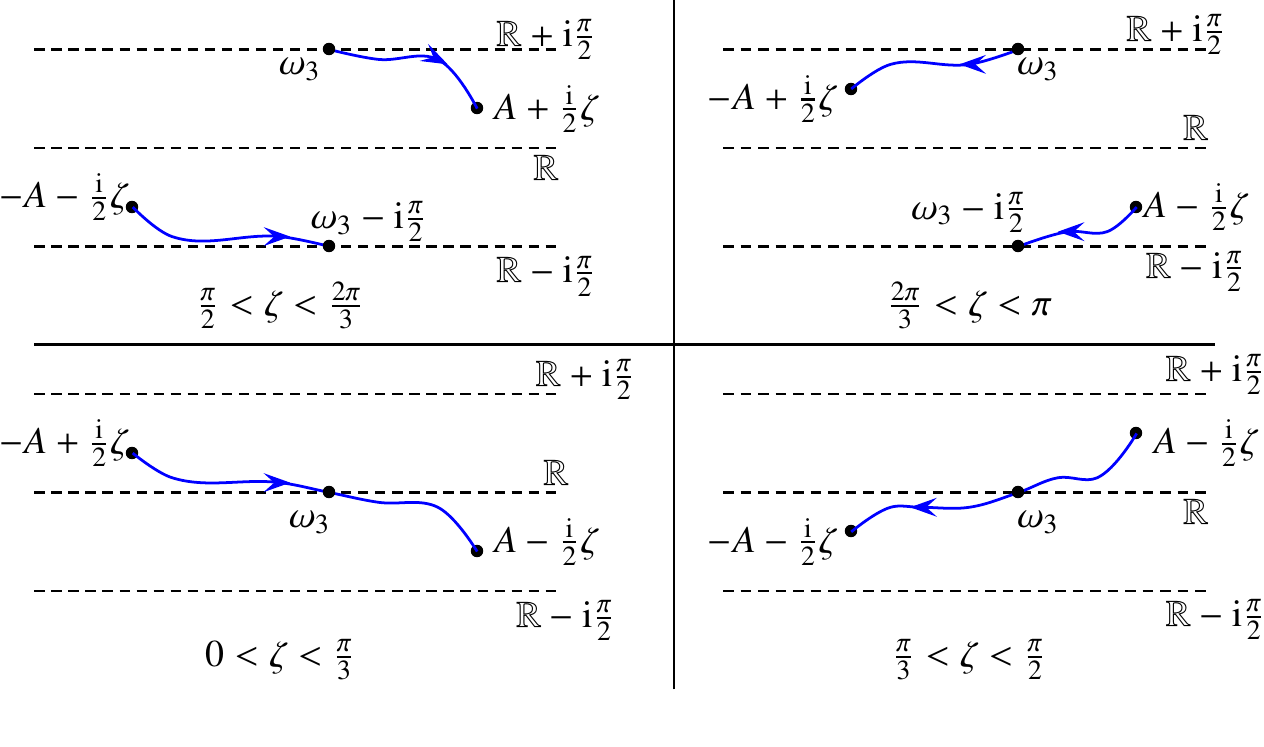}
\caption{  The contours $\msc{C}_{3;A}$, for various values of $\zeta$ and throughout the regime  $0<\op{v}<\op{v}_{\infty}$. 
\label{Figure contour C3A pour v entre 0 et  v infty} }
\end{center}

\end{figure}

\begin{prop}
The sum of three integrals can be decomposed as follows
\bem
\mc{I}_{3,0,0} \, + \, \mc{I}_{1,1,0} \, + \, \mc{I}_{0,0,1} \; = \; \f{1}{6} \hspace{-6mm}\Int{ \msc{C}_{1;A} \supset  \msc{C}_{1;A} \supset \msc{C}_{1;A}  }{} \hspace{-6mm}  \f{ \dd^3 \nu }{   (2\pi)^3 } J_{3,0,0}(\nu_1,\nu_2,\nu_3) \\
\; + \; \f{1}{2}\Int{  \msc{C}_{1;A} }{} \f{ \dd \nu_1 }{ 2\pi } \Int{ \msc{C}_{2;A} }{}  \f{ \dd \nu_2 }{ 2\pi }   J_{1,1,0}(\nu_1,\nu_2)  
\; + \; \Int{  \msc{C}_{3;A}^{(\e{mod})} }{} \f{ \dd \nu }{ 2\pi }   J_{0,0,1}(\nu)  \;. 
\end{multline}
Above 
\beq
\msc{C}_{3;A}^{(\e{mod})} \; = \; \msc{C}_{3;A} \; + \; \f{1}{3}\bs{1}_{ \intoo{0}{\f{\pi}{4}}\cup  \intoo{\f{3\pi}{4}}{\pi} }(\zeta)  \msc{J}_{A;\op{v}} \;, 
\enq
in which the contour $\msc{C}_{3;A}$ is as depicted in Figure \ref{Figure contour C3A pour v plus grand que v infty}, in what concerns the regime $\op{v}>\op{v}_{\infty}$ 
and Figure \ref{Figure contour C3A pour v entre 0 et  v infty},  in what concerns the regime $0<\op{v} <\op{v}_{\infty}$. Furthermore,  
$ \msc{J}_{A;\op{v}}$ refers to the below union of oriented segments
\beq
 \msc{J}_{A;\op{v}} \; =\; \Big[ -A + \i \tau_{\op{v};L}\, (\tfrac{\pi}{2}-\zeta_{\mf{p}})  \, ; \,   -A + \i \tau_{\op{v};L}   \,  \zeta_{\mf{p}}   \Big] \bigcup 
 \Big[    A + \i \tau_{\op{v};R} \,  \zeta_{\mf{p}} \, ; \,     A + \i \tau_{\op{v};R} \, (\tfrac{\pi}{2}-\zeta_{\mf{p}})  \Big] \;, 
\label{definition interval final J A v}
\enq
in which $\tau_{\op{v};\a}$ is as defined in \eqref{definition parametres tau de v et alpha} while $\zeta_{\mf{p}}=\e{min}(\zeta,\pi-\zeta)$. Finally, $\msc{C}_{2;A} $ stands for 
the integration curve depicted in Figure \ref{Figure contour des 2-cordes apres reduction regime  0 zeta Pi sur 2} for $0<\zeta<\tfrac{\pi}{2}$ and Figure
\ref{Figure contour des 2-cordes apres reduction regime  Pi sur 2 zeta pi} for $\tfrac{\pi}{2} < \zeta < \pi $. 
 
\end{prop}

\Proof

\subsubsection*{$\bullet$ The integrals $\mc{I}_{1,1,0}$ and $\mc{I}_{0,0,1}$}

Upon splitting the contour $\msc{C}_{1}$ to the contours $\msc{C}_{1;A}$ and $\Ga_{A}$ depicted in Figure \ref{Figure contour des particules apres compactification en var reelle A},
one gets 
\beq
\mc{I}_{1,1,0} \; = \; \mc{I}_{1,1,0}^{(\e{reg})} \, + \, \mc{I}_{1,1,0}^{(\e{res})}
\enq
where, upon making explicit that  $\msc{C}_2=\mf{s}_2 \R$, one has 
\beq
\mc{I}_{1,1,0}^{(\e{reg})} \; = \;  \mf{s}_2 \Int{ \msc{C}_{1;A} }{} \f{ \dd \nu_1 }{ 2\pi } \Int{ \R }{}  \f{ \dd \nu_2 }{ 2\pi }   J_{1,1,0}(\nu_1,\nu_2) \quad, \qquad 
\mc{I}_{1,1,0}^{(\e{res})} \; = \; \mf{s}_2 \Int{ \Ga_{A} }{} \f{ \dd \nu_1 }{ 2\pi } \Int{ \R }{}  \f{ \dd \nu_2 }{ 2\pi }   J_{1,1,0}(\nu_1,\nu_2) \; . 
\label{definition des integrales I110 reg et I110 res}
\enq

One may then partition $\mc{I}_{1,1,0}^{(\e{res})}$ as
\beq
\mc{I}_{1,1,0}^{(\e{res})} \; = \; \mf{s}_2 \sul{ \a \in \{L,R\} }{}  \Int{ \R }{}  \f{ \dd \nu_2 }{ 2\pi }  \Int{ \Ga_{A;\a} }{} \f{ \dd \nu_1 }{ 2\pi } J_{1,1,0}(\nu_1,\nu_2) 
\enq
what allows to take the $\Ga_{A;\a}$ contour integrals by residues. One may recast the integrand as
\beq
J_{1,1,0}\big( \nu_1, \nu_2   \big)  \;  = \;     \f{ \sin^2(\zeta) }{ \sin(2\zeta) }  \cdot 
\f{ \sinh\big[\nu_{12}-\i\tfrac{\zeta}{2} \big] \cdot  \sinh\big[ \nu_{12}+\i \tfrac{\zeta}{2} \big]  }{ \sinh\big[ \nu_{12}-\i \tfrac{3\zeta}{2} \big] \cdot \sinh \big[ \nu_{12} + \i \tfrac{3\zeta}{2} \big]  } \cdot 
\wt{J}_{3,0,0} \Big(\nu_1,\nu_2+\i \tfrac{\zeta}{2},\nu_2- \i \tfrac{\zeta}{2} \Big) \;. 
\enq

Let
\beq
\big( \tfrac{3 \zeta }{ 2 } \big)_{\mf{p}} \; = \; \left\{ \ba{cc}  \tfrac{3\zeta }{2}    &  0 \, < \, \zeta \,  < \, \tfrac{\pi}{3}  \vspace{2mm} \\
 \pi \,  -  \, \tfrac{3\zeta }{2}    &  \tfrac{\pi}{3} \, < \, \zeta \,  < \, \tfrac{2 \pi}{3}     \vspace{2mm} \\ 
  \tfrac{3\zeta }{2} \, - \,  \pi     &  \tfrac{2\pi}{3} \, < \, \zeta \,  < \,  \pi     \ea   \right. 
\qquad \e{so}\; \e{that} \qquad  \big( \tfrac{3 \zeta }{ 2 } \big)_{\mf{p}} \in \intoo{0}{\tfrac{\pi}{2}} \; .  
\enq
 Then, the poles of $J_{1,1,0}$ in the variable $\nu_1$  that will be present inside of $\Ga_{A;\a}$ will be located at the points 
\beq
\nu_1=\nu_2 + \i \tau_{\op{v};\a}  \big( \tfrac{3 \zeta }{ 2 } \big)_{\mf{p}}  \; ,\quad \e{this} \, \e{provided}\, \e{that} \; \;  \nu_2 \in \mf{I}_{\a} \;, 
\enq
  with 
\beq
 \mf{I}_{L} \; = \; \intoo{-\infty}{-A} \quad, \quad   \mf{I}_{R} \; = \; \intoo{A}{+\infty} \;.
\label{definition des intervalles I Left et Right}
\enq
The corresponding residue will come with the sign $\tau_{\op{v};\a}$ stemming from the orientation of $\Ga_{A;\a}$. One obtains 
\beq
\mc{I}_{1,1,0}^{(\e{res})} \; = \;  \mf{s}_2  \sul{ \a \in \{L,R\} }{} \mf{c}_{\a} \Int{ \R }{}  \f{ \dd \nu }{ 2\pi } 
\underbrace{ \wt{J}_{3,0,0} \Big(\nu\, + \, \i \tau_{\op{v};\a}  \big( \tfrac{3 \zeta }{ 2 } \big)_{\mf{p}}  ,\nu + \i \tfrac{\zeta}{2}, \nu- \i \tfrac{\zeta}{2} \Big) }_{ \equiv  \wt{f}_{3,0,0}(\nu) }
\cdot \bs{1}_{\mf{I}_{\a}}(\nu)
\enq
where 
\beq
\mf{c}_{\a} \; = \; \i  \f{ \sin^2(\zeta) }{ \sin(2\zeta) } \cdot   \f{  \pl{\eps=\pm}{} \sinh\Big[\i \tau_{\op{v};\a}  \big( \tfrac{3 \zeta }{ 2 } \big)_{\mf{p}} + \eps \i\tfrac{\zeta}{2} \Big]   }
{ \tau_{\op{v};\a} \cdot  \sinh\Big[ 2\i \tau_{\op{v};\a}  \big( \tfrac{3 \zeta }{ 2 } \big)_{\mf{p}}  \Big]  } \; = \; - \mf{s}_{3} \f{     \sin^3(\zeta) }{  \sin(3\zeta)}
\enq
and, upon using the $\i\pi$-periodicity and the symmetry in respect to the last two variables, 
\beqa
\wt{f}_{3,0,0}(\nu)   & = &  \left\{\ba{cc}  \wt{J}_{3,0,0} \Big(\nu\, + \, \i \tau_{\op{v};\a}  \tfrac{3 \zeta }{ 2 }    ,\nu + \i \tau_{\op{v};\a} \tfrac{\zeta}{2}, \nu - \i \tau_{\op{v};\a} \tfrac{\zeta}{2} \Big)   \;,  
										&  0 < \zeta < \f{\pi}{3}   \; \; \e{or} \; \;  \f{2\pi}{3} < \zeta < \pi   \vspace{2mm}  \\
  \wt{J}_{3,0,0} \Big(\nu\, - \, \i \tau_{\op{v};\a}  \tfrac{3 \zeta }{ 2 }    ,\nu - \i\tau_{\op{v};\a}  \tfrac{\zeta}{2},\nu +  \i \tau_{\op{v};\a}\tfrac{\zeta}{2} \Big)    
										&    \f{\pi}{3} < \zeta  < \f{2\pi}{3}
\ea  \right. \\
& = & \f{ \sin(3\zeta) }{ \sin^{3}(\zeta) } \cdot \wt{J}_{0,0,1} \Big( \nu\, + \, \i \mf{s}_3\tau_{\op{v};\a}  \tfrac{ \zeta }{ 2 } \Big) \;. 
\eeqa
Thus, one gets 
\beq
\mc{I}_{1,1,0}^{(\e{res})} \; = \; - \mf{s}_3 \mf{s}_2 \sul{ \a \in \{L,R\} }{}  \Int{ \mf{I}_{\a} }{}  \f{ \dd \nu  }{ 2\pi }  \wt{J}_{0,0,1} \Big( \nu\, + \, \i \mf{s}_3\tau_{\op{v};\a}  \tfrac{ \zeta }{ 2 } \Big)
\enq
The above integral is already in good shape so as to combine it with the contributions stemming from $\mc{I}_{0,0,1}$.

Indeed, one gets that 
\beq
\mc{I}_{1,1,0}^{(\e{res})} \, + \, \mc{I}_{0,0,1} 
\; = \;   \Int{ \msc{C}_{0,0,1}  }{}  \f{ \dd \nu  }{ 2\pi }  \wt{J}_{0,0,1} \big( \nu \big)
\enq
where $\msc{C}_{0,0,1}$ corresponds to the below union of oriented curves
\beq
\msc{C}_{0,0,1} \, = \, \Big\{ \mf{s}_3 \mf{s}_2 \R  \, + \,  \i \tfrac{\pi}{2} \bs{1}_{ \intoo{ \f{\pi}{2} }{ \pi } }(\zeta)   \Big\} \bigcup
\Big\{  - \mf{s}_3 \mf{s}_2 \mf{I}_L \, + \,  \i \mf{s}_3 \tau_{\op{v};L}  \tfrac{ \zeta }{ 2 } \Big\} \bigcup 
\Big\{  - \mf{s}_3 \mf{s}_2 \mf{I}_{R} \, + \,  \i \mf{s}_3 \tau_{\op{v};R}  \tfrac{ \zeta }{ 2 } \Big\} \;. 
\label{ecriture courbes de integrale 001 avant deformation}
\enq
By using the $\i\pi$ periodicity in $\nu$ of  $\wt{J}_{0,0,1} \big( \nu \big)$ and its analyticity in the domain $\Re(z)>\tf{A}{2}$, one may deform 
the integration curve $\mf{s}_3 \mf{s}_2 \R  \, + \,  \i \tfrac{\pi}{2} \bs{1}_{ \intoo{ \f{\pi}{2} }{ \pi } }(\zeta) $
to the curve 
\beq
\msc{C}_{3;A} \bigcup
\Big\{    \mf{s}_3 \mf{s}_2 \mf{I}_L \, + \,  \i \mf{s}_3 \tau_{\op{v};L}  \tfrac{ \zeta }{ 2 } \Big\} \bigcup 
\Big\{    \mf{s}_3 \mf{s}_2 \mf{I}_{R} \, + \,  \i \mf{s}_3 \tau_{\op{v};R}  \tfrac{ \zeta }{ 2 } \Big\} \;,  
\label{ecriture courbes de integrale 001 apres deformation}
\enq
where $\msc{C}_{3;A}$ is as given in Figure \ref{Figure contour C3A pour v plus grand que v infty} for the regime $\op{v}>\op{v}_{\infty}$ 
and Figure \ref{Figure contour C3A pour v entre 0 et  v infty} for the regime $0<\op{v} <\op{v}_{\infty}$. 
Then, since the last two curves in \eqref{ecriture courbes de integrale 001 apres deformation} have an opposite
orientation to the last two curves in \eqref{ecriture courbes de integrale 001 avant deformation}, these contributions cancel so that, eventually, 
\beq
\mc{I}_{1,1,0}^{(\e{res})} \, + \, \mc{I}_{0,0,1} 
\; = \;   \Int{ \msc{C}_{3;A}  }{}  \f{ \dd \nu  }{ 2\pi }  \wt{J}_{0,0,1} \big( \nu \big) \;. 
\enq
All-in-all, one gets 
\beq
\mc{I}_{1,1,0}  \, + \, \mc{I}_{0,0,1} 
\; = \; \mc{I}_{1,1,0}^{(\e{reg})} \; + \;   \Int{ \msc{C}_{3;A}  }{}  \f{ \dd \nu  }{ 2\pi }  \wt{J}_{0,0,1} \big( \nu \big) \;, 
\enq
in which $\mc{I}_{1,1,0}^{(\e{reg})}$ is as defined in \eqref{definition des integrales I110 reg et I110 res}.

\subsubsection*{$\bullet$ The integral $\mc{I}_{3,0,0}$}

For a given $\eps>0$ and small enough, it is easy to convince oneself that the following decomposition of the integration contour holds:
\bem
\big( \msc{C}_{1} \big)^3 \; = \;  
\Big\{ \msc{C}_{1;A+2\eps} \times  \msc{C}_{1;A+\eps} \times \msc{C}_{1;A} \Big\} \bigcup 
 \Big\{ \Ga_{A+2\eps} \times  \msc{C}_{1;A+\eps} \times \msc{C}_{1;A} \Big\} 
 \bigcup  \Big\{ \msc{C}_{1;A+\frac{\eps}{2}} \times  \Ga_{A + \eps}  \times \msc{C}_{1;A} \Big\}  \\
 \bigcup  \Big\{ \Ga_{A+\frac{\eps}{2}} \times  \Ga_{A + \eps}  \times \msc{C}_{1;A} \Big\} 
 \bigcup  \Big\{ \msc{C}_{1;A-\frac{\eps}{2}} \times  \msc{C}_{1;A-\eps}\times \Ga_{A} \Big\}  
 \bigcup  \Big\{ \Ga_{A-\frac{\eps}{2}} \times  \msc{C}_{1;A-\eps}\times \Ga_{A} \Big\}  \\
 \bigcup  \Big\{  \msc{C}_{1;A-2\eps} \times  \Ga_{A-\eps}\times \Ga_{A} \Big\} 
 \bigcup \Big\{  \Ga_{A-2\eps} \times  \Ga_{A-\eps}\times \Ga_{A} \Big\}  \; . 
\end{multline}
Thus, by taking the $\eps\tend 0^+$ limit of this partitioning of contours, and upon using the symmetry of the integrand, it is readily seen that $\mc{I}_{3,0,0}$ can be decomposed as 
\beq
\mc{I}_{3,0,0} \; = \;  \mc{I}_{3,0,0}^{(\e{reg})} \, + \, \mc{I}_{3\hookrightarrow 1 \times 2}^{(\e{res})}\, + \, \mc{I}_{3\hookrightarrow 2\times 1}^{(\e{res})}
\, + \, \mc{I}_{3\hookrightarrow 0\times 3}^{(\e{res})} \;. 
\label{ecriture decomposition integrale 3 particules sur integrales plus simples}
\enq
where 
\beq
 \mc{I}_{3,0,0}^{(\e{reg})} \; = \; \f{1}{3!}  \hspace{-5mm} \Int{  \msc{C}_{1;A} \supset  \msc{C}_{1;A} \supset \msc{C}_{1;A}  }{} \hspace{-6mm} \f{ \dd^3 \nu }{    (2\pi)^3 } J_{3,0,0}(\nu_1,\nu_2,\nu_3) 
\qquad , \qquad 
 \mc{I}_{3\hookrightarrow 1 \times 2}^{(\e{res})}  \; = \; \f{1}{2}\hspace{-5mm}  \Int{  \msc{C}_{1;A} \times  ( \Ga_{A} \supset    \Ga_{A} ) }{} \hspace{-6mm} \f{ \dd^3 \nu }{    (2\pi)^3 } J_{3,0,0}(\nu_1,\nu_2,\nu_3) 
\enq
as well as 
\beq
\mc{I}_{3\hookrightarrow 2\times 1}^{(\e{res})} \; = \;  \f{1}{2}\hspace{-5mm}  \Int{  (\msc{C}_{1;A} \supset  \msc{C}_{1;A} ) \times \Ga_{A}  }{} \hspace{-6mm} \f{ \dd^3 \nu }{    (2\pi)^3 } J_{3,0,0}(\nu_1,\nu_2,\nu_3) 
\qquad \e{and} \qquad
 \mc{I}_{3\hookrightarrow 0\times 3}^{(\e{res})} \; = \; \f{1}{3!}\hspace{-4mm}  \Int{  \Ga_{A} \supset  \Ga_{A} \supset \Ga_{A}  }{} \hspace{-5mm} \f{ \dd^3 \nu }{    (2\pi)^3 } J_{3,0,0}(\nu_1,\nu_2,\nu_3) \;. 
\enq
The integrations appearing above should be understood as encased integrals which can be realised by taking the limits
\beqa
\msc{C}_{1;A} \supset  \msc{C}_{1;A} \supset \msc{C}_{1;A} & \equiv  &  \lim_{\eps \tend 0^+}\Big\{ \msc{C}_{1;A+2\eps} \times  \msc{C}_{1;A+\eps} \times \msc{C}_{1;A} \Big\} \; , 
\nonumber \\
\Ga_{A} \supset  \Ga_{A} \supset \Ga_{A} & \equiv  & \lim_{\eps \tend 0^+}\Big\{ \Ga_{ A + 2\eps } \times  \Ga_{ A + \eps } \times \Ga_{ A } \Big\}  \; , \nonumber  \\
  \msc{C}_{1;A} \times  ( \Ga_{A} \supset    \Ga_{A} ) & \equiv  &     \lim_{\eps \tend 0^+} \Big\{ \msc{C}_{1;A} \times \Ga_{ A + 2\eps } \times  \Ga_{ A +\eps }  \Big\}  \; ,  \nonumber  \\ 
 (\msc{C}_{1;A} \supset  \msc{C}_{1;A} ) \times \Ga_{A}  & \equiv &    \lim_{\eps \tend 0^+} \Big\{  \msc{C}_{1;A-2\eps} \times  \msc{C}_{1;A-\eps} \times  \Ga_{ A  }  \Big\} \;. 
\label{definition des contours emboites pour integrale a trois particules}
\eeqa
The use of the $\eps\tend 0^+$ prescription produces intermediate integrals that are all well-defined. It remains to investigate separately each 
integral arising in the decomposition \eqref{ecriture decomposition integrale 3 particules sur integrales plus simples}.

The encased structure of the contours arising in the integral $\mc{I}_{3\hookrightarrow 2\times1}^{(\e{res})}$ allows one to conclude that all the poles in the $\nu_3$ variable
are not surrounded by $\Ga_{A}$. As a consequence
\beq
\mc{I}_{3\hookrightarrow 2\times 1}^{(\e{res})}  \; = \;   0 \;. 
\enq

\subsubsection*{$\bullet$ The integrals $\mc{I}_{3\hookrightarrow 1 \times 2 }^{(\e{res})}  + \mc{I}_{1,1,0}^{(\e{reg})}$}

I first focus on the reduction of the number of integrations in $\mc{I}_{3\hookrightarrow 1 \times 2 }^{(\e{res})}$. For that purpose, 
it is convenient to recall the definition 
\beq
\Phi_{11}(x) \; = \; \f{ \sinh(x) }{ \sinh(x-\i\zeta) } \qquad \e{and} \; \e{agree} \; \e{upon} \qquad 
\Phi_{11}(x_1,\dots, x_n) \; = \; \pl{a=1}{n} \Phi_{11}(x_a)
\enq
so that 
\beq
\Phi_{11}\big(\nu,-\nu\big) \; = \;  \f{  \sinh^2[\nu] }{ \sinh\big[ \nu - \i \zeta_{\mf{p}} \big] \cdot \sinh\big[ \nu + \i \zeta_{\mf{p}} \big] } \;. 
\enq

The integration over $\nu_2$ in $\mc{I}_{3\hookrightarrow 1 \times 2 }^{(\e{res})}$ can be computed by residues. The encased structure of the integration curves \eqref{definition des contours emboites pour integrale a trois particules}
ensures that only the poles in $\nu_2$ of the factor $\Phi_{11}\big(\nu_{23},\nu_{32}\big)$ will be located inside of the contour $\Ga_A$ for the $\nu_2$-variables. 

Observe that for $\a \in \{L,R\}$, the contour  $\Ga_{A;\a}$ is located in the domain delimited by $0\leq \Im(z) \leq \tf{\pi}{2}$ when $\tau_{\op{v};\a}=1$ and 
in the domain delimited by $- \tf{\pi}{2} \leq \Im(z) \leq 0$ when $\tau_{\op{v};\a}=-1$. Also, the orientation of $\Ga_{A;\a}$, in respect to the counterclockwise one, is 
given precisely by $\tau_{\op{v};\a}$.

\begin{figure}[ht]
\begin{center}
\includegraphics{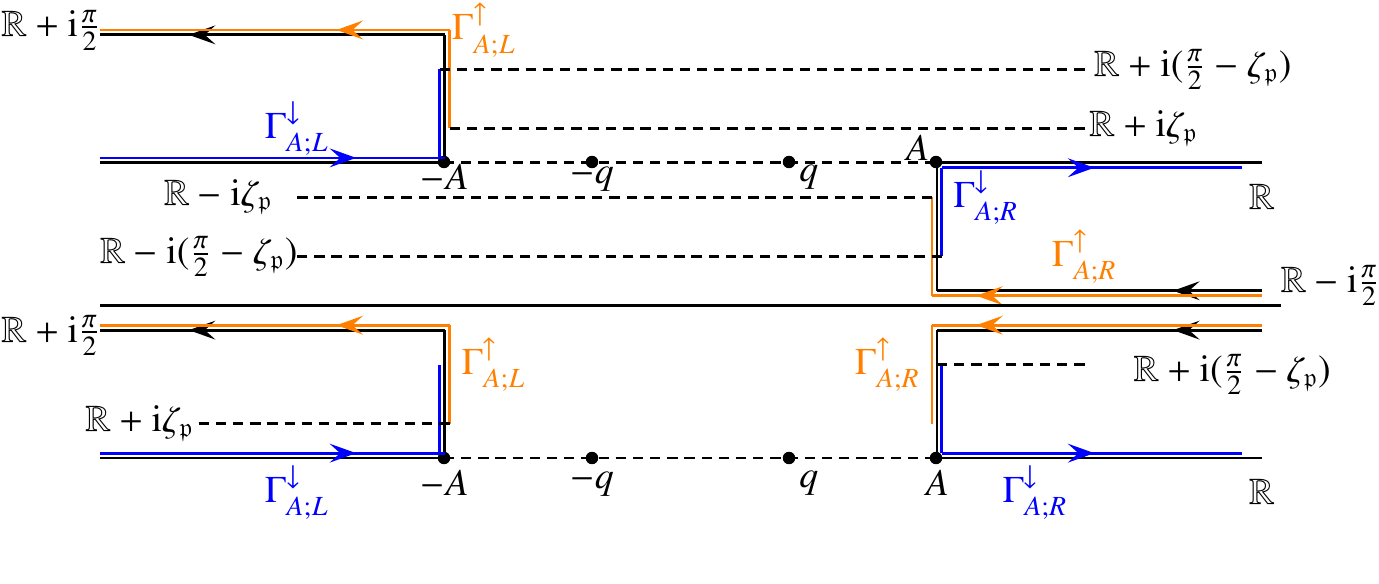}
\caption{  The contours $\Ga_{A}$ -in black- along with the partial contours $\Ga_{A;L/R}^{\ua}$ -in orange- and $\Ga_{A;L/R}^{\da}$ -in blue-. These contours depend on the value of the velocity 
$\op{v}=\tf{m}{t}$. The two situations that are depicted correspond to, from bottom to top,  $\op{v}>\op{v}_{\infty}$ and  $\op{v}_{\infty}>\op{v}>0$. 
\label{Figure contour Gamma Ext left et right pour integrale I300 vers I110} }
\end{center}

\end{figure}

Thus, for $\nu_{2} \in \Ga_{A;\a}$, if $\tau_{\op{v};\a}=1$, then there will be simple poles at
\beq
\nu_2 \; = \; \left\{ \ba{ccccc}  \nu_3 \, + \, \i \zeta_{\mf{p}} & \e{when} &  \nu_3 \in \mf{I}_{\a} \bigcup \big\{\ups_{\a} A + \i \intff{ 0 }{ \tfrac{\pi}{2}-\zeta_{\mf{p}} }   \big\}  & i.e. &    \nu_3 \in \Ga_{A;\a}^{\da}   \vspace{2mm} \\
			      \nu_3 \, -  \, \i \zeta_{\mf{p}} & \e{when} &  \nu_3 \in \Big\{ \mf{I}_{\a} + \i\tfrac{\pi}{2} \Big\} \bigcup \Big\{\ups_{\a} A + \i \intff{ \zeta_{\mf{p}} }{ \tfrac{\pi}{2} }   \Big\}  & i.e. &    \nu_3 \in \Ga_{A;\a}^{\ua}   
\ea \right.  
\enq
where $\mf{I}_{\a}$ is as introduced in \eqref{definition des intervalles I Left et Right} and $\ups_{L}=-1$ while $\ups_{R}=1$. 
Also, for the sake of further convenience, I have made use of the contours $\Ga_{A;\a}^{\da/\ua}$ which are  depicted in Figure \ref{Figure contour Gamma Ext left et right pour integrale I300 vers I110}. 

In the case where  $\tau_{\op{v};\a}=-1$, then there will be simple poles at
\beq
\nu_2 \; = \; \left\{ \ba{ccccc}  \nu_3 \, - \, \i \zeta_{\mf{p}} & \e{when} &  \nu_3 \in \mf{I}_{\a} \bigcup \Big\{\ups_{\a} A + \i \intff{ \zeta_{\mf{p}} - \tfrac{\pi}{2} }{ 0 }   \Big\}  & i.e. &    \nu_3 \in \Ga_{A;\a}^{\da} \vspace{2mm}   \\
			      \nu_3 \, + \, \i \zeta_{\mf{p}} & \e{when} &  \nu_3 \in \Big\{ \mf{I}_{\a} - \i\tfrac{\pi}{2} \Big\} \bigcup \Big\{\ups_{\a} A + \i \intff{ -\tfrac{\pi}{2} }{ -\zeta_{\mf{p}} }   \Big\}  & i.e. &    \nu_3 \in \Ga_{A;\a}^{\ua}   
\ea \right.   \;. 
\enq
Thus, all-in-all, the poles will be located at 
\beq
\nu_2 \; = \; \left\{  \ba{ccc}   \nu_3 \, + \, \i \tau_{\op{v};\a} \zeta_{\mf{p}}    &  \e{provided} \, \e{that}  & \nu_3 \in  \Ga_{A;\a}^{\da}   \vspace{2mm}  \\
				    \nu_3 \, - \, \i \tau_{\op{v};\a} \zeta_{\mf{p}}    & \e{provided} \, \e{that}  & \nu_3 \in  \Ga_{A;\a}^{\ua}   \ea \right.  \;. 
\enq
Thus, taking into account the  orientation $\tau_{\op{v};\a}$ of $\Ga_{A;L/R}$ in respect to the counterclockwise-one, one gets that 
\beq
\mc{I}_{3\hookrightarrow 1 \times 2 }^{(\e{res})} \; = \;  \f{1}{2} \sul{ \a \in \{L,R\} }{}   \Int{ \msc{C}_{1;A} }{} \f{\dd \nu_1}{2\pi} \bigg\{ \Int{  \Ga_{A;\a}^{\ua}   }{}  \f{\dd \nu_3}{2\pi}
w_{\a}(\nu_1,\nu_3,\zeta_{\mf{p}} ) \cdot \wt{J}_{3}\Big( \nu_1, \nu_3 \, - \, \i \tau_{\op{v};\a} \zeta_{\mf{p}}, \nu_3 \Big)\; + \,  (\ua , \zeta_{\mf{p}}) \hookrightarrow (\da , -\zeta_{\mf{p}}) \bigg\} \;. 
\enq
The prefactor takes the form 
\beqa
w_{\a}(\nu_1,\nu_3, \pm \zeta_{\mf{p}} )& = &  \i \tau_{\op{v};\a}  \f{ \sinh^2\big[ \mp \i \tau_{\op{v};\a}  \zeta_{\mf{p}} \big]  }{  \sinh \big[ \mp 2 \i \tau_{\op{v};\a}  \zeta_{\mf{p}} \big]  }
\cdot \Phi_{11}\big( \nu_{13}, \nu_{31} \big) \cdot  \Phi_{11}\Big( \nu_{13} \, \pm \, \i \tau_{\op{v};\a} \zeta_{\mf{p}}, \nu_{31} \, \mp \, \i \tau_{\op{v};\a} \zeta_{\mf{p}} \Big) \\
& = &  \pm \mf{s}_2 \cdot \f{ \sin^2(\zeta) }{ \sin(2\zeta)  } \cdot \f{  \sinh[\nu_{13}]\cdot \sinh[\nu_{13}\, \pm \, \i \mf{s}_2 \tau_{\op{v};\a} \zeta ]  }
{ \sinh[\nu_{13}\, \mp \,  \i \mf{s}_2 \tau_{\op{v};\a} \zeta]\cdot \sinh[\nu_{13}\, \pm \, 2 \i \mf{s}_2 \tau_{\op{v};\a} \zeta ]   } \;. 
\eeqa
Thus, upon using the identities \eqref{ecriture relation J110 et J300}, one gets that 
\beq
\mc{I}_{3\hookrightarrow 1 \times 2 }^{(\e{res})} \; = \;  \f{ \mf{s}_2 }{2} \sul{ \a \in \{L,R\} }{}   \Int{ \msc{C}_{1;A} }{} \f{\dd \nu_1}{2\pi} \bigg\{ \Int{  \Ga_{A;\a}^{\ua}   }{}  \f{\dd \nu_3}{2\pi}
  J_{1,1}\Big( \nu_1, \nu_3 \, - \, \i \mf{s}_2 \tau_{\op{v};\a} \tfrac{\zeta}{2} \Big)\; - \;  
  \Int{  \Ga_{A;\a}^{\da}   }{}  \f{\dd \nu_3}{2\pi}  J_{1,1}\Big( \nu_1, \nu_3 \, + \, \i \mf{s}_2 \tau_{\op{v};\a} \tfrac{\zeta}{2} \Big)\bigg\} \;. 
\enq
Then, since for $\nu_1 \in \msc{C}_{1;A} $ the integrand is analytic for $\big| \Re(\nu_3) \big| \geq A$, one can deform the integration  
contours $ \Ga_{A;\a}^{\ua/\da} $ as
\beq
 \Ga_{A;\a}^{\ua} \hookrightarrow   \i \tau_{\op{v};\a} \zeta_{\mf{p}} \, - \, \mf{I}_{\a} \qquad \e{and} \qquad 
  \Ga_{A;\a}^{\da} \hookrightarrow   \i \tau_{\op{v};\a} \big( \tfrac{\pi}{2} \, - \, \zeta_{\mf{p}} \big) \, + \, \mf{I}_{\a} 
\enq
where the sign in front of $\mf{I}_{\a}$ denotes the interval's orientation. Then, since
\beq
\zeta_{\mf{p}}-\mf{s}_2 \frac{\zeta}{2} \; = \; \frac{\mf{s}_2  }{2}\zeta \; + \; \pi \bs{1}_{\intoo{ \tfrac{\pi}{2} }{\pi}}(\zeta) \;,
\enq
the $\i\pi$ periodicity of the integrand then allows one to recast the integral as 
\beq
\mc{I}_{3\hookrightarrow 1 \times 2 }^{(\e{res})} \; = \;  - \f{ \mf{s}_2 }{2}    \Int{ \msc{C}_{1;A} }{} \f{\dd \nu_1}{2\pi}  \Int{  \mf{J}  }{}  \f{\dd \nu_2}{2\pi}
  J_{1,1}\big( \nu_1, \nu_2 \big)     \;, 
\enq
where 
\beq
\mf{J} \; = \; \bigg\{  \Big\{ \mf{I}_{L} + \i \mf{s}_2 \tau_{\op{v};L} \tfrac{\zeta}{2} \Big\}\cup  \Big\{ \mf{I}_{R} + \i \mf{s}_2 \tau_{\op{v};R} \tfrac{\zeta}{2} \Big\} \bigg\}\bigcup 
\bigg\{  \Big\{ \mf{I}_{L} + \i \mf{s}_2 \tau_{\op{v};L} \tfrac{\pi-\zeta}{2} \Big\}\cup  \Big\{ \mf{I}_{R} + \i \mf{s}_2 \tau_{\op{v};R} \tfrac{\pi-\zeta}{2} \Big\} \bigg\}
\enq
Since
\beq
\mc{I}_{1,1,0}^{(\e{reg})}\; = \;     \mf{s}_2      \Int{ \msc{C}_{1;A} }{} \f{\dd \nu_1}{2\pi}  \Int{  \R  }{}  \f{\dd \nu_2}{2\pi}
  J_{1,1}\big( \nu_1, \nu_2 \big)    \;,
\enq
 it is easy to see that, by using contour deformation, one can cancel the integration along $\mf{J}_{\a}$ occurring in both integrals, 
and that, upon incorporating the sign prefactor in the orientation of the curves,  the central part of the integration curve may be deformed to the contours $\msc{C}_{2;A}$ as depicted in Figures
\ref{Figure contour des 2-cordes apres reduction regime  0 zeta Pi sur 2} and \ref{Figure contour des 2-cordes apres reduction regime  Pi sur 2 zeta pi}. 
Thus, 
\beq
\mc{I}_{3\hookrightarrow 1 \times 2 }^{(\e{res})} \, +  \, \mc{I}_{1,1,0}^{(\e{reg})} \; = \; \Int{ \msc{C}_{1;A} }{} \f{\dd \nu_1}{2\pi}  \Int{   \msc{C}_{2;A}  }{}  \f{\dd \nu_2}{2\pi}
  J_{1,1}\big( \nu_1, \nu_2 \big) 
\enq

\subsubsection*{$\bullet$ The integral $\mc{I}_{3\hookrightarrow 0 \times 3 }^{(\e{res})}  $}

One may, again, compute the $\nu_1$ integral by taking the residues. There are poles at 
\beq
\nu_1 \; = \; \left\{  \ba{ccc}   \nu_2 \, + \, \i \tau_{\op{v};\a} \zeta_{\mf{p}}    &  \e{for}  & \nu_2 \in  \Ga_{A;\a}^{\da}   \vspace{2mm}  \\
				    \nu_2 \, - \, \i \tau_{\op{v};\a} \zeta_{\mf{p}}    & \e{for}  & \nu_2 \in  \Ga_{A;\a}^{\ua}   \ea \right.  \quad \e{and} \quad 
\nu_1 \; = \; \left\{  \ba{ccc}   \nu_3 \, + \, \i \tau_{\op{v};\a} \zeta_{\mf{p}}    &  \e{for}  & \nu_3 \in  \Ga_{A;\a}^{\da}   \vspace{2mm}  \\
				    \nu_3 \, - \, \i \tau_{\op{v};\a} \zeta_{\mf{p}}    & \e{for}  & \nu_3 \in  \Ga_{A;\a}^{\ua}   \ea \right. \;. 
\enq
Analogous calculations to the previous situation lead to 
\bem
\mc{I}_{3\hookrightarrow 0 \times 3 }^{(\e{res})} \; = \;  \f{ \mf{s}_2 }{6} \sul{ \a \in \{L,R\} }{}  \Bigg\{  
\Int{ \Ga_{A;\a}^{\ua} \subset \Ga_{A} }{} \hspace{-3mm} \f{\dd \nu_2 \dd \nu_3 }{ (2\pi)^2 }   J_{1,1}\Big( \nu_3, \nu_2 \, - \, \i \mf{s}_2 \tau_{\op{v};\a} \tfrac{\zeta}{2} \Big)\; +  \hspace{-3mm} 
		  \Int{ \Ga_{A} \subset \Ga_{A;\a}^{\ua}    }{} \hspace{-3mm} \f{\dd \nu_2 \dd \nu_3 }{ (2\pi)^2 }  J_{1,1}\Big( \nu_2, \nu_3 \, - \, \i \mf{s}_2 \tau_{\op{v};\a} \tfrac{\zeta}{2} \Big)  \\
\, -  \hspace{-3mm}  \Int{ \Ga_{A;\a}^{\da} \subset \Ga_{A} }{} \hspace{-3mm} \f{\dd \nu_2 \dd \nu_3 }{ (2\pi)^2 }   J_{1,1}\Big( \nu_3, \nu_2 \, + \, \i \mf{s}_2 \tau_{\op{v};\a} \tfrac{\zeta}{2} \Big)
  \, - \hspace{-3mm} \Int{ \Ga_A \subset  \Ga_{A;\a}^{\da}   }{} \hspace{-3mm} \f{\dd \nu_2 \dd \nu_3 }{ (2\pi)^2 }   J_{1,1}\Big( \nu_2, \nu_3 \, + \, \i \mf{s}_2 \tau_{\op{v};\a} \tfrac{\zeta}{2} \Big)\Bigg\} \\
\; = \;  \f{ \mf{s}_2 }{6} \sul{ \a \in \{L,R\} }{}  \Bigg\{  
\Int{ \substack{ \Ga_{A} \subset \Ga_{A;\a}^{\ua}  \\ \Ga_{A} \supset \Ga_{A;\a}^{\ua}  } }{} \hspace{-3mm} \f{\dd \nu_2 \dd \nu_3 }{ (2\pi)^2 }   J_{1,1}\Big( \nu_2, \nu_3 \, - \, \i \mf{s}_2 \tau_{\op{v};\a} \tfrac{\zeta}{2} \Big)
\, -  \hspace{-3mm} 
\Int{ \substack{ \Ga_{A} \subset \Ga_{A;\a}^{\ua}  \\ \Ga_{A} \supset \Ga_{A;\a}^{\ua}  }  }{} \hspace{-3mm} \f{\dd \nu_2 \dd \nu_3 }{ (2\pi)^2 }   J_{1,1}\Big( \nu_2, \nu_3 \, + \, \i \mf{s}_2 \tau_{\op{v};\a} \tfrac{\zeta}{2} \Big)
\Bigg\} \;. 
\end{multline}
One may push the residue calculation one step further by taking the remaining integrals over $\Ga_{A}$.  In doing so, one needs to distinguish between the case
$ \Ga_{A} \subset \Ga_{A;\a}^{\ua/\da}$ and $\Ga_{A;\a}^{\ua} \subset \Ga_{A}$ since the associated integrals generate slightly different contributions. 

Recall that 
\beq
  J_{1,1}\Big( \nu_2, \nu_3 \, \mp \, \i \mf{s}_2 \tau_{\op{v};\a} \tfrac{\zeta}{2} \Big) \; = \; 
 \f{ \sin^2(\zeta) }{ \sin(2\zeta)  } \cdot \f{  \sinh[\nu_{23}]\cdot \sinh[\nu_{23}\, \pm \, \i \mf{s}_2 \tau_{\op{v};\a} \zeta_{\mf{p}} ]  \cdot  J_{3,0,0}\Big( \nu_2, \nu_3, \nu_3 \, \mp  \, \i  \tau_{\op{v};\a} \zeta_{\mf{p}}  \Big) }
{ \sinh[\nu_{23}\, \mp \,  \i \mf{s}_2 \tau_{\op{v};\a} \zeta_{\mf{p}}]\cdot \sinh[\nu_{23}\, \pm \, 2 \i \mf{s}_2 \tau_{\op{v};\a} \zeta_{\mf{p}} ]   } 
 \;. 
\enq
Thus, the integrand has poles at 
\beq
\nu_2 \; = \; \left\{   \ba{cccc}     \nu_3 + \i \tau_{\op{v};\a} \zeta_{\mf{p}}    &    \e{when}    &  
			      \Ga_{A} \supset \Ga_{A;\a}^{\ua}   \quad \e{and} \quad \nu_3 \in \Big\{ \ups_{\a} A + \ups_{\a} \big[\i \tau_{\op{v};\a} (\tfrac{\pi}{2}\, - \, \zeta_{\mf{p}} \big)  \, ; \,  \i \tau_{\op{v};\a} \zeta_{\mf{p}}  \big] \Big\} \vspace{2mm}  \\ 
					 \nu_3 - 2 \i \tau_{\op{v};\a} \zeta_{\mf{p}}    &    \e{when}    &  
			      \Ga_{A} \supset \Ga_{A;\a}^{\ua} \; \e{or} \; \Ga_{A} \subset \Ga_{A;\a}^{\ua}  \quad \e{and} \quad \nu_3 \in -\mf{I}_{\a} + \i \tau_{\op{v};\a} \tfrac{\pi}{2} \vspace{2mm}  \\
			      
			      &\e{or} &  \Ga_{A} \supset \Ga_{A;\a}^{\ua} \quad \e{and} \quad 
			      \nu_3 \in \Big\{ \ups_{\a} A + \ups_{\a} \big[ \i \tau_{\op{v};\a} \tfrac{\pi}{2} \, ; \,  2 \i \tau_{\op{v};\a}  \zeta_{\mf{p}}   \big] \Big\}   \;
\ea\right. 
\enq
Note however, that the first pole will only exist if also the subsidiary condition holds $\tfrac{\pi}{2}-\zeta_{\mf{p}}> \zeta_{\mf{p}}$, while the second one when  $2 \zeta_{\mf{p}} < \tfrac{\pi}{2}  $. 
Both conditions amount to 
\beq
\zeta \in   \big] 0 \, ; \,   \tfrac{\pi}{4} \big[ \cup  \big] \tfrac{3\pi}{4}  \, ; \, \pi \big[   \;. 
\enq
The remaining poles exist under the same subsidiary condition and take the form 
\beq
\nu_2 \; = \; \left\{   \ba{cccc}     \nu_3 - \i \tau_{\op{v};\a} \zeta_{\mf{p}}    &    \e{when}    &  
			    \Ga_{A}\supset  \Ga_{A;\a}^{\da}    \quad \e{and} \quad \nu_3 \in \Big\{ \ups_{\a} A + \ups_{\a}\big[ \i \tau_{\op{v};\a} (\tfrac{\pi}{2}\, - \, \zeta_{\mf{p}} \big)  \, ; \,  \i \tau_{\op{v};\a} \zeta_{\mf{p}}\big] \Big\} \vspace{2mm}  \\ 
					 \nu_3 + 2 \i \tau_{\op{v};\a} \zeta_{\mf{p}}    &    \e{when}    &  
			      \Ga_{A}\supset  \Ga_{A;\a}^{\da} \; \e{or} \; \Ga_{A}\subset  \Ga_{A;\a}^{\da}   \quad \e{and} \quad \nu_3 \in  \mf{I}_{\a}   \vspace{2mm}  \\
			      
			      & \e{or} &  \Ga_{A}\supset  \Ga_{A;\a}^{\da}  \quad \e{and} \quad 
			      \nu_3 \in \Big\{ \ups_{\a} A + \ups_{\a} \big[ 0 \, ; \, \i \tau_{\op{v};\a} \big( \tfrac{\pi}{2} - 2 \zeta_{\mf{p}} \big)  \big] \Big\}   \;
\ea\right. 
\enq
A straightforward computation then shows that 
\beq
  2\i\pi \tau_{\op{v};\a} \e{Res} \bigg( J_{1,1}\Big( \nu_2, \nu_3 \, \mp \, \i \mf{s}_2 \tau_{\op{v};\a} \tfrac{\zeta}{2} \Big)  \cdot \f{ \dd \nu_2 }{ 2\pi }  \, , \, \nu_2=\nu_3 \pm \i \tau_{\op{v};\a} \zeta_{\mf{p}} \bigg) \; = \; 
\mp \mf{s}_2    J_{0,0,1}\big( \nu_3 \big) \;, 
\enq
while 
\beq
  2\i\pi \tau_{\op{v};\a} \e{Res} \bigg( J_{1,1}\Big( \nu_2, \nu_3 \, \mp \, \i \mf{s}_2 \tau_{\op{v};\a} \tfrac{\zeta}{2} \Big)  \cdot \f{ \dd \nu_2 }{ 2\pi }  \, , \, \nu_2=\nu_3 \mp 2 \i \tau_{\op{v};\a} \zeta_{\mf{p}} \bigg) \; = \; 
\pm \mf{s}_2    J_{0,0,1}\big( \nu_3 \, \mp \, \i \mf{s}_2 \tau_{\op{v};\a} \zeta \big) \;. 
\enq
Thus, all-in-all, the residue calculation yields
\beq
\mc{I}_{3\hookrightarrow 0 \times 3 }^{(\e{res})} \; = \;  \f{ \mf{s}_2 }{6}   \cdot \bs{1}_{ \big] 0 \, ; \,   \tfrac{\pi}{4} \big[ \cup  \big] \tfrac{3\pi}{4}  \, ; \, \pi \big[  }(\zeta) \cdot 
\Big\{ \msc{S}^{(\ua)} \, + \,  \msc{S}^{(\da)} \Big\}
\enq
where, using $\ups_{L}=-1$ and $\ups_{R}=1$, one has  
\beqa
\msc{S}^{(\ua)}  &  = &  -\mf{s}_2 \sul{ \a \in \{L,R\} }{} \ups_{\a}\hspace{-3mm} \Int{ \ups_{\a} A + \i \tau_{\op{v};\a}( \tfrac{\pi}{2} - \zeta_{\mf{p}} )  }{  \ups_{\a} A + \i \tau_{\op{v};\a}  \zeta_{\mf{p}}   }  \hspace{-5mm}  \f{\dd \nu }{ 2\pi } J_{0,0,1}(\nu) 
 \; + \, \mf{s}_2 \sul{ \a \in \{L,R\} }{} \ups_{\a} \hspace{-3mm}  \Int{ \ups_{\a} A + \i \tau_{\op{v};\a}\tfrac{\pi}{2}   }{  \ups_{\a} A + 2 \i \tau_{\op{v};\a}  \zeta_{\mf{p}}   }  \hspace{-5mm}  \f{\dd \nu }{ 2\pi } J_{0,0,1}\Big(\nu \, - \,\i \tau_{\op{v};\a} \mf{s}_2 \zeta \Big) \\
&& \; + \; 2 \mf{s}_2  \sul{ \a \in \{L,R\} }{}\hspace{-2mm}  \Int{ -\mf{I}_{\a}+ \i \tau_{\op{v};\a} \tfrac{\pi}{2}  }{ } \hspace{-5mm} \f{\dd \nu }{ 2\pi }  J_{0,0,1}\Big(\nu \, - \,\i \tau_{\op{v};\a} \mf{s}_2 \zeta \Big)  \\
\msc{S}^{(\ua)}  & = & 2 \mf{s}_2  \sul{ \a \in \{L,R\} }{} \hspace{-1mm}  \Int{ -\mf{I}_{\a}+ \i \tau_{\op{v};\a} ( \tfrac{\pi}{2} - \zeta_{\mf{p}}) }{ } \hspace{-5mm} \f{\dd \nu }{ 2\pi } J_{0,0,1}\big(\nu  \big) 
\eeqa
and similarly, 
\beqa
\msc{S}^{(\da)}  &  = &  -\mf{s}_2 \sul{ \a \in \{L,R\} }{}  \ups_{\a} \hspace{-3mm} 
\Int{  \ups_{\a} A + \i \tau_{\op{v};\a}  ( \tfrac{\pi}{2} - \zeta_{\mf{p}} )  }{ \ups_{\a} A + \i \tau_{\op{v};\a} \zeta_{\mf{p}}  }  \hspace{-5mm}  \f{\dd \nu }{ 2\pi } J_{0,0,1}(\nu) 
 \; + \, \mf{s}_2 \sul{ \a \in \{L,R\} }{}  \ups_{\a} \hspace{-3mm}  
 \Int{  \ups_{\a} A +   \i \tau_{\op{v};\a}  (\tfrac{\pi}{2}-2\zeta_{\mf{p}} )  }{ \ups_{\a} A     }  \hspace{-5mm}  \f{\dd \nu }{ 2\pi } J_{0,0,1}\Big(\nu \, + \,\i \tau_{\op{v};\a} \mf{s}_2\zeta \Big) \\
&& \; + \; 2 \mf{s}_2  \sul{ \a \in \{L,R\} }{}   \Int{  \mf{I}_{\a} }{ }  \f{\dd \nu }{ 2\pi }  J_{0,0,1}\Big(\nu \, + \,\i \tau_{\op{v};\a} \mf{s}_2 \zeta \Big)  \\
\msc{S}^{(\da)}  & = & 2 \mf{s}_2  \sul{ \a \in \{L,R\} }{}   \Int{ -\mf{I}_{\a}+ \i \tau_{\op{v};\a}  \zeta_{\mf{p}}  }{ } \hspace{-5mm} \f{\dd \nu }{ 2\pi } J_{0,0,1}\big(\nu  \big) \;. 
\eeqa
Thus, one gets 
\beq
\mc{I}_{3\hookrightarrow 0 \times 3 }^{(\e{res})} \; = \;  \f{ 1 }{3}   \cdot \bs{1}_{ \big] 0 \, ; \,   \tfrac{\pi}{4} \big[ \cup  \big] \tfrac{3\pi}{4}  \, ; \, \pi \big[  }(\zeta) \cdot 
\Int{ \msc{J}_{\e{eff}}  }{ } \f{\dd \nu }{ 2\pi } J_{0,0,1}\big(\nu  \big) 
\enq
where 
\beq
\msc{J}_{\e{eff}} \; = \; \bigg\{ \cup_{\a \in \{L,R\} } \Big\{  -\mf{I}_{\a} + \i\tau_{\op{v};\a}  (\tfrac{\pi}{2}- \zeta_{\mf{p}} )   \Big\}  \bigg\} \bigcup
 \bigg\{ \cup_{\a \in \{L,R\} } \Big\{  \mf{I}_{\a} + \i\tau_{\op{v};\a}   \zeta_{\mf{p}}    \Big\}  \bigg\} \;. 
\enq
Note that the intervals $\mf{I}_{\a}$ appear with opposite orientations, so that one can deform the contours to the curves $ \msc{J}_{A;\op{v}}$ as defined in \eqref{definition interval final J A v} and get 
\beq
\mc{I}_{3\hookrightarrow 0 \times 3 }^{(\e{res})} \; = \;  \f{ 1 }{3}   \cdot \bs{1}_{ \big] 0 \, ; \,   \tfrac{\pi}{4} \big[ \cup  \big] \tfrac{3\pi}{4}  \, ; \, \pi \big[  }(\zeta) \cdot 
\Int{ \msc{J}_{ A; \op{v} }  }{ } \f{\dd \nu }{ 2\pi } J_{0,0,1}\big(\nu  \big)  \;. 
\enq
Upon putting all the intermediate rewritings together, the claim follows. \qed

\subsection{A conjectural contour deformation in the general case}
\label{Sous Section conjecture generale sur forme contours deformes}

It is clear that if the "minimal structure property" holds then, for $|\op{v}|<\op{v}_{\infty}$, the portion of the curve $\msc{C}_{2;A}$, resp.  $\msc{C}_{3;A}$,
that is located in the neighbourhood of $\om_2$, resp. $\om_3$, may be chosen so that it coincides with the curve
\beq
\msc{J}_r \, = \, \om_r \; + \;  \mf{s}_r (-1)^{\sg_r} \cdot h_r^{-1}\Big(    \intoo{ -\ex{ \i\f{\pi}{4} \veps_{r} }\eta }{ \ex{ \i\f{\pi}{4} \veps_{r} } \eta } \Big) \;, \quad r=2 \; \e{or} \; 3 \,, 
\enq
in which $h_r, \veps_r$ are as introduced in \eqref{definition ur et vepsr r corde generale}.

In fact, should more saddle-points exist for $r=2$ or $3$
then one may deform the curves $\msc{C}_{3;A}$ to pass through the points $\om_{r}^{(a)}$ and to coincide, in the neighbourhood of the $\om_{r}^{(a)}$
with the curves 
\beq
\msc{J}_r^{(a)} \, = \, \om_r^{(a)} \; + \;  \mf{s}_r (-1)^{\sg_r} \cdot \big(h_r^{(a)})^{-1}\Big(    \intoo{ -\ex{ \i\f{\pi}{4} \veps_{r}^{(a)} }\eta }{ \ex{ \i\f{\pi}{4} \veps_{r}^{(a)} } \eta } \Big) \;,  
\enq
$r=2$ or $3$ and such that $\e{inf}_{\la \in \msc{C}_{r;A}\setminus \cup_{a}  \msc{J}_r^{(a)} } \big\{ \Im[u_r(\la;\op{v})] \big\} >0$.

Based on the results for the $n=2$ and $n=3$ sectors, I formulate the below conjecture. 
  
\begin{conj}
\label{Conjecture deformation contour integration secteur n excitation massive}

Assume that the "minimal structure property" holds. Then, the combination of integrals $\mc{I}_{\e{tot} }^{(n)}$ introduced in \eqref{definition I tot n}
 may be recast in the form 
\beq
\mc{I}_{\e{tot}}^{(n)}\; = \; \sul{ \bs{n} \in \mf{S}_n }{} \wt{\mc{I}}_{\bs{n}}  \qquad \e{where} \qquad 
\wt{\mc{I}}_{\bs{n}} \; =  \hspace{-4mm} \Int{   \msc{C}_{1;A} \supset \cdots \supset  \msc{C}_{1;A} }{} \hspace{-5mm}  \f{ \dd^{n_1}\nu^{(1)} }{ (2\pi)^{n_1} \cdot n_1! } 
\pl{r \in \mf{N}_m }{ } \bigg\{ \Int{ \big( \msc{C}_{r;A}^{(\e{mod})} \big)^{n_{r}} }{} \hspace{-4mm}  \f{ \dd^{n_r}\nu^{(r)} }{ (2\pi)^{n_r} \cdot n_r! } \bigg\} \cdot J_{\bs{n}}\big( \bs{\nu}_{\bs{n}})\;. 
\enq
 For $r\in \mf{N}_{\e{st}}$, there exists compact curves  $\msc{J}_r,  \msc{J}_{r;1}, \dots,  \msc{J}_{r;m_r}$ in $\Cx$ such that the new contours can be expressed as 
\beq
\msc{C}_{r;A}^{(\e{mod})} \, = \, \bs{1}_{\intoo{ - \op{v}_{\infty} }{ \op{v}_{\infty} }  }(\op{v}) \msc{J}_r \, + \,
\sul{a=1}{m_r} \mf{c}_{r;a} \msc{J}_{r;a}
\enq
where,  given $h_r$ as introduced in \eqref{definition ur et vepsr r corde generale}, 
\begin{itemize}

 \item $\msc{J}_r \, = \,  \om_r \,  + \, \mf{s}_r (-1)^{\sg_r} \cdot h_r^{-1}\Big(    \big] -\eta \ex{ \i\f{\pi}{4} \veps_{r} } \, ; \, \eta \ex{ \i\f{\pi}{4} \veps_{r} } \big[  \Big)$ 
 where $\veps_{r}=\e{sgn}\Big( u_r^{\prime\prime}(\om_r, \op{v}) \Big)$ and $\eta>0$ is small enough;
 \item $\inf_{\la \in  \msc{J}_{r;1}\cup \dots\cup  \msc{J}_{r;m_r} } \Big( \Im(u_r(\la,\op{v})) \Big) \geq  c_r >0$ . 

\end{itemize}

\end{conj}

The conjecture has to be slightly modified in the general case

\begin{conj}
\label{Conjecture deformation contour integration secteur n excitation massive cas general}

The combination of integrals $\mc{I}_{\e{tot} }^{(n)}$ introduced in \eqref{definition I tot n}
 may be recast in the form 
\beq
\mc{I}_{\e{tot}}^{(n)}\; = \; \sul{ \bs{n} \in \mf{S}_n }{} \wt{\mc{I}}_{\bs{n}}  \qquad \e{where} \qquad 
\wt{\mc{I}}_{\bs{n}} \; =  \hspace{-4mm} \Int{   \msc{C}_{1;A} \supset \cdots \supset  \msc{C}_{1;A} }{} \hspace{-5mm}  \f{ \dd^{n_1}\nu^{(1)} }{ (2\pi)^{n_1} \cdot n_1! } 
\pl{r \in \mf{N}_m }{ } \bigg\{ \Int{ \big( \msc{C}_{r;A}^{(\e{mod})} \big)^{n_{r}} }{} \hspace{-4mm}  \f{ \dd^{n_r}\nu^{(r)} }{ (2\pi)^{n_r} \cdot n_r! } \bigg\} \cdot J_{\bs{n}}\big( \bs{\nu}_{\bs{n}})\;. 
\enq
 For $r\in \mf{N}_{\e{st}}$, there exists compact curves  $\msc{J}_r,  \msc{J}_{r;1}, \dots,  \msc{J}_{r;m_r}$ in $\Cx$ such that the new contours can be expressed as 
\beq
\msc{C}_{r;A}^{(\e{mod})} \, = \, \bs{1}_{\intoo{ - \op{v}_{r}^{(M)} }{ \op{v}_{r}^{(M)}  }}(\op{v}) \sul{a=1}{\vsg_r} \msc{J}_r^{(a)} \, + \,
\sul{a=1}{m_r} \mf{c}_{r;a} \msc{J}_{r;a}
\label{ecriture decomposition de la courbe C r A mod}
\enq
where,  given $h_r^{(a)}$ as introduced in \eqref{definition ur et vepsr r corde case le plus general de point selle}, 
\begin{itemize}

 \item $\msc{J}_r^{(a)} \, = \,  \om_r^{(a)} \,  + \, \mf{s}_r (-1)^{\sg_r} \cdot (h_r^{(a)})^{-1}\Big(    \big] -\eta \ex{ \i\f{\pi}{4} \veps_{r}^{(a)} } \, ; \, \eta \ex{ \i\f{\pi}{4} \veps_{r}^{(a)} } \big[  \Big)$ 
 where $\veps_{r}^{(a)}=\e{sgn}\Big( u_r^{\prime\prime}(\om_r^{(a)}, \op{v}) \Big)$;
 \item $\inf_{\la \in  \msc{J}_{r;1}\cup \dots\cup  \msc{J}_{r;m_r} } \Big( \Im(u_r(\la,\op{v})) \Big) \geq  c_r >0$ . 

\end{itemize}

\end{conj}

One could surely apply the strategy outlined in Section \ref{Section deformation des contours dans integrales auxiliaires} to
establish the conjecture for sectors with $n$ greater than 3. However, such an analysis would probably not bring more structure to the problem
and it would rather be desirable to unravel a more general combinatorial structure allowing one to simplify the pole computation and thus to prove the conjecture. 
It appears suggestive to interpret $\mc{I}^{(n)}_{\e{tot}}$ as some multidimensional residue. Such a presentation should allow for a more geometric
proof of Conjecture \ref{Conjecture deformation contour integration secteur n excitation massive}. However, so far, the effective way of doing so 
has eluded me.

\section{ The steepest descent analysis of the form factor series} 
\label{Section steepest descent analysis}

In this section, I will carry out the steepest descent analysis of the series. I shall mostly focus on the structurally simplest situation, namely the 
one grasped by the range of parameters where the "minimal structure property" holds, \textit{c.f.} Subsection \ref{Sous Section preprietes pour locus saddel pts} for the details. 
I will treat the general case in the last subsection.

\begin{figure}[ht]
\begin{center}

\includegraphics{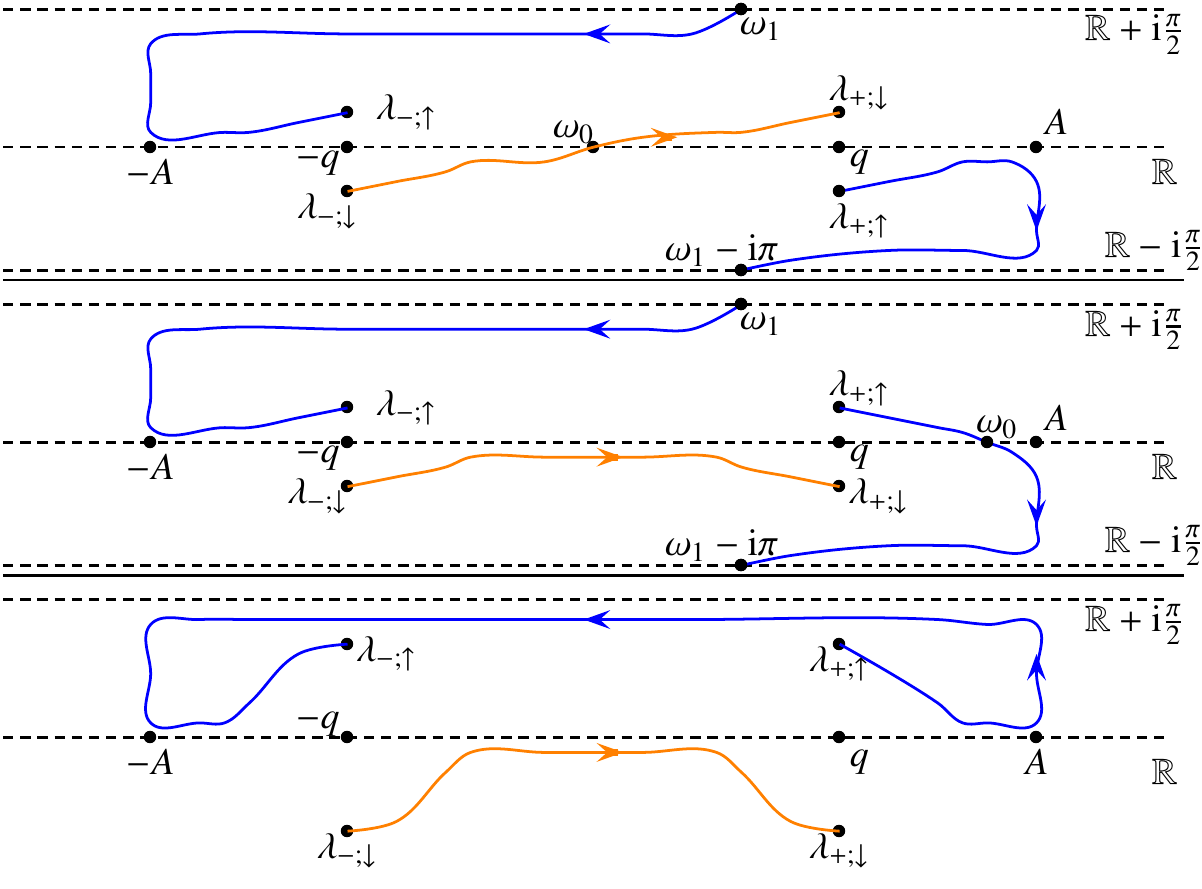}
\caption{  The particle $\msc{C}_{1;A}^{(\e{tot})}$ -in blue- and the hole $\msc{C}_{h}^{(\e{tot})}$ -in orange- contours after deformations towards a steepest descent path. The contours are plotted for three regimes of the velocity 
$\op{v}=\tf{m}{t}$ appearing from bottom to top $\op{v}>\op{v}_{\infty}$, $\op{v}_{\infty}>\op{v}>\op{v}_F$ and $\op{v}_F> \op{v} >0$. 
The shape of the curves in the regimes associated with negative $\op{v}$s can be deduced by symmetry.  
\label{Figure contour steepest descent pour particules et trous} }
\end{center}

\end{figure}

Upon inserting the decomposition of contours provided by Conjecture \ref{Conjecture deformation contour integration secteur n excitation massive}, 
the building blocks $\mc{S}_{\bs{n}}(m,t)$ of the  massless form factor series expansion can be recast as 
\beq
\mc{S}_{\bs{n}}(m,t) \; = \;  \pl{ r \in  \mf{N}_{\e{st}} }{} \;\Bigg\{  \Int{ \big( \msc{C}_{r;A}^{(\e{mod})} \big)^{n_r} }{} \hspace{-3mm} \f{ \dd^{n_r}\nu^{(r)} }{ n_r! \cdot (2\pi)^{n_r}  }   \Bigg\} 
\cdot \hspace{-4mm} \Int{   \msc{C}_{1;A} \supset \cdots \supset  \msc{C}_{1;A} }{} \hspace{-5mm}  \f{ \dd^{n_1}\nu^{(1)} }{ (2\pi)^{n_1} \cdot n_1! } 
\cdot \hspace{-3mm} \Int{ \big( \msc{C}_{h} \big)^{n_h}   }{} \hspace{-3mm} \f{ \dd^{n_h}\mu  }{ n_h! \cdot (2\pi)^{n_h}  } \cdot
   \mc{F}^{(\ga)}_{\e{tot}} \big( \mf{Y} \big)  \;, 
\enq
where
\beq
 \mc{F}^{(\ga)}_{\e{tot}} \big( \mf{Y} \big) \; = \; \f{   \mc{F}^{(\ga)}\big( \mf{Y} \big) \cdot    \ex{\i     m \msc{U}(\mf{Y},\op{v})  }    }{  \pl{\ups=\pm}{}  \big[  -  \i (\ups m - \op{v}_F t)   \big]^{   \vth_{\ups}^2(\mf{Y}) }  }  
\cdot  \bigg( 1+ \mf{r}_{\de,m, t}\big( \mf{Y} \big) \bigg) \;. 
\enq
By invoking the analyticity of the integrand for particle rapidities in the vicinity of  $\msc{C}_{1;A}$ and hole rapidities in the vicinity of $\msc{C}_{h}$, 
one deforms the particle contours $ \msc{C}_{1;A}$ and the hole contour  $\msc{C}_{h}$ to the steepest descent contours $\msc{C}_{1;A}^{(\e{tot})}$ and $\msc{C}_{h}^{(\e{tot})}$
depicted in Figure \ref{Figure contour steepest descent pour particules et trous}. It is clear that, in doing so, one may choose the shape of the
curves in the vicinity of the saddle-points  $\om_1$ (or any of its $\i\pi$-translates) and $\om_0$ to be given by 
\beq
\msc{J}_1 \; = \; \om_1 \, - \, h_1^{-1}\Big(    \big] -\eta \ex{ \i\f{\pi}{4} \veps_{1} } \, ; \, \eta \ex{ \i\f{\pi}{4} \veps_{1} } \big[  \Big) \qquad \e{and} \qquad 
\msc{J}_0 \; = \;  \om_0 \, + \,   h_0^{-1}\Big(    \big] -\eta \ex{ \i\f{\pi}{4} \veps_{0} } \, ; \, \eta \ex{ \i\f{\pi}{4} \veps_{0} } \big[  \Big) \;. 
\enq
There $\veps_{0} = \e{sgn}\big[ u_1^{\prime\prime}(\om_0,\op{v})\big] = -1$ and $\veps_{1} = \e{sgn}\big[ u_1^{\prime\prime}(\om_1,\op{v})\big] = 1$. 
The parameterisation of the curves in the vicinity of the points $\la_{\pm;\ua/\da}$ may also be chosen as desired. It is convenient to make the following choice
of the local parameterisation
\beq
\msc{J}_{1;\eps} \;  = \; u_1^{-1}\Big( u_{1}(\la_{\eps;\ua},\op{v}) \,+\, \eps \intff{0}{\i\eta} , \op{v}   \Big) \qquad \e{and} \qquad 
\msc{J}_{0;\eps} \;  = \; u_1^{-1}\Big( u_{1}(\la_{\eps;\da},\op{v}) \, - \, \eps \intff{0}{-\i\eta} , \op{v}   \Big)
\label{definition des courbes integration voisinage zone Fermi}
\enq
where I remind that  $u_{1}(\la_{\eps;\ua},\op{v})  = \i \de +  u_{1}( \eps q,\op{v})$ while $u_{1}(\la_{\eps;\ua},\op{v}) = - \i \de +  u_{1}( \eps q,\op{v})$, 
\textit{c.f.} Figure \ref{Figure contour des particules et trous delta deformes}. Thus, it follows that, analogously to the decomposition \eqref{ecriture decomposition de la courbe C r A mod}, it holds
\beqa
\msc{C}^{ (\e{tot}) }_{1;A} & = & \bs{1}_{ \mf{I}_{\e{SL}} } (\op{v}) \msc{J}_0 \; + \; \bs{1}_{\mf{I}_{\infty} }(\op{v}) \msc{J}_{1}
\, + \, \sul{\eps=\pm}{} \msc{J}_{1;\eps} \; + \; \sul{a=1}{m_1} \msc{J}_{1;a} \;,  \\
\msc{C}^{ (\e{tot}) }_{h} & = & \bs{1}_{ \mf{I}_{\e{TL}} } (\op{v}) \msc{J}_0  
\, + \, \sul{\eps=\pm}{} \msc{J}_{0;\eps} \; + \; \sul{a=1}{m_1} \msc{J}_{0;a}  \;, 
 \eeqa
where I have introduced 
\beq
\mf{I}_{\infty} \; = \; \intoo{-\op{v}_{\infty} }{ \op{v}_{\infty}} \;\; , \quad 
\mf{I}_{\e{SL}}=\intoo{-\op{v}_{\infty} }{ \op{v}_{\infty}} \setminus \intff{-\op{v}_F}{\op{v}_F} \quad \e{and} \quad \mf{I}_{\e{TL}}= \intoo{-\op{v}_F}{\op{v}_F} \;.
\label{definition I infty et SL et TL}
\enq

\subsection{The algebraically dominant part of  $\mc{S}_{\bs{n}}(m,t)$}
\label{Sous section partie algebriquement dominante de Sn}

The curves $\msc{C}_{r;A}^{(\e{mod})}$ and  $\msc{C}_{h}^{(\e{mod})}$ split into various pieces. 
 Some of these are only responsible for generating exponentially small corrections in $m$ to the 
 building block $\mc{S}_{\bs{n}}(m,t)$ of the form factor series while other produce algebraically falling-off corrections. 
 In this subsection, I  decompose  $\mc{S}_{\bs{n}}(m,t)$ in two parts according to their type of contribution to the 
 asymptotic expansion.  
 
\subsubsection{Vicinity of the endpoints of the Fermi zone}
\label{Sous Section voisinage des points de Fermi}

Recall that $\mc{S}_{\bs{n}}(m,t)$ depends on a regularising parameter $\de>0$, which is arbitrary but should be taken strictly positive $\de>0$. 
In particular, the remainder depends on $\de$ and is controlled as given in \eqref{ecriture du controle sur le reste}. 
This means that, in order to be able to neglect the contributions of the remainder in the large $(m,t)$ limit, one 
should choose $\de$ to go to zero in an $(m,t)$-dependent way. For instance, choosing 
\beq
\de= \f{ 1 }{ m^{1-\frac{\tau}{2}} } \quad \e{yields} \quad  \mf{r}_{\de,m, t}\big( \mf{Y} \big) \; = \; \e{O}\Big(  m^{\tau-1 }  \Big)
\label{ecriture controle simplifie reste}
\enq
where $\tau>0$ can be taken small enough. Still, the curves $\msc{J}_{a;\eps}$ \eqref{definition des courbes integration voisinage zone Fermi}, $a=0,1$,
which represent the behaviour of the particle and hole contours in the vicinity of the endpoints $\eps q$ of the Fermi zone, also depend on $\de$.  
In particular, since $\de \tend 0$ with $m$ the endpoints of these curves will collapse to $\eps q$ with rate $\de$. Thus, one should show that 
these curves continue to produce only exponentially small contributions in $m$ to the asymptotic behaviour of $\mc{S}_{\bs{n}}(m,t)$ 
even if $\de \tend 0$ with $m$. 

By using the symmetry of the integrand under any permutations of the rapidities $\{\nu_a^{(1)}\}$ or $\{\mu_a\}$ and the decomposition 
\beq
\msc{C}^{(\e{tot})}_{1;A} \; = \; \msc{J}_{1;+}\cup\msc{J}_{1;-} \cup \msc{C}^{(\e{mod})}_{1;A} \;, \qquad
\msc{C}^{(\e{tot})}_{h} \; = \; \msc{J}_{0;+}\cup\msc{J}_{0;-} \cup \msc{C}^{(\e{mod})}_{h}
\enq
one may recast  $\mc{S}_{\bs{n}}(m,t)$ in the form 
\bem
\mc{S}_{\bs{n}}(m,t) \; = \; \sul{ \substack{  n_{h;+}, n_{h;-}, \wt{n}_{h} \geq 0  \\ n_{h;+}+n_{h;-}+\wt{n}_{h}=n_h} }{}  \;  \sul{ \substack{  n_{1;+}, n_{1;-}, \wt{n}_{1} \geq 0  \\ n_{1;+}+n_{1;-}+\wt{n}_{1}=n_1} }{} \; 
\pl{ r \in  \mf{N}_{\e{st}} }{} \;\Bigg\{  \Int{ \big( \msc{C}_{r;A}^{(\e{mod})} \big)^{n_r} }{} \hspace{-3mm} \f{ \dd^{n_r}\nu^{(r)} }{ n_r! \cdot (2\pi)^{n_r}  }   \Bigg\}  \\
\times  \hspace{-8mm} \Int{   \msc{C}_{1;A}^{(\e{mod})} \supset \cdots \supset  \msc{C}_{1;A}^{(\e{mod})} }{} \hspace{-7mm}  \f{ \dd^{ \wt{n}_{1}}\nu^{(1)} }{ (2\pi)^{ \wt{n}_{1}} \cdot \wt{n}_{1}! } 
\cdot \hspace{-3mm} \Int{ \big( \msc{C}_{h}^{(\e{mod})} \big)^{\wt{n}_{h}}   }{} \hspace{-3mm} \f{ \dd^{\wt{n}_{h}}\mu  }{ \wt{n}_{h}! \cdot (2\pi)^{\wt{n}_{h}}  } \cdot
\msc{S}_{\bs{n}_{\e{tot}} } \big( \mf{Y}_{\e{out}} \big) \;. 
\label{ecriture partition Sn en sous integrale proches zone de Fermi}
\end{multline}
There I agree upon 
\beqa
\bs{n}_{\e{tot}} & = & \Big( \ell_+, \ell_-\, ; \, \big( n_{h;+}, n_{h;-}, \wt{n}_{h}\big), \big( n_{1;+}, n_{1;-}, \wt{n}_{1} \big), n_{r_2},\dots, n_{r_k}, \dots  \Big) \\
 \mf{Y}_{\e{out}} & = & \Big\{ -\op{s}_{\ga};  \big\{ \ell_{\ups}\big\} \; \big| \; \big\{ \mu_{a} \big\}_{1}^{ \wt{n}_{h} } \, ; \, \big\{ \nu_{a}^{(r_1)} \big\}_{a=1}^{ \wt{n}_1}; \dots ;  \big\{ \nu_{a}^{(r_k)} \big\}_{a=1}^{n_{r_k}}, \dots  \Big\}
\eeqa
compare with \eqref{definition vecteur entier pour nombres excitations} and  \eqref{definition rapidites massives reduites}. Also, I have introduced 
\beq
 \msc{S}_{\bs{n}_{\e{tot}} } \big( \mf{Y}_{\e{out}} \big)  \; = \;\pl{\eps=\pm }{} 
 \Bigg\{ \Int{ \big( \msc{J}_{0;\eps} \big)^{n_{h;\eps}}   }{} \hspace{-3mm} \f{ \dd^{ n_{h;\eps} }\mu  }{ n_{h;\eps}! \cdot (2\pi)^{ n_{h;\eps} }  } \; \cdot  \hspace{-3mm}
 \Int{ \big( \msc{J}_{1;\eps} \big)^{n_{1;\eps}}   }{} \hspace{-3mm} \f{ \dd^{ n_{1;\eps} }\nu^{(1;\eps)}  }{ n_{1;\eps}! \cdot (2\pi)^{ n_{1;\eps} }  } \Bigg\} \cdot  \mc{F}^{(\ga)}_{\e{tot}} \big( \mf{Y}_{\e{tot}} \big)
\enq
in which, for $\mf{Y}_{\e{out}}$ as introduced above, it is understood that 
\beq
\mf{Y}_{\e{tot}} \; = \; \bigg\{ -\op{s}_{\ga};  \big\{ \ell_{\ups}\big\} \; \big| \; 
\Big\{ \big\{ \mu_{a} \big\}_{1}^{ \wt{n}_{h} }  \cup \big\{ \mu_{a}^{(+)} \big\}_{1}^{ n_{h;+} } \cup \big\{ \mu_{a}^{(-)} \big\}_{1}^{ n_{h;-} }    \Big\} 
\, ; \, \Big\{ \big\{ \nu_{a}^{(1)} \big\}_{1}^{ \wt{n}_{1} }  \cup \big\{ \nu_{a}^{(1;+)} \big\}_{1}^{ n_{1;+} } \cup \big\{ \nu_{a}^{(1;-)} \big\}_{1}^{ n_{1;-} }    \Big\} \, ; \, 
\big\{ \nu_{a}^{(r_2)} \big\}_{a=1}^{ n_2}; \dots   \bigg\} \;. 
\nonumber
\enq

The local behaviour of the form factor density $\mc{F}^{(\ga)} \big( \mf{Y}_{\e{tot}} \big)$ when some of the particle or hole rapidities approach the Fermi boundary 
is given in \eqref{ecriture comportement local FF smooth a rapidite coincidantes}-\eqref{ecriture fonction singuliere D}. Thus, it is easy to convince oneself 
that one has the factorisation 
\beq
\mc{F}^{(\ga)}_{\e{tot}} \big( \mf{Y}_{\e{tot}} \big) \; = \; \ex{\i m \msc{U}(\mf{Y}_{\e{out}},\op{v}) }\cdot  \pl{\eps= \pm }{}
\bigg\{ \mc{W}_{\eps}\Big(  \big\{ \mu_{a}^{(+)} \big\}_{1}^{ n_{h;+} } \, ; \, \big\{ \nu_{a}^{(1;+)} \big\}_{1}^{ n_{1;+} } \Big)\big[ \vth_{\e{tot}} \big] \bigg\} \cdot \mc{F}^{(\ga)}_{\e{red}} \big( \mf{Y}_{\e{tot}} \big) \;. 
\enq
There, $\mc{F}^{(\ga)}_{\e{red}} \big( \mf{Y}_{\e{tot}} \big)$ is a smooth function in respect to the $\pm$ type particle or hole rapidities located in a neighbourhood of the Fermi zone $\pm q$. Also,
the sole dependence on $(m,t)$ of this reduced form factor density arises on the level of the remainder. The latter is still controlled as in \eqref{ecriture controle simplifie reste}, 
and this control is differentiable in respect to the rapidities contained in  $\mf{Y}_{\e{tot}}$. 
The prefactor containing $\msc{U}(\mf{Y}_{\e{out}},\op{v}) $ contains all the oscillatory in $m$ behaviour relatively to the rapidities 
contained in $\mf{Y}_{\e{out}}$. The functions $\vth_{\e{tot}} $ are defined as 
\beq
\vth_{\e{tot}} \Big( x \mid \big\{ \mu_{a}^{(+)} \big\}_{1}^{ n_{h;+} } \, ; \, \big\{ \nu_{a}^{(1;+)} \big\}_{1}^{ n_{1;+} }  \Big) \; = \; 
\vth\big(x \mid  \mf{Y}_{\e{tot}} \big)  \;, 
\enq
with $\vth$ as introduced in \eqref{definition phase habilee totale excitation}. Finally, the prefactors $\mc{W}_{\eps}$ gather all the singularities of the integrand
in the vicinity of the endpoints of the Fermi zone. Given a set of rapidities
\beq
\Ups \; = \; \Big\{  \big\{ \mu_{a} \big\}_{1}^{ n_{h} } \, ; \, \big\{ \nu_{a} \big\}_{1}^{ n_{p} } \Big\}  \quad \e{and} \,\e{a} \, \e{function} \quad  \vp( x\mid \Ups)
\enq
this prefactor is defined as 
\bem
\mc{W}_{\eps}\big( \Ups \big)\big[\vp\big] \; = \; \pl{a=1}{n_p} \bigg\{ \ex{\i m u_1(\nu_a,\op{v})}  \Big[ \eps \, \Big( u_1(\nu_a,\op{v}) \, - \, u_1(\eps q,\op{v})  \Big) \Big]^{-2 \eps \vp(\nu_a\mid \Ups)} 
\cdot u_1^{\prime}(\nu_a,\op{v}) \bigg\}  \cdot \pl{a=1}{n_h} \Big\{ u_1^{\prime}(\mu_a,\op{v}) \Big\}  \\
\times  \pl{a=1}{n_h} \bigg\{ \ex{- \i m u_1(\mu_a,\op{v})}  \Big[ \eps \, \Big( u_1(\mu_a,\op{v}) \, - \, u_1(\eps q,\op{v})  \Big) \Big]^{2 \eps \vp(\mu_a\mid \Ups)}  \bigg\} \cdot 
\f{ \pl{ a < b }{ n_p } \Big[ u_1(\nu_a,\op{v}) \, - \, u_1(\nu_b,\op{v})  \Big]^2 \cdot  \pl{ a < b }{ n_h } \Big[ u_1(\mu_a,\op{v}) \, - \, u_1(\mu_b,\op{v})  \Big]^2   }
{ \pl{ a=1 }{ n_p } \pl{ b = 1 }{ n_h } \Big[ u_1(\nu_a,\op{v}) \, - \, u_1(\mu_b,\op{v})  \Big]^2  } \;. 
\end{multline}
The large-$m$ behaviour of the class of integrals arising in the definition of $ \msc{S}_{\bs{n}_{\e{tot}} } \big( \mf{Y}_{\e{out}} \big)$ is analysed in Lemma \ref{Lemme cptm gd m integrales modele bord zone Fermi}
below. It allows one  to conclude that, at least for bounded $n_h$ and $n_p$, when $\de=m^{-1+\tf{\tau}{2}}$, it holds 
\beq
 \msc{S}_{\bs{n}_{\e{tot}} } \big( \mf{Y}_{\e{out}} \big) \; = \; \e{O}\Big( \ex{ - m^{ \frac{1}{2}\tau } }  \Big)
\qquad \e{whenever} \qquad n_{h;\eps}\not= 0  \; \e{or} \;  n_{1;\eps}\not= 0
\enq
for some $\eps\in \{+,-\}$. Thus, it follows that one has 
\beq
\mc{S}_{\bs{n}}(m,t)\; = \; \mc{S}_{\bs{n}}^{(\e{mod})}(m,t) \; + \; \e{O}\Big( \ex{ - m^{ \frac{1}{2} \tau } }  \Big)
\enq
where 
\beq
\mc{S}_{\bs{n}}^{(\e{mod})}(m,t) \; = \;  \pl{ r \in  \mf{N}_{\e{st}} }{} \;\Bigg\{  \Int{ \big( \msc{C}_{r;A}^{(\e{mod})} \big)^{n_r} }{} \hspace{-3mm} \f{ \dd^{n_r}\nu^{(r)} }{ n_r! \cdot (2\pi)^{n_r}  }   \Bigg\} 
\cdot \hspace{-4mm} \Int{   \msc{C}_{1;A}^{(\e{mod})} \supset \cdots \supset  \msc{C}_{1;A}^{(\e{mod})} }{} \hspace{-5mm}  \f{ \dd^{n_1}\nu^{(1)} }{ (2\pi)^{n_1} \cdot n_1! } 
\cdot \hspace{-3mm} \Int{ \big( \msc{C}_{h}^{(\e{mod})} \big)^{n_h}   }{} \hspace{-3mm} \f{ \dd^{n_h}\mu  }{ n_h! \cdot (2\pi)^{n_h}  } \cdot
   \mc{F}^{(\ga)}_{\e{tot}} \big( \mf{Y} \big)  \;.
\enq
This representation only involves integrations over curves that are located uniformly away from the endpoints of the Fermi zone for the particle and hole rapidities.

\begin{lemme}
 \label{Lemme cptm gd m integrales modele bord zone Fermi}
 Let $\mc{G}(\Ups)$ be  smooth and symmetric in each subset of rapidities building up $\Ups \; = \; \Big\{  \big\{ \mu_{a} \big\}_{1}^{ n_{h} } \, ; \, \big\{ \nu_{a} \big\}_{1}^{ n_{p} } \Big\} $. 
Then, the integral 
\beq
\mc{M}^{(\eps)}_{n_h,n_p} \; = \;   \Int{ \big( \msc{J}_{0;\eps} \big)^{n_{h}}   }{} \hspace{-3mm}  \dd^{ n_{h} }\mu   \; \cdot  
 \Int{ \big( \msc{J}_{1;\eps} \big)^{n_{p}}   }{} \hspace{-3mm}   \dd^{ n_{p} }\nu  \;   \mc{W}_{\eps}\big( \Ups \big)\big[\vp\big] \cdot \mc{G}(\Ups) \;, 
\enq
with $\msc{J}_{a;\eps}$ as introduced in \eqref{definition des courbes integration voisinage zone Fermi}, has the following behaviour when $m\tend + \infty$, uniformly in 
$\de  m \tend + \infty$:
\bem
\mc{M}^{(\eps)}_{n_h,n_p} \; = \;  \f{  \ex{-\de m (n_h+n_p)} \cdot \de^{2 \veps \vp_{\eps} (n_h-n_p)}   }{ \big[-\i m  \big]^{ (n_p-n_h)^2 } \cdot  [ 2m \de ] ^{2 n_p n_h}  }
\cdot \eps^{n_h+n_p} \big[-\i \eps \big]^{2 \eps (n_h+n_p) \vp_{\eps} } \cdot G^2(1+n_p)\cdot G^{2}(1+n_h) \\ 
\times   \bigg(  \mc{G}\Big(   \{\eps q \}_1^{n_h} ; \{\eps q \}_1^{n_p} \Big)  \, + \, \e{O}\Big( \f{ 1 }{ \de m } + \f{\de \ln m}{  m } + \f{\ln m }{ m }  \Big) \bigg) \;. 
\end{multline}
 Above, $G$ stands for the Barnes function \cite{BarnesDoubleGaFctn1} and 
\beq
\vp_{\eps} \; = \; \vp \Big( \eps q \mid  \{\eps q \}_1^{n_h} ; \{\eps q \}_1^{n_p} \Big)\;. 
\label{definition vp eps}
\enq
\end{lemme}

\Proof 

One starts by making the change of variables 
\beq
\nu_a \; = \; u_1^{-1}\Big( \i \de + u_{1}(\eps q ,\op{v} ) + \i \tfrac{k_a}{m} , \op{v} \Big) \quad \e{and} \quad 
\mu_a \; = \; u_1^{-1}\Big( \i \de + u_{1}(\eps q , \op{v} ) -\i  \tfrac{t_a}{m}  , \op{v} \Big)  \;. 
\enq
This recast the integral as
\bem
\mc{M}^{(\eps)}_{n_h,n_p} \; = \;  \f{  [ \eps \ex{-m\de} ]^{n_p+n_h}  }{ [-\i m ]^{ (n_p-n_h)^2} }  \Int{ 0   }{ \eta }    \dd^{ n_{h} }t    
 \Int{ 0  }{ \eta }  \dd^{ n_{p} }k   \;   \pl{a=1}{n_p}\bigg\{ \ex{-  k_a} \cdot  \Big[ \f{\eps}{m} (\i m \de + \i k_a )\Big]^{-2\eps \wt{\vp}(k_a\mid \wt{\Ups}) } \bigg\}   \\ 
\times  \pl{a=1}{n_h}\bigg\{ \ex{-  t_a} \cdot \Big[ \f{\eps}{m}  (-\i m \de -\i t_a )\Big]^{ 2\eps \wt{\vp}(t_a\mid \wt{\Ups}) } \bigg\}
\cdot \f{ \pl{a<b}{ n_p} \big(k_a-k_b\big)^2 \cdot \pl{a<b}{ n_h} \big(t_a-t_b\big)^2  }{ \pl{a=1}{ n_p} \pl{b=1}{ n_h} \big( 2\i m \de + \i k_a + \i t_b \big)^2  }  \cdot \wt{\mc{G}}(\wt{\Ups}) \;. 
\label{ecriture integrale modele bord apres changement de variables}
\end{multline}
The new notation arising above correspond to 
\beq
\wt{\Ups}=\Big\{ \{k_a\}_1^{n_p} \, ; \,  \{t_a\}_1^{n_h} \Big\} \; , \qquad  \wt{\mc{G}}(\wt{\Ups})=\mc{G}(\Ups_{\e{im}})
\qquad \e{and} \qquad  \wt{\vp}(x\mid \wt{\Ups}) = \wt{\vp}\, ( \wt{x} \mid \Ups_{\e{im}}) \; .
\enq
There, I agree upon 
\beq
\Ups_{\e{im}} \; = \; \bigg\{   \Big\{ u_1^{-1}\big(\, \wt{t}_a,\op{v} \big) \Big\}_1^{n_h} \, ; \,  \Big\{ u_1^{-1}\big( \wt{k}_a,\op{v}\big) \Big\}_1^{n_p}  \bigg\} \qquad \e{with} \qquad
\left\{ \ba{ccc} \wt{k}_a  & = & \i \de \, + \, \i \tfrac{k_a}{m} \,+\,  u_1( \eps q,\op{v}) \vspace{2mm} \\ 
 \wt{t}_a  & = & -\i \de \, - \, \i \tfrac{t_a}{m} \,+\,  u_1( \eps q,\op{v})   \ea \right. \; . 
\enq
To get the leading large $m$ behaviour, since $m\de \tend + \infty$, one may simplify the $m$ dependent prefactors 
\beq
\wt{\mc{G}}(\wt{\Ups}) \; = \; \mc{G}\big(\{\eps q\}_1^{n_h} \, ; \, \{\eps q\}_1^{n_p} \big)  \;  + \; \e{O}\bigg(\de \, + \,  \sul{a=1}{n_p}\f{ k_a }{m}  \, + \, \sul{a=1}{n_h}\f{ t_a }{m} \bigg)
\enq
and, likewise,  
\beq
\wt{\vp}(k_a\mid \wt{\Ups}) \; = \; \vp(\eps q \mid \{\eps q\}_1^{n_h} \, ; \, \{\eps q\}_1^{n_p} ) \;  + \; \e{O}\bigg(\de \, + \,  \sul{a=1}{n_p}\f{ k_a }{m}  \, + \, \sul{a=1}{n_h}\f{ t_a }{m} \bigg) \;. 
\enq
Analogous simplifications also holds for the products, leading all-in-all to 
\bem
\mc{M}^{(\eps)}_{n_h,n_p} \; = \;  \f{  [\i\eps \ex{-m\de} ]^{n_p+n_h}  \cdot \de^{2\eps \vp_{\eps} (n_h-n_p) }  }{ [-\i m ]^{ (n_p-n_h)^2} \cdot [2m \de]^{2n_p n_h} }
\cdot \big[ -\i\eps \big]^{2\eps \vp_{\eps} (n_h+n_p)}
 \Int{ 0  }{m  \eta }  \dd^{ n_{p} }k   \;    \pl{a=1}{n_p}  \ex{-  k_a}  \cdot   \pl{a<b}{ n_p} \big(k_a-k_b\big)^2      \\
\times \Int{ 0   }{ m \eta }    \dd^{ n_{h} }t     \pl{a=1}{n_h}  \ex{-  t_a} \cdot    \pl{a<b}{ n_h} \big(t_a-t_b\big)^2  
\cdot \Bigg\{ \mc{G}\big(\{\eps q\}_1^{n_h} \, ; \, \{\eps q\}_1^{n_p} \big)  + \e{O}\bigg(\de \ln m  \, + \,  \f{1}{\de m } \, + \,  \ln m \sul{a=1}{n_p}\f{ k_a }{m}  \, + \, \sul{a=1}{n_h}\f{ t_a }{m} \bigg) \Bigg\}\;. 
\end{multline}
Here $\vp_{\eps}$ is as defined in \eqref{definition vp eps}.
At this stage, one may already send $m  \eta \tend +\infty $ in the integrals for the price of exponentially small corrections. Then, it remains to invoke the explicit expression for the 
multiple integral
\beq
\Int{0}{+\infty}  \dd^n w \; \pl{a=1}{n} \ex{-w_a}  \cdot \pl{a<b}{ n_p} \big(w_a-w_b\big)^2   \; = \; G^2(1+n) \;,
\enq
so as to conclude. \qed

\subsubsection{Reduction to the steepest descent paths}
\label{Sous section reduction au chemin steepest descent}

By virtue of the results of the previous section, it is enough to focus on the integral $\mc{S}_{\bs{n}}^{(\e{mod})}(m,t)$ when one wants to access to the power-law in $(m,t)$ part of the large-$m$ behaviour 
of  $\mc{S}_{\bs{n}}(m,t)$. Several more simplifications are still possible. By virtue of $\msc{J}_{r;a}$ for $a\geq 1$ and any $r \in \mf{N}\cup\{0\}$ being compact and satisfying 
\beq
\inf_{\la \in \msc{J}_{r;a} } \Big(  \Im[ u_r(\la,\op{v}) ]  \Big) \, \geq \, c_r >0 \; \e{for} \; \; r \in \mf{N} \qquad \e{and} \qquad 
\sup_{\la \in \msc{J}_{0;a} } \Big(  \Im[ u_r(\la,\op{v}) ]  \Big) \, \leq \, -c_0 <0 
\enq
for some constants  $c_r$, as soon as any rapidity $\nu_{b}^{(r)}$ is integrated over a curve $\msc{J}_{r;a}$, the corresponding contribution will generate exponentially 
small corrections in $m$ to the asymptotics of $\mc{S}_{\bs{n}}^{(\e{ply})}(m,t)$. In fact, the only curves building up $\msc{C}_{r;A}^{(\e{mod})}$
or $\msc{C}_{h}^{(\e{mod})}$ that may generate power-law corrections in $m$ are the curves $\msc{J}_{r}$ for $r \geq 0$. These are present 
only if $|\op{v}|<\op{v}_{\infty}$. By carrying out an analogous partitioning of the integration domains as in 
\eqref{ecriture partition Sn en sous integrale proches zone de Fermi}, one may readily conclude that, up to exponentially small corrections,
it is allowed to make the substitutions in the integration contours occurring in $\mc{S}_{\bs{n}}^{(\e{mod})}(m,t)$;
\beq
\msc{C}_{r;A}^{(\e{mod})} \; \hookrightarrow \;   \bs{1}_{ \mf{I}_{\infty} }(\op{v}) \cdot \msc{J}_{r} \; \; \e{for} \;\; r \in \mf{N}_{\e{st}}
\quad \e{and} \quad 
\left\{ \ba{ccc}  \msc{C}_{1;A}^{(\e{mod})} & \hookrightarrow &   \bs{1}_{ \mf{I}_{\infty}  }(\op{v}) \cdot   \msc{J}_{1} \; + \;  \bs{1}_{ \mf{I}_{\e{SL}} }(\op{v})\cdot \msc{J}_{0}  \vspace{2mm} \\
                          \msc{C}_{h}^{(\e{mod})} & \hookrightarrow &     \bs{1}_{\mf{I}_{\e{TL}} }(\op{v})\cdot  \msc{J}_{0}         \ea \right. 
\enq
\textit{c.f.} \eqref{definition I infty et SL et TL} for the definition of the intervals $\mf{I}_{*}$. Namely, it holds 
\beq
\mc{S}_{\bs{n}}^{(\e{mod})}(m,t)\; = \; \bs{1}_{ \mf{J}_{\infty} }(\op{v}) \mc{S}_{\bs{n}}^{(\e{sd})}(m,t) \; + \; \e{O}\Big(  \ex{-c m} \Big)
\enq
for some $c>0$ and where 
\beq
\mc{S}_{\bs{n}}^{(\e{sd})}(m,t) \; = \;  \pl{ r \in  \mf{N}_{\e{st}} }{} \;\Bigg\{  \Int{ \big( \msc{J}_{r}  \big)^{n_r} }{} \hspace{-3mm} \f{ \dd^{n_r}\nu^{(r)} }{ n_r! \cdot (2\pi)^{n_r}  }   \Bigg\} 
\;\; \cdot \hspace{-8mm} \Int{   \big( \msc{J}_{1} \; + \;  \bs{1}_{ \mf{I}_{\e{SL}} }(\op{v})\cdot \msc{J}_{0} \big)^{n_1} }{} \hspace{-5mm}  \f{ \dd^{n_1}\nu^{(1)} }{ (2\pi)^{n_1} \cdot n_1! } 
\;\;\cdot \hspace{-6mm} \Int{ \big(  \bs{1}_{\mf{I}_{\e{TL}} }(\op{v})\cdot  \msc{J}_{0} \big)^{n_h}   }{} \hspace{-6mm} \f{ \dd^{n_h}\mu  }{ n_h! \cdot (2\pi)^{n_h}  } \cdot
   \mc{F}^{(\ga)}_{\e{tot}} \big( \mf{Y} \big)  \;.
\enq
When $\op{v}_F < |\op{v}| < \op{v}_{\infty}$ the integration over particle rapidities $\nu^{(1)}$ runs over two disjoint steepest-descent paths, and the integral needs to be decomposed further. 
For doing so, observe that for any symmetric function $f$, it holds 
\beq
  \hspace{-5mm} \Int{ \big( \msc{J}_{1} \cup \msc{J}_0 \big)^{ n} }{} \hspace{-5mm} \f{ \dd^{n} \nu }{ n! \cdot (2\pi)^{n}   }  
 f(\nu_1,\dots, \nu_n)   
 \; =  \hspace{-2mm} 
\sul{ n=n_0+n_1 }{} \pl{r=0}{1} \Bigg\{ \Int{ \big( \msc{J}_{r} \big)^{n_r} }{} \hspace{-3mm} \f{ \dd^{n_r} \nu^{(r)} }{ n_r! \cdot (2\pi)^{n_r}   }  \Bigg\} \cdot 
   f\Big(\nu_1^{(0)},\dots,\nu_{n_0}^{(0)}, \nu_1^{(1)},\dots,\nu_{n_1}^{(1)}\Big)  \;. 
\enq
This entails the representation 
\bem
\mc{S}_{\bs{n}}^{(\e{sd})}(m,t)  \; = \;   \sul{n_1=n_0+\wt{n}_1}{} \Big\{ \bs{1}_{\mf{I}_{\e{TL}} }(\op{v}) \Big\}^{n_h} \cdot \Big\{ \bs{1}_{\mf{I}_{\e{TL}} }(\op{v}) \Big\}^{ \wt{n}_1 }
\pl{ r \in  \mf{N}_{\e{st}}  }{} \;\Bigg\{  \Int{ \big( \msc{J}_{r}  \big)^{n_r} }{} \hspace{-3mm} \f{ \dd^{n_r}\nu^{(r)} }{ n_r! \cdot (2\pi)^{n_r}  }  \Bigg\}  \cdot 
\Bigg\{  \Int{ \big( \msc{J}_{1} \big)^{\wt{n}_1} }{} \hspace{-3mm} \f{ \dd^{ \wt{n}_{1} }\nu^{(1)} }{ \wt{n}_1! \cdot (2\pi)^{\wt{n}_{1}}  }   \Bigg\}  \\
\times 
\Bigg\{  \Int{ \big( \msc{J}_{0} \big)^{n_0} }{} \hspace{-3mm} \f{ \dd^{n_{0} }\nu^{(0)} }{ n_0! \cdot (2\pi)^{n_0}   }   \Bigg\} 
\cdot  \Bigg\{  \Int{ \big( \msc{J}_{0} \big)^{n_h} }{} \hspace{-3mm} \f{ \dd^{n_h} \mu }{ n_h! \cdot (2\pi)^{n_h}  }  \Bigg\}  \cdot 
 \mc{F}^{(\ga)}_{\e{tot}} \big( \mf{Y}_{\e{part}} \ \big)   \;.   
\label{decomposition Ssd sur ctr part et trous SD}
\end{multline}
There, it is undercurrent that$ \Big\{ \bs{1}_{\mf{I} }(\op{v}) \Big\}^{0}\equiv 1$ irrespectively of the value of $n$. Also, the integration variables 
$\mf{Y}$ take the form 
\beq
\mf{Y}_{\e{part}} \; = \; \bigg\{ -\op{s}_{\ga};  \big\{ \ell_{\ups}\big\} \; \big| \; 
 \big\{ \mu_{a} \big\}_{1}^{ n_{h} }    
\, ; \, \Big\{ \big\{ \nu_{a}^{(1)} \big\}_{1}^{ \wt{n}_{1} }  \cup \big\{ \nu_{a}^{(0)} \big\}_{1}^{ n_{0} }    \Big\} \, ; \, 
\big\{ \nu_{a}^{(r_2)} \big\}_{a=1}^{ n_2}; \dots   \bigg\} 
\label{ecriture configurations rapidites partitionnes sur ctrs saddle point}
\enq
The presence of the indicator functions in \eqref{decomposition Ssd sur ctr part et trous SD} ensures that, in the summation over the excitation numbers \eqref{ecriture serie FF massless sg ga sg ga dagger}, 
only configurations such that $n_h=0$ will contribute to the power-law part of the asymptotics in the space like regime $\op{v}_F\, < \,  \op{v} \, < \, \op{v}_{\infty}$
while, one should always keep $n_0=0$ in \eqref{decomposition Ssd sur ctr part et trous SD}, in the time-like regime. 
In order to write down the subordinate expansion of the two-point function, it its convenient, for further purposes, to 
recall that given a set of rapidities $\mf{Y}_{\e{part}}$ as given in \eqref{ecriture configurations rapidites partitionnes sur ctrs saddle point}, one has the factorisation
over the kinematical zeroes of the form factor density:
\beq
 \mc{F}^{(\ga)}\big( \mf{Y}_{\e{part}} \big)  \; = \; \pl{r \in \mf{N}_{\e{st}}\cup\{0\} }{}  \pl{a<b}{n_r} \Big\{ \sinh^{2}\big[\nu_a^{(r)} - \nu_b^{(r)} \big]  \Big\} \,  \cdot
 \pl{a<b}{\wt{n}_1} \Big\{ \sinh^{2}\big[\nu_a^{(1)} - \nu_b^{(1)} \big]  \Big\}  \cdot 
 \pl{a<b}{n_h} \Big\{  \sinh^{2}\big[\mu_a - \mu_b \big]  \Big\}    \cdot    \mc{F}^{(\ga)}_{\e{off}}\big( \mf{Y}_{\e{part}} \big) \;. 
\enq

All these considerations allow to recast the two-point function as 
\beq
\big< \big(\sg_1^{\ga} (t)\big)^{\dagger} \sg_{m+1}^{\ga}(0) \big>_{\e{c}} \; = \;   (-1)^{m \op{s}_{\ga} }  \sul{ \bs{n} \in \mf{S}_{\op{v}} }{} \ex{\i m p_{F}(\ell_+-\ell_-)} \cdot \mc{S}_{\bs{n}}^{(\e{ply})}(m,t)  \;  +  \; \e{O}\Big(m^{-\infty}\Big) \;. 
\enq
There, the summation runs over all configurations of the excitation integers which contribute to the power-law part of the asymptotics:
\beq
\mf{S}_{\op{v}} \; = \; \left\{ \ba{cc} 
\bigg\{ \bs{n} = \Big( \ell_+, \ell_-;    n_{0}  , n_{1}  , n_{r_2}, \dots, n_{r_k}, \dots \Big) \; : \; 
	      \op{s}_{\ga} \, = \, \sul{\ups=\pm}{} \ell_{\ups} \, + \,  n_{0}   \, + \,  n_{1} \, + \, \sul{ r \in \mf{N}_{\e{st}} }{} r n_r  \bigg\}   & \e{if}  \;\;   \op{v}_F < |\op{v}| <  \op{v}_{\infty} \vspace{3mm}  \\
\bigg\{ \bs{n} = \Big( \ell_+, \ell_-;  n_h     , n_{1} , n_{r_2}, \dots, n_{r_k}, \dots \Big) \; : \; 
 n_h  +\op{s}_{\ga}  \, = \, \sul{\ups=\pm}{} \ell_{\ups}    \, + \,  n_{1} \, + \, \sul{ r \in \mf{N}_{\e{st}} }{} r n_r  \bigg\}  & \e{if}  \;\;   0 \leq |\op{v}| <  \op{v}_{F} 
\ea \right. \;. 
\enq

The summand  $\mc{S}_{\bs{n}}^{(\e{ply})}(m,t)$ is built up from the integrals passing through the saddle-point contours of the various excitations. It is expressed as 
\bem
\mc{S}_{\bs{n}}^{(\e{ply})}(m,t)  \; = \;   
\pl{ r \in  \mf{N}  }{} \;\Bigg\{  \Int{ \big( \msc{J}_{r}  \big)^{n_r} }{} \hspace{-3mm} \f{ \dd^{n_r}\nu^{(r)} }{ n_r! \cdot (2\pi)^{n_r}  } \pl{a<b}{n_r} \Big\{ \sinh^2\big[ \nu_a^{(r)}-\nu_b^{(r)}\big]\Big\} \cdot 
\pl{a=1}{n_r} \ex{ \i m u_r(\nu_a^{(r)},\op{v}) } \Bigg\} \\
\times 
\Bigg\{  \Int{ \big( \msc{J}_{0} \big)^{n_0} }{} \hspace{-3mm} \f{ \dd^{n_{0} }\nu^{(0)} }{ n_0! \cdot (2\pi)^{n_0}   } \pl{a<b}{ n_0 } \Big\{ \sinh^2\big[ \nu_a^{(0)}-\nu_b^{(0)}\big]\Big\} \cdot 
\pl{a=1}{n_0}   \ex{ \i m u_1(\nu_a^{(0)},\op{v}) }  \Bigg\}^{ \bs{1}_{\intoo{ \op{v}_F }{ \op{v}_{\infty} } }(|\op{v}|) } \\
\times \Bigg\{  \Int{ \big( \msc{J}_{0} \big)^{n_h} }{} \hspace{-3mm} \f{ \dd^{n_h} \mu }{ n_h! \cdot (2\pi)^{n_h}  } \pl{a<b}{n_h} \Big\{ \sinh^2\big[ \mu_a-\mu_b\big]\Big\} \cdot 
\pl{a=1}{n_h} \ex{ -\i m u_1(\mu_a,\op{v}) } \Bigg\}^{ \bs{1}_{ \intoo{0}{\op{v}_F} }(|\op{v}|) }
\cdot \f{   \mc{F}^{(\ga)}_{\e{off}}\big( \mf{Y}_{\op{v}} \big)   \cdot  \Big( 1+ \mf{r}_{\de,m, t}\big( \mf{Y}_{\op{v}} \big) \Big)   }{  \pl{\ups=\pm}{}  \big[  -  \i (\ups m - \op{v}_F t)   \big]^{   \vth_{\ups}^2(\mf{Y}_{\op{v}}) }  }    \;.   
\label{ecriture S ply 1ere forme}
\end{multline}
There, the indicator functions means that the corresponding factors are absent when the indicator is zero. Finally, the integration variables are given as
\beq
\mf{Y}_{\op{v}} \; = \; \left\{ \ba{ccc}  \bigg\{ -\op{s}_{\ga};  \big\{ \ell_{\ups}\big\} \; \big| \; 
  \big\{ \mu_{a} \big\}_{1}^{ n_{h} }      
\, ; \, \Big\{ \big\{ \nu_{a}^{(1)} \big\}_{1}^{ n_{1} }       \Big\} \, ; \, 
\big\{ \nu_{a}^{(r_2)} \big\}_{a=1}^{ n_2}; \dots   \bigg\}    			&\e{for} &   | \op{v} | \, < \, \op{v}_F \\ 
\bigg\{ -\op{s}_{\ga};  \big\{ \ell_{\ups}\big\} \; \big| \; 
\emptyset 
\, ; \, \Big\{ \big\{ \nu_{a}^{(1)} \big\}_{1}^{ n_{1} }  \cup \big\{ \nu_{a}^{(0)} \big\}_{1}^{ n_{0} }    \Big\} \, ; \, 
\big\{ \nu_{a}^{(r_2)} \big\}_{a=1}^{ n_2}; \dots   \bigg\} 						&\e{for} &  \op{v}_F \, < \,  |\op{v}| \, < \, \op{v}_{\infty}   
\ea \right. \; . 
\enq

\subsection{The conformal regime : $|\op{v}|>\op{v}_{\infty}$ }
\label{Sous section regime conforme}

Recall that the two-point function is expressed as 
\beq
\big< \big(\sg_1^{\ga} (t)\big)^{\dagger} \sg_{m+1}^{\ga}(0) \big>_{\e{c}} \; = \;   (-1)^{m \op{s}_{\ga} }  \sul{ \bs{n} \in \mf{S}  }{}  \mc{S}_{\bs{n}} (m,t)   \;. 
\label{ecriture dvpment fct 2 pts sur integrales FF locaux}
\enq
The previous analysis shows that, in the regime $|\op{v}|>\op{v}_{\infty}$ where the saddle-points are away from the original integration curves
$\msc{C}_{r}$ and $\msc{C}_h$, whenever $\bs{n} \in \mf{S}$ has non-zero hole or $r$-string components, \textit{viz}. $n_h\not=0$ or $n_r\not=0$ 
for some $r \in \mf{N}$ then, effectively the contribution of the associated  $\mc{S}_{\bs{n}} (m,t)$ coefficient will be $\e{O}( m^{-\infty})$. 
Thus, the only algebraic contributions to the two point functions issue from summations in \eqref{ecriture dvpment fct 2 pts sur integrales FF locaux} 
over the configurations of integers
\beq
\bs{n} \; = \; \big( \ell + \op{s}_{\ga} , -\ell; 0,0,\dots \big) \quad \ell \in \mathbb{Z} \;, 
\enq
\textit{viz}. over excited states having no massive excitations $0=n_h=n_{r_1}=\dots=n_{r_k}=\dots $ and built up only from 
Umklapp excitations with opposite left/right Umklapp deficiencies. The right Umklapp deficiency is shifted by the operator's spin. 

This, yields
\beq
\big< \big(\sg_1^{\ga} (t)\big)^{\dagger} \sg_{m+1}^{\ga}(0) \big>_{\e{c}} \; = \;   (-1)^{m \op{s}_{\ga} } \sul{ \ell \in \mathbb{Z} }{} 
\,  \f{  \mc{F}^{(\ga)}\big(\mf{Y}_{\ell} \big)\cdot \ex{2\i m  \ell p_{F} }  }{  \pl{\ups=\pm}{}  [ -\i (\ups m - \op{v}_F t)  ]^{  \vth_{\ups}^2(\mf{Y}_{\ell})  } }
\cdot \Bigg\{ 1 \; + \; \sul{\ups=\pm }{} \e{O}\bigg( \,\f{ 1 }{m^{1-0^+}}  \bigg)  \Bigg\} \;. 
\enq
where 
\beq
\mf{Y}_{\ell} \; = \; \Big\{ -\op{s}_{\ga} ; \{\ell_+=\ell+\op{s}_{\ga}, \ell_- = - \ell \} \mid \emptyset; \emptyset; \cdots  \Big\} \;. 
\enq
 Thus, upon the identification 
\beq
  \mc{F}^{(\ga)}_{\ell} \; = \; \mc{F}^{(\ga)}\big(\mf{Y}_{\ell} \big) \quad \e{and} \quad 
\De_{\ups,\ell} \, = \, \vth_{\ups}(\mf{Y}_{\ell})
\enq
one recovers the asymptotic expansion given in \eqref{ecriture DA cas purement conforme}. The explicit form of the critical exponents \eqref{ecriture explicite exposant Delta ups ell dans regime conforme}
follows from \eqref{ecriture identites entre phase et charge habilles}-\eqref{definition shifted sfift function} in the appendix.

\subsection{The geniuine asymptotic regime : $|\op{v}|<\op{v}_{\infty}$ } 
\label{Sous section regime profond asymptotiques}

For the sake of compactness of notations, it is convenient to introduce 
\beq
\varkappa_{\op{v}} \; = \; \left\{  \ba{ccc}  1 & \e{if} &   \op{v}_F < |\op{v}| <  \op{v}_{\infty}  \\ 
	    -1 & \e{if} &  0 \leq |\op{v}| <  \op{v}_{F} \ea  \right. 
\enq
and change the notations $(n_h,\mu_a) \hookrightarrow (n_{0}, \nu_{a}^{(0)} ) $ for the hole rapidities in the regime $ 0 \leq |\op{v}| <  \op{v}_{F}$. 
Then, one can recast $\mc{S}_{\bs{n}}^{(\e{ply})}(m,t)$ as given in \eqref{ecriture S ply 1ere forme}, in the form
\bem
\mc{S}_{\bs{n}}^{(\e{ply})}(m,t)  \; = \;   
\pl{ r \in  \mf{N}  }{} \;\Bigg\{  \Int{ \big( \msc{J}_{r}  \big)^{n_r} }{} \hspace{-3mm} \f{ \dd^{n_r}\nu^{(r)} }{ n_r! \cdot (2\pi)^{n_r}  } \pl{a<b}{n_r} \Big\{ \sinh^2\big[ \nu_a^{(r)}-\nu_b^{(r)}\big]\Big\} \cdot 
\pl{a=1}{n_r} \ex{ \i m u_r(\nu_a^{(r)},\op{v}) } \Bigg\} \\
\times 
   \Int{ \big( \msc{J}_{0} \big)^{n_0} }{} \hspace{-3mm} \f{ \dd^{n_{0} }\nu^{(0)} }{ n_0! \cdot (2\pi)^{n_0}   } \pl{a<b}{ n_0 } \Big\{ \sinh^2\big[ \nu_a^{(0)}-\nu_b^{(0)}\big]\Big\} \cdot 
\pl{a=1}{n_0}   \ex{ \i m \varkappa_{\op{v}}  u_1(\nu_a^{(0)},\op{v}) }     
\cdot \f{   \mc{F}^{(\ga)}_{\e{off}}\big( \mf{Y}_{\op{v}} \big)   \cdot  \Big( 1+ \mf{r}_{\de,m, t}\big( \mf{Y}_{\op{v}} \big) \Big)   }{  \pl{\ups=\pm}{}  \big[  -  \i (\ups m - \op{v}_F t)   \big]^{   \vth_{\ups}^2(\mf{Y}_{\op{v}}) }  }       
\end{multline}
where the integration variables are now expressed as
\beq
\mf{Y}_{\op{v}} \; = \; \left\{ \ba{ccc}  \bigg\{ -\op{s}_{\ga};  \big\{ \ell_{\ups}\big\} \; \big| \; 
\big\{ \nu_{a}^{(0)} \big\}_{1}^{ n_{0} }      
\, ; \,   \big\{ \nu_{a}^{(1)} \big\}_{1}^{ n_{1} }        \, ; \, 
\big\{ \nu_{a}^{(r_2)} \big\}_{a=1}^{ n_2}; \dots   \bigg\}    			&\e{for} &   | \op{v} | \, < \, \op{v}_F \\ 
\bigg\{ -\op{s}_{\ga};  \big\{ \ell_{\ups}\big\} \; \big| \; 
\emptyset 
\, ; \, \Big\{ \big\{ \nu_{a}^{(1)} \big\}_{1}^{ n_{1} }  \cup \big\{ \nu_{a}^{(0)} \big\}_{1}^{ n_{0} }    \Big\} \, ; \, 
\big\{ \nu_{a}^{(r_2)} \big\}_{a=1}^{ n_2}; \dots   \bigg\} 						&\e{for} &  \op{v}_F \, < \,  |\op{v}| \, < \, \op{v}_{\infty}   
\ea \right. \; . 
\enq

Then, the definition of the integration contours
\beq
\msc{J}_{r} \; = \; \om_r \, + \, \mf{s}_r (-1)^{\sg_{r}} \cdot h_{r}^{-1}\Big(  \intoo{-\eta \ex{\i\frac{\pi}{4}\veps_{r}} }{ \eta \ex{\i\frac{\pi}{4}\veps_{r}}} \Big)  \; \e{for} \; \; r \in \mf{N} \quad \e{and} \quad 
\msc{J}_{0} \; = \; \om_0 \, + \,   h_{1}^{-1}\Big(  \intoo{-\eta \ex{\i \varkappa_{\op{v}} \veps_{0} \frac{\pi}{4} } }{ \eta \ex{\i \varkappa_{\op{v}} \veps_{0} \frac{\pi}{4}}} \Big) 
\enq
leads to the change variables
\beq
\nu_a^{(r)} \; = \;\om_r\, + \,  h_{r}^{-1}\bigg(  \f{ x_a^{(r)} }{  \sqrt{m} }   \cdot   \ex{\i\frac{\pi}{4}\veps_{r}}  \bigg) \; \e{for} \; \; r \in \mf{N} \quad \e{and} \quad 
\nu_a^{(0)} \; = \;\om_0\, + \,  h_{0}^{-1}\bigg(  \f{ x_a^{(0)} }{  \sqrt{m} }   \cdot   \ex{\i \varkappa_{\op{v}} \veps_{0} \frac{\pi}{4} }  \bigg)  \;. 
\enq
This recasts $\mc{S}_{\bs{n}}^{(\e{ply})}(m,t)$ in the form 
\bem
\mc{S}_{\bs{n}}^{(\e{ply})}(m,t)  \; = \;   
\pl{ r \in  \mf{N}  }{} \;\Bigg\{ \f{ \big[ (-1)^{\sg_r}\mf{s}_r\big]^{n_r} }{ n_r! \cdot (2\pi)^{n_r}  } \bigg( \f{ 2 \ex{ \i \frac{\pi}{2} \veps_r } }{ m | u_r^{\prime\prime}(\om_r; \op{v}) |     }  \bigg)^{ \tfrac{1}{2} n_r^2 }
\pl{a=1}{ n_r} \Big\{ \ex{\i m u_r(\om_r;\op{v}) } \Big\} \cdot 
\Int{- \eta \sqrt{m}  }{ \eta \sqrt{m}   }   \dd^{n_r} x^{(r)}  \pl{a<b}{n_r} \big[ x_a^{(r)}  -   x_b^{(r)} \big]^2 \cdot \pl{a=1}{n_r} \ex{ -   (x_a^{(r)})^2 }   \Bigg\}  \\
\times \Bigg\{ \f{  \prod_{a=1}^{ n_0} \Big\{ \ex{\i m  \varkappa_{\op{v}} u_0(\om_0;\op{v}) } \Big\}  }{ n_0! \cdot (2\pi)^{n_0}  }
\bigg( \f{ 2 \ex{ \varkappa_{\op{v}} \i \frac{\pi}{2} \veps_0 } }{ m  | u_1^{\prime\prime}(\om_0; \op{v}) |     }  \bigg)^{ \tfrac{1}{2} n_0^2 }
 \cdot 
\Int{ - \eta \sqrt{m}    }{  \eta \sqrt{m}  }   \dd^{n_0} x^{(0)}  \pl{a<b}{n_0} \big[ x_a^{(0)}  -   x_b^{(0)} \big]^2 \cdot \pl{a=1}{n_0} \ex{ -    (x_a^{(0)})^2 }   \Bigg\} \\
\times 
\f{  \mc{P}_{\bs{n}}\big( \mf{Y}_{\op{v};m} \big) \cdot  \mc{F}^{(\ga)}_{\e{off}}\big( \mf{Y}_{\op{v};m} \big)    } 
{  \pl{\ups=\pm}{}  \big[  -  \i (\ups m - \op{v}_F t)   \big]^{   \vth_{\ups}^2(\mf{Y}_{\op{v};m}) }  }
 \cdot  \Big( 1+ \mf{r}_{\de,m, t}\big( \mf{Y}_{\op{v};m} \big) \Big) \;.  
\end{multline}
Above, I made use of 
\bem
\mc{P}_{\bs{n}}\big( \mf{Y}_{\op{v};m} \big) \; = \; \pl{r \in \mf{N}\cup \{0\} }{} \left\{ \bigg( \f{2  \ex{\i \varkappa_{\op{v}}^{(r)} \veps_{r} \frac{\pi}{4}}   }{ |u^{\prime\prime}_{r}(\om_r;\op{v})| } \bigg)^{\f{n_r}{2}} \cdot 
\pl{a=1}{n_r}  \bigg\{    \f{1}{h_r^{\prime}\Big[h_{r}^{-1}\Big(    x_a^{(r)}\cdot   \ex{\i \varkappa_{\op{v}}^{(r)} \veps_{r} \frac{\pi}{4}} \cdot m^{-\frac{1}{2}}       \Big) \Big] }  \bigg\}   \right. \\
\left. \times \pl{a<b}{n_r} \Bigg(  \f{ \sinh\Big[ h_{r}^{-1}\Big(     x_a^{(r)}  \cdot \ex{\i \varkappa_{\op{v}}^{(r)} \veps_{r} \frac{\pi}{4}  }  \cdot m^{-\frac{1}{2}}        \Big) 
\, - \, h_{r}^{-1}\Big(   x_b^{(r)}  \cdot \ex{\i \varkappa_{\op{v}}^{(r)} \veps_{r} \frac{\pi}{4}  }  \cdot m^{-\frac{1}{2}}     \Big) \Big] }
{ \big( x_a^{(r)} - x_b^{(r)} \big) \ex{\i \varkappa_{\op{v}}^{(r)} \veps_{r} \frac{\pi}{4}}  }   \cdot \sqrt{ \f{ m |u^{\prime\prime}_{r}(\om_r;\op{v})| }{2} } \Bigg)^2       \right\} \;, 
\end{multline}
in which $\varkappa_{\op{v}}^{(r)}=\varkappa_{\op{v}}$ and $\varkappa_{\op{v}}^{(r)}=1$ for $r \in  \mf{N}$. 
Finally,  the collection of sets $\mf{Y}_{\op{v};m}$ is expressed in terms of the $x_a^{(r)}$s as follows
\beq
\mf{Y}_{\op{v};m} \; = \;   \Bigg\{ -\op{s}_{\ga};  \big\{ \ell_{\ups}\big\} \; \big| \; 
\bigg\{ \om_0 + h_{0}^{-1}\bigg(   \f{  x_a^{(0)}  \cdot \ex{\i \varkappa_{\op{v}} \veps_{0} \frac{\pi}{4}  } }{ m^{1/2}  }        \bigg)  \;  \bigg\}_{1}^{ n_{0} }      
\, ; \,  \bigg\{ \om_1 + h_{1}^{-1}\bigg(   \f{  x_a^{(1)}  \cdot \ex{\i   \veps_{1} \frac{\pi}{4}  } }{ m^{1/2}  }        \bigg)  \;  \bigg\}_{1}^{ n_{1} }         \, ; \, 
 \bigg\{ \om_{r_2} + h_{r_2}^{-1}\bigg(   \f{  x_a^{(r_2)}  \cdot \ex{\i   \veps_{r_2} \frac{\pi}{4}  } }{ m^{1/2}  }        \bigg)  \;  \bigg\}_{1}^{ n_{r_2} }       ; \dots   \Bigg\}    	
\enq
for $| \op{v} | \, < \, \op{v}_F$ and 
\beq
\mf{Y}_{\op{v};m} \; = \;  \bigg\{ -\op{s}_{\ga};  \big\{ \ell_{\ups}\big\} \; \big| \; 
\emptyset 
\, ; \, \Bigg\{ \bigg\{ \om_0 + h_{0}^{-1}\bigg(   \f{  x_a^{(0)}  \cdot \ex{\i \varkappa_{\op{v}} \veps_{0} \frac{\pi}{4}  } }{ m^{1/2}  }        \bigg)  \;  \bigg\}_{1}^{ n_{0} }      \bigcup   \bigg\{ \om_1 + h_{1}^{-1}\bigg(   \f{  x_a^{(1)}  \cdot \ex{\i   \veps_{1} \frac{\pi}{4}  } }{ m^{1/2}  }        \bigg)  \;  \bigg\}_{1}^{ n_{1} }  \Bigg\} \, ; \, 
 \bigg\{ \om_{r_2} + h_{r_2}^{-1}\bigg(   \f{  x_a^{(r_2)}  \cdot \ex{\i   \veps_{r_2} \frac{\pi}{4}  } }{ m^{1/2}  }        \bigg)  \;  \bigg\}_{1}^{ n_{r_2} }       ; \dots   \Bigg\}    	
\enq
for $\op{v}_F \, < \,  |\op{v}| \, < \, \op{v}_{\infty}$.

Then, by following standard considerations of the saddle-point analysis, to get the large-$m$ asymptotic expansion of $\mc{S}_{\bs{n}}^{(\e{ply})}(m,t) $ it is enough to expand the integrand
in the neighbourhood of $x_a^{(r)}=0$. By symmetry, the linear in the $x_a^{(r)}$ correction will vanish and hence, the first subdominant correction will issue from the quadratic term
of the expansion. Thus, the corrections to the leading asymptotics will go as $\e{O}\big( m^{-1}\cdot \ln^2m \big)$. 
Also, after making this expansion, one may extend the integration domains to $\R^{n_r}$ for the price of exponentially small corrections. 
Then, it remains to recall the explicit form of the Gaudin-Mehta integral 
\beq
\Int{ \R^n }{} \pl{a=1}{n} \Big\{ \ex{-y_a^2} \Big\} \pl{a<b}{n} (y_a-y_b)^2 \cdot \dd^n y  \; = \; \Big( \f{1}{2} \Big)^{ \tfrac{1}{2} n^2 } \cdot (2\pi)^{\tfrac{n}{2}} \cdot  G(2+n)  \;,  
\enq
so as to get 
\bem
\mc{S}_{\bs{n}}^{(\e{ply})}(m,t)  \; = \;   
\pl{ r \in  \mf{N}  }{} \;\Bigg\{ \f{ \big[ (-1)^{\sg_r}\mf{s}_r\big]^{n_r} }{   (2\pi)^{ \tf{ n_r }{ 2 } }  }  \cdot  G(1+n_r) \cdot \bigg( \f{  \i  }{  u_r^{\prime\prime}(\om_r; \op{v})      }  \bigg)^{ \tfrac{1}{2} n_r^2 }
\pl{a=1}{ n_r} \Big\{ \ex{\i m u_r(\om_r,\op{v}) } \Big\}  \Bigg\}  \\
\times \Bigg\{ \f{ G(1+n_0) }{     (2\pi)^{ \tf{ n_0 }{ 2 }  }  } \bigg( \f{    \varkappa_{\op{v}} \i  }{   u_1^{\prime\prime}(\om_0, \op{v})       }  \bigg)^{ \tfrac{1}{2} n_0^2 }
\pl{a=1}{ n_0} \Big\{ \ex{\i m  \varkappa_{\op{v}} u_0(\om_0;\op{v}) } \Big\}    \Bigg\} \\
\times  
\f{  \cdot \mc{F}^{(\ga)}_{\e{off}}\big( \mf{Y}_{\op{v};\infty} \big)    } 
{  \pl{\ups=\pm}{}  \big[  -  \i (\ups m - \op{v}_F t)   \big]^{   \vth_{\ups}^2(\mf{Y}_{\op{v};\infty}) }  } 
\cdot 
\Bigg( 1+   \mf{r}_{\de,m, t}\big( \mf{Y}_{\op{v};\infty} \big) \; + \;  \e{O}\bigg( \f{ (\ln m )^2}{ m } \bigg) \Bigg)  \; . 
\end{multline}
There, we agree upon 
\beq
\mf{Y}_{\op{v};\infty} \; = \; \left\{ \ba{cc}  
\bigg\{ -\op{s}_{\ga} \, ;\,  \big\{ \ell_{\ups} \big\} \; \big| \; \emptyset \, ; \,   \Big\{ \big\{ \om_0 \big\}^{n_0} \cup   \big\{ \om_1 \big\}^{n_1}   \Big\} ;  \big\{ \om_{r_2} \big\}^{n_{r_2}} ; \dots    \bigg\}  
&  \qquad  \op{v}_F < |\op{v}| < \op{v}_{\infty}   \vspace{3mm} \\
\bigg\{  -\op{s}_{\ga} \, ;\,   \big\{ \ell_{\ups} \big\}\; \big| \; \big\{ \om_0 \big\}^{n_0}  ;    \big\{ \om_1 \big\}^{n_1};  \big\{ \om_{r_2} \big\}^{n_{r_2}} ; \dots     \bigg\}   
&  \qquad     |\op{v}| < \op{v}_{F}   \ea \right.  \;. 
\enq
 By using that the remainder is controlled as $\mf{r}_{\de,m, t}\big( \mf{Y}_{\op{v};\infty} \big) = \e{O}(m^{\tau-1})$, one gets the result announced in Subsection \eqref{Sous Section resultats principaux}.

 \subsection{The asymptotics in the general case}
\label{SousSection asymptotiques cas general} 

Dealing with the most general case of the structure of possible saddle-points may be done by following the above scheme of analysis, by invoking Conjecture 
\ref{Conjecture deformation contour integration secteur n excitation massive cas general} and for the price of slightly more combinatorics. 
Here, I only give the final results since the strategy to get there is quite clear. 

First let $\op{v}_{\e{max}}=\e{sup}_{r\in \mf{N}}\big[ \op{v}_{r}^{(M)} \big]$. It is unclear if $\op{v}_{\e{max}}$ may be infinite in some situations, but it seems unlikely
and non-physical, so that I shall assume that  $\op{v}_{\e{max}}<+\infty$. This will set the velocity scale at which the separation between the conformal and genuine asymptotics arises. 
If $\op{v}_{\e{max}}=+\infty$ that would simply mean that the region of conformal-type asymptotics is simply reduced to the long-distance asymptotics of the static correlators. 
Thus, for $|\op{v}|>\op{v}_{\e{max}}$, the asymptotics are still given by the previously obtained answer in the conformal regime. 

For $|\op{v}|<\op{v}_{\e{max}}$, define 
\beq
\mf{N}_{\e{sp}}=\{t_2,t_3\dots,  \}\subset \mf{N}_{\e{st}}
\label{definition entiers de Nsp}
\enq
to be the collection of allowed $r$-strings lengths at given $\zeta$
and $\op{v}$ such that $|\op{v}|<\op{v}_r^{(M)}$ if and only if $r \in \mf{N}_{\e{sp}}$. Denote by $\om_r^{(1)},\dots, \om_r^{(\vsg_r)}$
the zeroes of $u_r^{\prime}(\la;\op{v})$ on $\msc{C}_r$. Finally, let $\om_0$ and $\om_1$ be the two zeroes on $\R\cup \R+\i\tf{\pi}{2}$ of $u_1^{\prime}(\la;\op{v})$. 
One has furthermore that $\om_0\in \intoo{-q}{q}$ if and only if $|\op{v}| < \op{v}_F$ and then $\om_1\in\big\{  \R \setminus \intff{-q}{q} \big\}\cup \{\R+\i\tf{\pi}{2}\}$. 
Finally, let 
\beq
\varkappa_{\op{v}} \; = \; \left\{ \ba{cc}   1  &  \op{v}_{1}^{(M)} > |\op{v}| > \op{v}_F \\ 
					   - 1  &  \op{v}_{F} > |\op{v}|  \ea \right.  \;. 
\enq
The asymptotics take the form 
\beq
\big< \big(\sg_1^{\ga} (t)\big)^{\dagger} \sg_{m+1}^{\ga}(0) \big>_{\e{c}} \; = \;   (-1)^{m \op{s}_{\ga} } \sul{ \bs{n} \in \mf{S}_{\op{v}} }{} 
\, \op{C}_{\bs{n}} \cdot \mc{F}_{ \bs{n} }^{(\ga)} \cdot  \pl{\ups=\pm}{} \Bigg\{  \f{  \ex{ \i  \ups \ell_{\ups} m   p_{F} }  }{    [ -\i (\ups m - \op{v}_F t)  ]^{  \De_{\ups,\bs{n}}^2   } } \Bigg\}
\cdot \f{ \ex{\i m \vp_{\bs{n}}(\op{v}) } }{  m^{\De_{\e{sp};\bs{n}}}  }
\cdot \Bigg\{ 1 \; + \; \e{O}\bigg( \,\f{1}{  m^{1-0^+} }  \bigg)  \Bigg\}
\enq

\begin{itemize}

 \item $\op{C}_{\bs{n}}$ represents the universal part of the amplitude. It is expressed in terms of the Barnes $G$ function \cite{BarnesDoubleGaFctn1} as 
\bem
\op{C}_{\bs{n}} \; = \;   \pl{a=0}{1} \Bigg\{ \f{  G(1+n_a)    \Big\{ \e{sgn}\big[ p^{\prime}_1(\om_a)\big]  \Big\}^{n_a}     }{ (2\pi)^{ \frac{n_a}{2} } }  \Bigg\} \cdot 
\bigg(  \f{  \i \varkappa_{\op{v}} }{ p^{\prime\prime}_1(\om_0) - \frac{1}{ \op{v} } \veps_1^{\prime\prime}(\om_0)  } \bigg)^{ \frac{1}{2}n_0^2 }  \cdot 
   \bigg(  \f{  \i   }{ p^{\prime\prime}_1(\om_10) - \frac{1}{ \op{v} } \veps_1^{\prime\prime}(\om_1)  } \bigg)^{ \frac{1}{2}n_1^2 }     \\
\pl{r \in \mf{N}_{\e{sp}}  }{} \pl{a=1}{\vsg_r} 
\Bigg\{  
  \f{ G\big( 1 + n_r^{(a)} \big) \cdot \Big\{ \e{sgn}\big[ p^{\prime}_r(\om_r^{(a)})\big]  \Big\}^{n_r^{(a)}}  }
  { (2\pi)^{ \frac{n_r^{(a)}}{2} }\cdot  \Big(-\i \big[p^{\prime\prime}_r(\om_r^{(a)}) - \frac{1}{ \op{v} } \veps_r^{\prime\prime}(\om_r^{(a)}) \big]\Big)^{ \frac{1}{2}(n_r^{(a)})^2 } }      \Bigg\} \;. 
\end{multline}

 \item The summation runs through all possible choices of vectors of integers belonging to the set 
\bem
\mf{S}_{\op{v}} \; = \; \bigg\{ \bs{n}=(\ell_+, \ell_-; n_0, n_1, \{ n_{t_2}^{(1)},\dots, n_{t_2}^{(\vsg_{t_2})} \} , \dots, \{ n_{t_k}^{(1)},\dots, n_{t_k}^{(\vsg_{t_k})} \} , \dots   )  \\
\; \, \e{such} \, \e{that} \,  \; n_r^{(a)} \in \mathbb{N}\; , \;  \ell_{\pm} \in \mathbb{Z}  \quad \e{and} \quad 
\op{s}_{\ga} = \sul{\ups=\pm}{} \ell_{\ups} \; + \; \varkappa_{\op{v}} n_0 +n_1 + \sul{r \in \mf{N}_{\e{sp}} }{} \sul{a=1}{\vsg_r} r n_r^{(a)}  \bigg\} \;. 
\end{multline}
There, the integers $t_k$ are as defined in \eqref{definition entiers de Nsp}.

\item  $\De_{\ups,\bs{n}}$ reads
\bem
\De_{\ups,\bs{n}}\; = \; - \, \ups \ell_{\ups} \, + \, \tfrac{ 1 }{ 2 } \op{s}_{\ga}  Z(  q ) \, - \varkappa_{\op{v}} n_0  \phi_1( \ups q , \om_{0} )    \, - \,  n_1  \phi_1( \ups q , \om_{1} )   \\
\; - \; \sul{r \in \mf{N}_{\e{sp}} }{} \sul{a=1}{\vsg_r}  n_r^{(a)}  \phi_{r}(\ups q , \om_r^{(a)} ) \; -\sul{ \ups^{\prime} \in \{ \pm \} }{}\ell_{ \ups^{\prime} }  \phi_1(\ups q , \ups^{\prime} q \, )  \;, 
\end{multline}
where $Z$ is the dressed charge \eqref{definition dressed charge} and  $\phi_{r}$ is the dressed phase \eqref{definition dressed phase} associated with an $r$-string excitation.
The critical exponent $\De_{\ups,\bs{n}}$ depends also on the location of the massive saddle-points, and thus on the ratio $\op{v}$ of $m$ and $t$.

 \item $ \vp_{\bs{n}}(\op{v}) $ is the oscillatory phase which takes its origin in the contributions of the massive excitations saddle-points to  the asymptotics. It takes the explicit form 
\beq
\vp_{\bs{n}}(\op{v}) \; = \; \varkappa_{\op{v}} n_0 u_{1}(\om_0,\op{v}) \; + \; n_1 u_{1}(\om_1,\op{v}) \; + \;   \sul{r \in \mf{N}_{\e{sp}} }{} \sul{a=1}{\vsg_r}  n_r^{(a)}  u_{r}(\om_r^{(a)},\op{v}) \;. 
\enq
Finally, $\De_{\e{sp};\bs{n}}$ captures the contribution of the massive modes to the power-law decay. It reads
\beq
\De_{\e{sp};\bs{n}} \; = \; \f{1}{2}\bigg\{  n_0^2  \,+ \,  n_1^2 \, + \,  \sul{r \in \mf{N}_{\e{sp}} }{} \sul{a=1}{\vsg_r}  (n_r^{(a)})^2 \bigg\} \;.  
\enq

 \item Last but not least, the amplitude $ \mc{F}_{ \bs{n} }^{(\ga)}$ represents the non-universal part of the asymptotics, and corresponds to a properly normalised in the volume
 thermodynamic limit of the form factor squared of the $\sg_{1}^{\ga}$ operator 
taken between the model's ground state and the lowest lying  excited state having left/right Umklapp deficiencies $\ell_{-/+}$
and,  in the thermodynamic limit, giving rise to hole, particle and $r$-string excitations subordinate to the below distribution of rapidities
\beq
\mf{Y}^{(\e{sp})}_{\bs{n}} \; = \; \left\{ \ba{cc}  
\bigg\{ -\op{s}_{\ga} \, ;\,  \big\{ \ell_{\ups} \big\} \; \big| \; \emptyset \, ; \,   \Big\{ \big\{ \om_0 \big\}^{n_0} \cup   \big\{ \om_1 \big\}^{n_1}   \Big\}  ;  \emptyset; \dots; \emptyset ; 
\Big\{ \big\{ \om_{t_k} \big\}^{n_{t_k}^{(a)}} \Big\}_{a=1}^{\vsg_{t_k}} ; \dots    \bigg\}  
&  \qquad  \op{v}_F < |\op{v}| < \op{v}_{1}^{(M)}   \vspace{3mm} \\
\bigg\{  -\op{s}_{\ga} \, ;\,   \big\{ \ell_{\ups} \big\}\; \big| \; \big\{ \om_0 \big\}^{n_0}  ;    \big\{ \om_1 \big\}^{n_1} ; ;  \emptyset; \dots; \emptyset ; 
\Big\{ \big\{ \om_{t_k} \big\}^{n_{t_k}^{(a)}} \Big\}_{a=1}^{\vsg_{t_k}} ; \dots      \bigg\}   
&  \qquad     |\op{v}| < \op{v}_{F}   \ea \right.  \;. 
\enq

\end{itemize}

Note that, in \eqref{ecriture saddle point configuration of roots}, the notation $\Big\{ \big\{ \om_0 \big\}^{n_0} \cup   \big\{ \om_1 \big\}^{n_1}   \Big\}$ associated with the regime $\op{v}_F < |\op{v}| < \op{v}_{1}^{(M)}$
refers to the set of particle rapidities built up from $n_0$ particles with rapidity $\om_0$ and $n_1$ particles with rapidity $\om_1$. There are no-hole excitations in that regime. 
Also, the only $r$-string roots that are present in $\mf{Y}^{(\e{sp})}_{\bs{n}} $ are those such that $r\in \mf{N}_{\e{sp}}$ and $n_r^{(a)}\not=0$ for some $a=1,\dots, \vsg_r$.

\section{Conclusion}
\label{Section conclusion}
 
In this paper, I have applied the saddle-point method so as to determine the long-distance and large-time asymptotic behaviour of dynamical two-point
functions in the massless regime of the XXZ chain at non-zero magnetic field. The analysis was carried on the level of the massless form factor series expansion for the model's
two-point functions which was obtained in \cite{KozMasslessFFSeriesXXZ}. All-in-all, my analysis provides one with the explicit form of the leading power-law asymptotics
associated with each oscillating harmonics and allows for a clear connection with the predictions 
\cite{GlazmanImambekovSchmidtReviewOnNLLuttingerTheory,HaldaneCritExponentsAndSpectralPropXXZ,LutherPeschelCriticalExponentsXXZZeroFieldLuttLiquid}
stemming from the use of the universality hypothesis and the conjectural correspondence with the physics of the Luttinger liquid model.

On top of providing an exact characterisation of the asymptotic
regimes, the present analysis allows one to determine the precise role played by the bound state excitations on the asymptotics. In particular, I have demonstrated that bound states
only contribute to exponentially small corrections in the distance to the equal-time correlators. 
Although this kind of result was conjectured to hold already long ago, it was never obtained, through exact manipulations. 
 While bound state do produce, in the end, exponentially small corrections to the asymptotics, the details of the analysis show that the presence of bound states
in the spectrum plays a very important role in the mechanism generating the exponentially small corrections. If this part of the spectrum was simply absent,
\textit{viz}. the excitation spectrum only boiling down to particles and holes, then there would arise other power-law corrections to the long-distance asymptotics, 
stemming from the portion of the spectrum parameterised by the particle rapidities  going to $\infty$. These corrections would, in particular, break the expected universal behaviour.

  The results of the paper are based on the conjecture of contour deformations, at least in what concerns the contribution of the sectors with higher than $3$ hole excitations. 
 It would be very interesting to find a proof of this conjecture for general $n$. Also, it would be interesting to get a better analytic grip on the properties of the 
 oscillating phase factors $u_r(\la,\op{v})$.

\section*{Acknowledgment}

K.K.K. acknowledges support from  CNRS and ENS de Lyon. The author is indebted to J.-S. Caux, F. Göhmann, J.M. Maillet, G. Niccoli and N.A. Slavnov for stimulating discussions
on various aspects of the project.

\appendix






\section{Observables in the infinite XXZ chain}
\label{Appendix Observables XXZ}

\subsection{The thermodynamic observables in the XXZ chain}
\label{Appendix Sols Line Int Eqns}

The observables describing the thermodynamic limit of the spin-$1/2$ XXZ chain are characterised by means of a collection of functions 
solving linear integral equations. These equations are driven by the integral kernel 
\beq
K(\la\mid \eta )  \, = \,   \f{ \sin(2\eta)   }{ 2\pi \sinh(\la + \i \eta)  \sinh(\la  - \i \eta)  }  \;. 
\label{ecriture fonction K de lambda et eta}
\enq

To introduce all of the functions of interest to this work, one starts by defining the $Q$-dependent dressed energy which allows one to construct the Fermi zone of the model. 
It is defined as the solution to the linear integral equation
\beq
\veps(\la\mid Q) \, + \, \Int{-Q}{Q} K\big(\la-\mu\mid \zeta \big) \, \veps(\la\mid Q)  \cdot \dd \mu \; = \;  h - 4 \pi J \sin(\zeta) K\big( \la \mid \tfrac{1}{2}\zeta \big)  \;. 
\label{definition energie habille et energie nue}
\enq
The endpoint of the Fermi zone is defined as the unique \cite{KozDugaveGohmannThermoFunctionsZeroTXXZMassless} positive solution $q$ to $\veps(q\mid q)=0$. 
Then, the function $\veps_1(\la)\equiv \veps(\la\mid q)$ corresponds to the dressed energy of the particle-hole excitations of the model. 
The functions 
\beq
\veps_r(\la)\; = \; r h - 4 \pi J \sin(\zeta) K\big( \la \mid \tfrac{r}{2}\zeta \big)   \, -\, \Int{-q}{q} K_{r}\big(\la-\mu \big) \veps_1(\mu)  \cdot \dd \mu 
\label{definition r energie habille}
\enq
with 
\beq
K_{r}(\la) \,  = \,  K\Big(\la \mid \tfrac{1}{2} \zeta(r+1) \Big) \, + \,  K\Big(\la \mid \tfrac{1}{2} \zeta(r-1) \Big)
\enq
correspond to the dressed energies of the $r$-bound state excitations.
For any $0<\zeta<\tf{\pi}{2}$ and under some additional constraints for $\tf{\pi}{2}< \zeta < \pi$, one can show \cite{KozProofOfStringSolutionsBetheeqnsXXZ}
that $\veps_{r}(\la+\i\de  \tf{\pi}{2}) > c_r>0$ for any $\la\in \R$, and $\de \in \{0, 1\}$. 
However, this lower bound should hold throughout the whole massless regime $0<\zeta<\pi$, irrespectively of some additional constraints. 
This property has been checked to hold by  numerical study of the solutions to \eqref{definition r energie habille}, \textit{c.f.} \cite{KozProofOfStringSolutionsBetheeqnsXXZ}. 

In order to introduce the dressed momenta of the $r$-bound states and of the particle-hole excitations, I first need to 
define the $r$-bound state bare phases $\theta_r$ :
\beq
\theta_r(\la) \, = \, 2\pi \Int{ \Ga_{\la}  }{}  K_r(\mu-0^+ ) \cdot \dd \mu \quad \e{for} \;\;  r \geq 2
\quad \e{and} \qquad
\theta_1(\la\mid \eta) \, = \, 2\pi \Int{ \Ga_{\la}  }{}  K(\mu-0^+\mid \eta ) \cdot \dd \mu  \;. 
\enq
The contour of integration corresponds to the union of two segments $ \Ga_{\la} \; = \; \intff{ 0 }{ \i \Im(\la) }\cup \intff{\i \Im(\la) }{ \la } $ and the $-0^+$ prescription indicates that the poles 
of the integrand at $\pm \i \eta +\i \pi \mathbb{Z}$ should be avoided from the left.

Then, the function  
\bem
p_r(\la)\; = \; \theta\big(\la\mid \tfrac{r}{2}\zeta \big)  \, -\, \Int{-q}{q} \theta_{r}\big(\la-\mu \big) p^{\prime}_1(\mu)  \cdot \f{ \dd \mu }{2\pi} \\
\, + \, \pi \ell_r(\zeta)-p_{F}m_r(\zeta)
-2p_{F} \sul{ \sg=\pm }{} \big(1\, - \, \de_{\sg,-}\de_{r,1} \big) \e{sgn}\Big( 1- \tfrac{2}{\pi} \cdot \wh{\tfrac{r+\sg}{2}\zeta} \Big) \cdot \bs{1}_{ \mc{A}_{r,\sg} } (\la)\;, 
\label{definition r moment habille}
\end{multline}
extended by $\i \pi$-periodicity to $\Cx$, corresponds to the dressed momentum of the $r$-bound states. Above, I have introduced
\beq
 \ell_r(\zeta)=1-r+2 \lfloor \tfrac{r \zeta}{2\pi} \rfloor  \qquad \e{and} \qquad
   m_r(\zeta)=2-r - \de_{r,1} + 2 \sul{ \ups = \pm }{} \lfloor \zeta \tfrac{r + \ups   }{2\pi} \rfloor  \;. 
\enq
Furthermore, I agree upon  
\beq
\wh{\eta} \, = \, \eta  - \pi \lfloor \tfrac{ \eta }{ \pi } \rfloor \quad \e{and} \qquad \mc{A}_{r,\sg} \, = \, 
\Big\{ \la \in \Cx \; : \; \tfrac{\pi}{2}\geq |\Im(\la) | \geq \e{min}\big(   \wh{\tfrac{r+\sg}{2}\zeta} ,  \pi- \wh{\tfrac{r+\sg}{2}\zeta} \big)  \Big\}  \;. 
\enq
In order to obtain $p_r$, one should first solve the linear integro-differential equation for $p_1$ and then 
use $p_1$ to define $p_r$ by \eqref{definition r moment habille}.  $p_1$ corresponds to the dressed momentum of the particle-hole excitations and  $p_F=p_1(q)$ corresponds to the Fermi momentum. 
$1$-strings have their rapidities $\la \in \big\{  \R \setminus \intff{-q}{q} \big\} \cup \big\{ \R+\i\tf{\pi}{2}\big\}$ while $r\geq 2$ strings, $r \in \mf{N}\setminus \{1\}$, 
have their rapidities $\la \in \R+\i \sg_{r}\tf{\pi}{2}$ for a $\sg_r=0$ or $1$, depending on the value of $r$ and $\zeta$. 

One can show \cite{KozProofOfStringSolutionsBetheeqnsXXZ} under similar conditions on $\zeta$ as for the dressed energy that, for any $\la \in \R$
\beq
\big| p^{\prime}_{r}\big(\la  +\i \de_{r} \tfrac{\pi}{2} \big) \big| \; > \; 0 \quad \e{when} \quad r \in \mf{N}\setminus \{1\}
\qquad \e{and} \qquad 
\e{min}\Big( p^{\prime}_{1}\big(\la\big) \, , \, - p^{\prime}_{1}\big(\la  + \i \tfrac{\pi}{2} \big)   \Big) \; > \; 0   \;. 
\label{ecriture equation positivite pr prime}
\enq
Again, a numerical investigation indicates that \eqref{ecriture equation positivite pr prime} does, in fact, hold irrespectively of the value of $\zeta$.

The $r$-bound dressed phase is defined as the solution to 
\beq
\phi_{r}(\la,\mu) \, = \, \f{ 1  }{ 2 \pi }  \theta_{r}\big( \la -\mu  \big) \, - \, \Int{-q}{q} K(\la-\nu)\,  \phi_{r}(\nu, \mu ) \cdot \dd \nu  \; + \; \f{m_{r}(\zeta)}{2}
\label{definition dressed phase}
\enq
and the dressed charge solves
\beq
Z(\la)\, + \,  \Int{-q}{q} K(\la-\mu)\,  Z(\mu ) \cdot \dd \mu   \, = \,  1 \;. 
\label{definition dressed charge}
\enq
The dressed charge is related to the dressed phase by the below identities \cite{KorepinSlavnovNonlinearIdentityScattPhase}:
\beq
\phi_1(\la,q) \, - \,   \phi_1(\la,-q) \, + \, 1 \; = \; Z(\la) \quad \e{and} \quad 1+\phi_1(q,q) - \phi_1(-q,q) \, = \, \f{1}{ Z(q) }  \;. 
\label{ecriture identites entre phase et charge habilles} 
\enq

Finally, the functions $\vth_{\ups}$ parameterising the exponents $\vth_{\ups}^2( \mf{Y} ) $ take the explicit form 
\beq
\vth_{\ups}( \mf{Y} ) \, = \, \, - \, \ups \ell_{\ups} \, + \, \tfrac{ 1 }{ 2 } \op{s}_{\ga}  Z(  q ) \, + \,   \sul{ a=1 }{ n_h } \phi_1( \ups q , \mu_{a} )   
\; - \; \sul{  r \in \mf{N}  }{  } \sul{ a=1 }{ n_r } \phi_{r}(\ups q , \nu_a^{(r)} ) 
\; - \hspace{-1mm} \sul{ \ups^{\prime} \in \{ \pm \} }{}\ell_{ \ups^{\prime} }  \phi_1(\ups q , \ups^{\prime} q \, )  \;. 
\label{definition shifted sfift function}
\enq
%
%
%






\subsection{The $r$-string states}
\label{Appendix Sous-section cordes}

The characterisation of bound states depends strongly on the value of $\zeta$ defining the anisotropy. 
It has been proven in \cite{KozProofOfStringSolutionsBetheeqnsXXZ} that for $\tf{\pi}{2}<\zeta< \pi$ an $r$ bound state  centred on $\R+\i \sg_{r} \tf{\pi}{2}$, $\sg_{r} = 0$ or $1$, exists if and only if the 
below constraints are all simultaneously satisfied:
\beq
(-1)^{\sg_{r}} \sin\big[ k \zeta \big] \cdot \sin\big[ (r-k)\zeta \big] \, > \, 0 \qquad \e{for} \; k=1,\dots, r-1. 
\label{ecriture equations existance corde Takahashi}
\enq
These are precisely the conditions argued earlier by Suzuki-Takahashi \cite{TakahashiSuzukiFiniteTXXZandStrings}
and, subsequently, in \cite{FowlerZotosConditionsOfStringExistenceXYZAndSineGordon,HidaConditionExistenceStringsXXZ,KorepinAnalysisofBoundStateConditionMassiveThirring}.

However, for $0< \zeta < \tf{\pi}{2}$, the work \cite{KozProofOfStringSolutionsBetheeqnsXXZ} proved that an $r$ bound state centred on $\R - \i \kappa_{r} \tf{\pi}{2}$, exists if and only if the 
below constraints are all simultaneously satisfied:
\beq
(-1)^{ \kappa_{r} } \cdot \sin\bigg(  \f{\pi \zeta}{\pi-\zeta} (k-p) \bigg) \cdot \sin\bigg(  \f{\pi \zeta}{\pi-\zeta} (r-k+p-\kappa_{r}-1) \bigg) \, > \, 0 
\label{ecriture des condition existence corde zero zeta pi sur 2}
\enq
for $k\, = \, 1,\dots, r-2$ and $k \in \intn{ w_{p}+1 }{ w_{p+1}-1 }$ with $p\geq 0$. 
Above, $\kappa_{r}=\lfloor (r-1)\tf{\zeta}{\pi} \rfloor$  and 
\beq
w_{p} \, = \, \Big\lfloor  \Big( p-\tfrac{\kappa_{r}}{2}+(r-1)\tfrac{\zeta}{2\pi}  \Big) \f{ \pi }{ \zeta } \Big\rfloor \;. 
\enq
While this has not been proven, it appears that the set of equations \eqref{ecriture des condition existence corde zero zeta pi sur 2} is, in fact, equivalent to 
\eqref{ecriture equations existance corde Takahashi}

\subsubsection{Properties of $2$-strings}
\label{Appendix sous section 2strings}

In that case, when $ \tfrac{\pi}{2}<\zeta < \pi$,  the equations for the existence of the $2$-string take the form $(-1)^{\sg_2} \sin^2(\zeta) \, > \, 0$, what 
entails that the $2$-string parity is $0$, \textit{viz}. the rapidity is real valued. 
As for the regime $0<\zeta < \tf{\pi}{2}$, the general analysis entails that the $2$-string also always exists and that its rapidity is real valued. 
Finally, since one has $|p^{\prime}_2(\la)|>0$ on $\R$, the asymptotics
\eqref{ecriture DA de p prime} along with the positivity of the constant $\a_{p}$ \eqref{positivite constantes alpha eps et alpha p}
allows one to infer that  $\e{sgn}[p^{\prime}_2]_{\mid \R}=\e{sgn}\big[ \sin(2\zeta) ]$. 

Thus, one has the array 
\beq
\ba{|c|c|c|c|}
\hline
							& 0< \zeta < \tf{\pi}{2}             &    \tf{\pi}{2} \, < \, \zeta  \,  <  \, \pi   \\ \hline
$2$-\e{string}\, \e{rapidity}  				 & \R						&   \R 						 \\ \hline 
\e{sgn}[p^{\prime}_2]_{\mid \R} 		&      +1					&  -1 					 \\
\hline
\ea
\enq

\subsubsection{Properties of $3$-strings}
\label{Appendix sous section 3strings}

In that case, when $ \tfrac{\pi}{2}<\zeta < \pi$, the equations for the existence of the $3$-string take the form $(-1)^{\sg_3} \sin(\zeta) \sin(2\zeta) \, > \, 0$. 
There is thus always a solution in the  regime: $\de_{3}=1$, \textit{viz}. the 3-string rapidities belong to $\R + \i \tf{\pi}{2}$. 
Further, one extracts from the asymptotics, as in the $2$-string case, that then 
\beq
\e{sgn}\big[ p_3^{\prime}\big]_{\mid \R + \i \frac{\pi}{2}} \; = \; - \e{sgn}\big[ \sin(3\zeta) \big] \;. 
\enq
In the regime $0<\zeta < \tf{\pi}{2}$, the formula for the string centre entails that the $3$ string rapidity is real valued and 
the existence condition for $3$-strings take the form 
\beq
\sin^2\Big( \f{ \pi \zeta }{ \pi-\zeta}\Big) \, >\, 0 \;. 
\enq
Again, from the asymptotics, one infers that 
\beq
\e{sgn}\big[ p_3^{\prime}\big]_{\mid \R } \; = \;  \e{sgn}\big[ \sin(3\zeta) \big]\;. 
\enq

Thus, one has the array 
\beq
\ba{|c|c|c|c|c|}
\hline
							& 0< \zeta < \tf{\pi}{3}        & \tf{\pi}{3}<\zeta < \tf{\pi}{2}     &    \tf{\pi}{2} \, < \, \zeta  \,  <  \, \tf{2\pi}{3}  &  \tf{2\pi}{3}<\zeta <   \pi   \\ \hline
$3$-\e{string}\, \e{rapidity}  				  & \R					&   \R 			&    		\R 		+\i\frac{\pi}{2}  & \R 		+\i\frac{\pi}{2}	 \\ \hline 
\e{sgn}[p^{\prime}_3]_{\mid \R+ \i \sg_3 \frac{\pi}{2}} &      \mf{s}_2 \mf{s}_3= 1		&  \mf{s}_2 \mf{s}_3=-1 	& \mf{s}_2 \mf{s}_3=1 		&  \mf{s}_2 \mf{s}_3=-1	 \\
\hline
\ea
\enq
and I employed the shorthand notation 
\beq
\mf{s}_{k}= \e{sgn}\big[ \sin(k \zeta) \big] \;. 
\enq

\subsubsection{Properties of $r$-strings}

One may obtain domains of existence and string parities for higher string lengths by inspecting the equations \eqref{ecriture des condition existence corde zero zeta pi sur 2}-\eqref{ecriture equations existance corde Takahashi}. 
It leads to the below results for $r\in \{4,\dots, 8\}$:
\beq
\ba{|c|c|c|c|c|}
\hline
							& 0< \zeta < \tfrac{\pi}{3}        &  \tfrac{2\pi}{3}<\zeta < \pi        \\ \hline
$4$-\e{string}\, \e{rapidity}  				  & \R		\;\; (\sg_{4}=0)			&   \R 	\;\; (\sg_{4}=0)		 	 \\ \hline 
\e{sgn}[p^{\prime}_4]_{\mid \R+ \i \sg_4 \frac{\pi}{2}} &      \mf{s}_4 		&  \mf{s}_4  	  \\
\hline
\ea
\enq
\beq
\ba{|c|c|c|c|c|}
\hline
							& 0< \zeta < \tfrac{\pi}{4}        & \tfrac{\pi}{3}<\zeta <\tfrac{\pi}{2}     &    \tfrac{\pi}{2} \, < \, \zeta  \,  <  \, \tfrac{2\pi}{3}  &  \tfrac{3\pi}{4}<\zeta <   \pi   \\ \hline
$5$-\e{string}\, \e{rapidity}  				  & \R		\;\; (\sg_{5}=0) 	&   \R 	+ \i\tfrac{\pi}{2} 	\;\; (\sg_{5}=1)	&    	\R 		\;\; (\sg_{5}=0)  & \R 		+\i\frac{\pi}{2}\;\; (\sg_{5}=1)	 \\ \hline 
\e{sgn}[p^{\prime}_5]_{\mid \R+ \i \sg_5 \frac{\pi}{2}} &      \mf{s}_5 		&   -\mf{s}_5             	&  \mf{s}_5 		&   -\mf{s}_5	 \\
\hline
\ea
\enq
\beq
\ba{|c|c|c|c|c|}
\hline
							& 0< \zeta < \tfrac{\pi}{5}        & \tfrac{4\pi}{5}<\zeta < \pi        \\ \hline
$6$-\e{string}\, \e{rapidity}  				  & \R		\;\; (\sg_{6}=0)			&   \R 	\;\; (\sg_{6}=0)		 	 \\ \hline 
\e{sgn}[p^{\prime}_6]_{\mid \R+ \i \sg_6 \frac{\pi}{2}} &      \mf{s}_6 		&  \mf{s}_6  	  \\
\hline
\ea
\enq
\beq
\ba{|c|c|c|c|c|c|c|}
\hline

						& 0< \zeta < \tfrac{\pi}{6}        & \tfrac{\pi}{4}<\zeta < \tfrac{\pi}{3}     &    \tfrac{\pi}{5} \, < \, \zeta  \,  <  \, \tfrac{\pi}{2}  &  \tfrac{\pi}{2}<\zeta <   \tfrac{3\pi }{5}   
										      &    \tfrac{2\pi}{3} \, < \, \zeta  \,  <  \, \tfrac{3\pi}{4}  &  \tfrac{5\pi}{6}<\zeta <  \pi  \\ \hline
$7$-\e{string}\, \e{rapidity}  				  & \R		 	&   \R 	+ \i\tfrac{\pi}{2} 	 	&    	\R 		   & \R 		+\i\frac{\pi}{2}   
						 &    	\R 		   & \R 		+\i\frac{\pi}{2}  \\ \hline 
\e{sgn}[p^{\prime}_7]_{\mid \R+ \i \sg_7 \frac{\pi}{2}} &      \mf{s}_7 		&   -\mf{s}_7              	& \mf{s}_7  		&   -\mf{s}_7 	& \mf{s}_7  		&   -\mf{s}_7  \\
\hline
\ea
\enq
\beq
\ba{|c|c|c|c|c|}
\hline
							& 0< \zeta < \tfrac{\pi}{7}        & \tfrac{\pi}{3}<\zeta < \tfrac{2\pi}{5}     &    \tfrac{3\pi}{5} \, < \, \zeta  \,  <  \, \tfrac{2\pi}{3}  &  \tfrac{6\pi}{7}<\zeta <   \pi   \\ \hline
$8$-\e{string}\, \e{rapidity}  				  & \R		\;\; (\sg_{8}=0) 	&   \R 	\;\; (\sg_{8}=0)	&    	\R 		\;\; (\sg_{8}=0)  & \R 		\;\; (\sg_{8}=0)	 \\ \hline 
\e{sgn}[p^{\prime}_8]_{\mid \R+ \i \sg_8 \frac{\pi}{2}} &      \mf{s}_8 		& \mf{s}_8             	&  \mf{s}_8 		&  \mf{s}_8	 \\
\hline
\ea
\enq
%
%
%






\subsection{The velocity of individual excitations}
\label{Appendix Section phase oscillante dpdte de la vitesse}

The velocity $\op{v}_{r}$ of an $r$-string excitation, if $r \in \mf{N}_{\e{st}}$, and the velocity $\op{v}_1$ of particle or hole excitation are all defined  by  the ratio
\beq
\op{v}_r(\la) \, = \; \f{ \veps_{r}^{\prime}(\la) }{ p_{r}^{\prime}(\la) } \; , \qquad  \e{in} \; \e{particular} \qquad  \op{v}_{F}=\f{ \veps_{1}^{\prime}(q) }{ p_{1}^{\prime}(q) }
\enq
is the Fermi velocity, namely the velocity of the particle-hole excitation on the right edge of the Fermi zone $\intff{-q}{q}$.

It is easy to see that for $|\Re(\la)|$ large enough it holds 
\beq 
p^{\prime}_r(\la) \; = \;  2 \sin(r\zeta) \,   \sul{k\geq 1}{} \ex{\mp 2 \la } \a_{p}^{(k)}    \qquad \e{and} \qquad
\veps_r(\la) \; = \;  r h -   \sin(r\zeta)  \sul{k\geq 1}{} \ex{\mp 2 \la } \a_{\veps}^{(k)}    \;. 
\label{ecriture DA de p prime}
\enq
The coefficients $\a_{p}^{(1)}$ and $\a_{\veps}^{(1)}$ take the explicit form 
\beq
\a_{p}^{(1)} \, = \, 2- \f{ 2  }{ \pi} \cos(\zeta) \Int{-q}{q} \cosh(2\mu) p^{\prime}_1(\mu) \dd \mu   \qquad \e{and} \qquad 
\a_{\veps}^{(1)} \, = \,   8 J \sin(\zeta) - \f{ 2  }{ \pi} \cos(\zeta) \Int{-q}{q} \sinh(2\mu) \veps^{\prime}_1(\mu) \dd \mu   \;. 
\enq
Both coefficients are real analytic functions of $\zeta \in \intoo{0}{\pi}$, they can thus vanish only at isolated points. Furthermore, 
as shown in \cite{KozDugaveGohmannThermoFunctionsZeroTXXZMassless}, one has that $\veps^{\prime}_1(\la)>0$ when $ \la \in \R^+$ and $p_1^{\prime}(\la)>0$ when $ \la \in \R$. 
This entails that the leading terms in the $\la \tend + \infty$ asymptotics of  $\veps^{\prime}_1$ and $p_1^{\prime}$ have to be strictly positive, and thus 
\beq
\a_{\veps}^{(1)} >0 \quad \e{and} \quad \a_{p}^{(1)}>0 \;. 
\label{positivite constantes alpha eps et alpha p}
\enq
Finally, the coefficients $\a_{p}^{(k)}$ and $\a_{\veps}^{(k)} $ are all bounded uniformly in $r$. 

This entails that the function $u_r(\la,\op{v})=p_r(\la)-\tfrac{1}{\op{v}}\veps_{r}(\la)$ has the following asymptotic behaviour 
\beq
u_{r}(\la ,\op{v}) \; = \; p_r^{(\pm)}(y)-\frac{r h }{\op{v}} \, \mp \, \a_{p}^{(1)}  \sin(r\zeta) \ex{\mp 2 \la } \bigg\{ \cdot \Big( 1 \mp \f{ \op{v}_{\infty} }{ \op{v}_F }\Big)
\; + \; \e{O}\Big( \ex{\mp 4 \la } \Big) \bigg\}
\enq
when $\la=x+\i y $ and $x \tend \pm \infty$. Finally, $p_r^{(\pm)}(y)$ are some explicitly computable piece-wise continuous functions of $y \in \intff{-\tf{\pi}{2} }{\tf{\pi}{2}}$. 
Finally, I have set 
\beq
\op{v}_{\infty} \, = \, \f{ \a_{\veps}^{(1)} }{ \a_p^{(1)} } \, = \,  \f{  8 \pi J \sin(\zeta)  \, - \,  2  \cos(\zeta) \int_{-q}^{q} \sinh(2\mu) \veps^{\prime}_1(\mu) \dd \mu   }
{2\pi \, - \,  2  \cos(\zeta) \int_{-q}^{q} \cosh(2\mu) p^{\prime}_1(\mu) \dd \mu  } \;. 
\label{definition cste V infty}
\enq

One may draw several consequences out of these asymptotics. First of all, there exists $A>0$ that is $r$ independent,  such that for $\la=x+\i y $ with $ \pm x>A$
\beq
\ba{|c|c|c|c|}
\hline
							& |\op{v}| \, > \,  \op{v}_{\infty}              &    0 \, < \, \op{v}  \,  <  \, \op{v}_{\infty} &  -\op{v}_{\infty} \, <  \, \op{v}  \, < \,  0 \\ \hline
\e{sgn}\Big[ \Im\Big( u_{r}(\la ,\op{v}) \Big)\Big]^{}    & \e{sgn}\big[ \sin(r\zeta) \sin(2y) \big]	&   \mp \e{sgn}\big[ \sin(r\zeta) \sin(2y) \big]   & \pm \e{sgn}\big[ \sin(r\zeta) \sin(2y) \big] \\
\hline
\ea
\enq

Furthermore, on the line $\R+\i \sg \tf{\pi}{2}$, it holds 
\beq
u_{r}^{\prime}( x +\i \sg \tfrac{\pi}{2} ,\op{v}) \; = \;(-1)^{ \sg } \,2 \a_{p}  \sin(r\zeta) \ex{ - 2 |x| } \cdot \Big( 1 \mp \f{ \op{v}_{\infty} }{ \op{v}  }\Big)
\; + \; \e{O}\Big( \ex{ -  4 |x| } \Big)  \quad \e{when} \; x \tend \pm \infty \;. 
\label{ecriture DA en infty de u r prime}
\enq
Since $u_{r}^{\prime}(\la,\op{v})$ is an analytic function in a small strip around $\R$ and $\R+\i\tfrac{\pi}{2}$, the large $x$ asymptotics given above ensures that it has in fact 
a finite number of zeroes on $\R+\i \sg \tfrac{\pi}{2}$. By comparing the signs of the asymptotics at $\pm \infty$, it also allows one to state on the parity of the number of these zeroes

More precisely, 
\begin{itemize}
 
 \item if $|\op{v}| \, > \,  \op{v}_{\infty} $, then $u_{r}^{\prime}(\la,\op{v})$ admits an even number of zeroes $2\kappa_{r}(\op{v},\sg)$ on $\R+\i\sg \tfrac{\pi}{2}$, with $\kappa_{r}(\op{v},\sg) \in \mathbb{N}$;

 \item if $|\op{v}| \,  < \,  \op{v}_{\infty} $, then $u_{r}^{\prime}(\la,\op{v})$ admits an odd number of zeroes $2\kappa_{r}(\op{v},\sg)+1$ on $\R+\i \sg \tfrac{\pi}{2}$, with $\kappa_{r}(\op{v},\sg) \in \mathbb{N}$. 
 
\end{itemize}
Also, since $p^{\prime}_r$ enjoys the lower bounds \eqref{ecriture equation positivite pr prime}, the asymptotics \eqref{ecriture DA en infty de u r prime} ensure that $\kappa_{r}(\op{v},\sg)=0$ when $\op{v}$ is large enough. 
Also, the properties of the functions $\veps_r^{\prime}$ ensure that $\kappa_{r}(\op{v},\sg)=0$ when $|\op{v}|$ is small enough. Indeed, one has the

\begin{lemme}
 \label{Lemme borne inf sur impuslion energie sur courbes excitations}
 
Let $r\in \mf{N}$ and $u_r(\la,\op{v})=p_r(\la)-\tfrac{1}{\op{v}}\veps_{r}(\la)$. Then, there exists $\op{v}_{r}^{(\mf{m})}< \op{v}_{\infty}$ 
and $\op{v}_{r}^{(M)}> \op{v}_{\infty}$ such that for $| \op{v} | >  \op{v}_{r}^{(M)} $ it holds,  for any $\la \in \R$, that 
\beq
\big| \Dp{\la} u_{r}(\la+\i\sg_{r} \tfrac{\pi}{2},\op{v}) \big| \; > \; 0 \quad \e{if} \quad r \in \mf{N}_{\e{st}} 
\qquad \e{and} \qquad 
\e{min}\Big( \Dp{\la} u_{1}\big(\la,\op{v}\big) \, , \, \Dp{\la} u_{1}\big(\la  + \i \tfrac{\pi}{2} ,\op{v}\big)   \Big) \; > \; 0 
\quad \e{for} \quad  r=1 \;   \;. 
\label{ecriture condition no saddle point for ur}
\enq
 Furthermore, for $| \op{v} | <  \op{v}_{r}^{(\mf{m})} $ and $r \in \mf{N}_{\e{st}}$ 
 the function $\la \mapsto \Dp{\la} u_{r}(\la+\i\sg_{r} \tfrac{\pi}{2},\op{v})$ admits a unique zero on $\R$. Likewise, 
the functions   $\la \mapsto \Dp{\la} u_{1}\big(\la,\op{v}\big)$ and $\la \mapsto , \Dp{\la} u_{1}\big(\la  + \i \tfrac{\pi}{2} ,\op{v}\big)$
also admit a unique zero on $\R$. 
 
\end{lemme}

\proof 

The large-$\la$ expansion of $p_r^{\prime}$ and $\veps_r^{\prime}$ building on \eqref{definition r energie habille}, \eqref{definition r moment habille} and $\veps_{1}(\pm q)=0$ yields
\beq
\Dp{\la}u_{r}\big( \la,\op{v} \big) \;\underset{ \Re(\la) \tend \pm \infty }{=} \; p^{\prime}_{r}(\la) \cdot \Big( 1  \mp \f{ \op{v}_{\infty} }{ \op{v} } \Big) \cdot \Big(1\, + \, \e{O}\big(  \ex{\mp 2\la }\big) \Big)
\enq
with $\op{v}_{\infty}$ as defined in \eqref{definition cste V infty}. Thence, there exists $\la_r>0$ large enough such that, for $|\la|>\la_r$, $\big|\Dp{\la}u_{r}\big( \la,\op{v} \big)\big|>0$ this provided that  $|\op{v}|> \op{v}^{(\infty)}$. 

However, owing to \eqref{ecriture equation positivite pr prime}, one has 
\beq
\big| p^{\prime}_{r}\big(\la  +\i \sg_{r} \tfrac{\pi}{2} \big) \big| \; \geq \; c_r>0 \quad \e{when} \quad r \in \mf{N}_{\e{st}} 
\qquad \e{and} \qquad 
\e{min}\Big( p^{\prime}_{1}\big(\la\big) \, , \, p^{\prime}_{1}\big(\la  + \i \tfrac{\pi}{2} \big)   \Big) \; > \; c_1 \geq 1  
\enq
for any $|\la|\leq \la_r$. Thence, since $\veps_{r}(\la)$ is bounded on $\R\cup \big\{ \R + \i \tfrac{\pi}{2} \big\}$, there exists $\op{v}_{r}^{(M)}$ large enough, in particular larger than 
$\op{v}^{(\infty)}$, such that for any $ |\op{v}| > \op{v}_{r}^{(M)}$ it holds 
\beq
1+\f{ \veps_{r}^{\prime}\big( \la + \i \sg \tfrac{\pi}{2} \big)  }{ \op{v} \cdot  p_{r}^{\prime}\big( \la + \i \sg \tfrac{\pi}{2} \big) } \geq \tilde{c_r} >0
\enq
for any $\la \in \R$ where $\sg=\sg_r$ if $r \in \mf{N}_{\e{st}}$ and where $\sg\in \{0,1\}$ is arbitrary if $r=1$. 

It remains to discuss the case of the lower-bound  $\op{v}_{r}^{(\mf{m})}$. One can conclude similarly by using that $\la \mapsto \veps_{r}^{\prime}\big( \la + \i \sg_r \tfrac{\pi}{2} \big) $
admits a unique zero on $\R$ and then applying a perturbative argument in $\op{v}$. \qed

 \vspace{3mm}

\begin{figure}[t]
 
$\ba{cc}
\includegraphics[width=.4\textwidth]{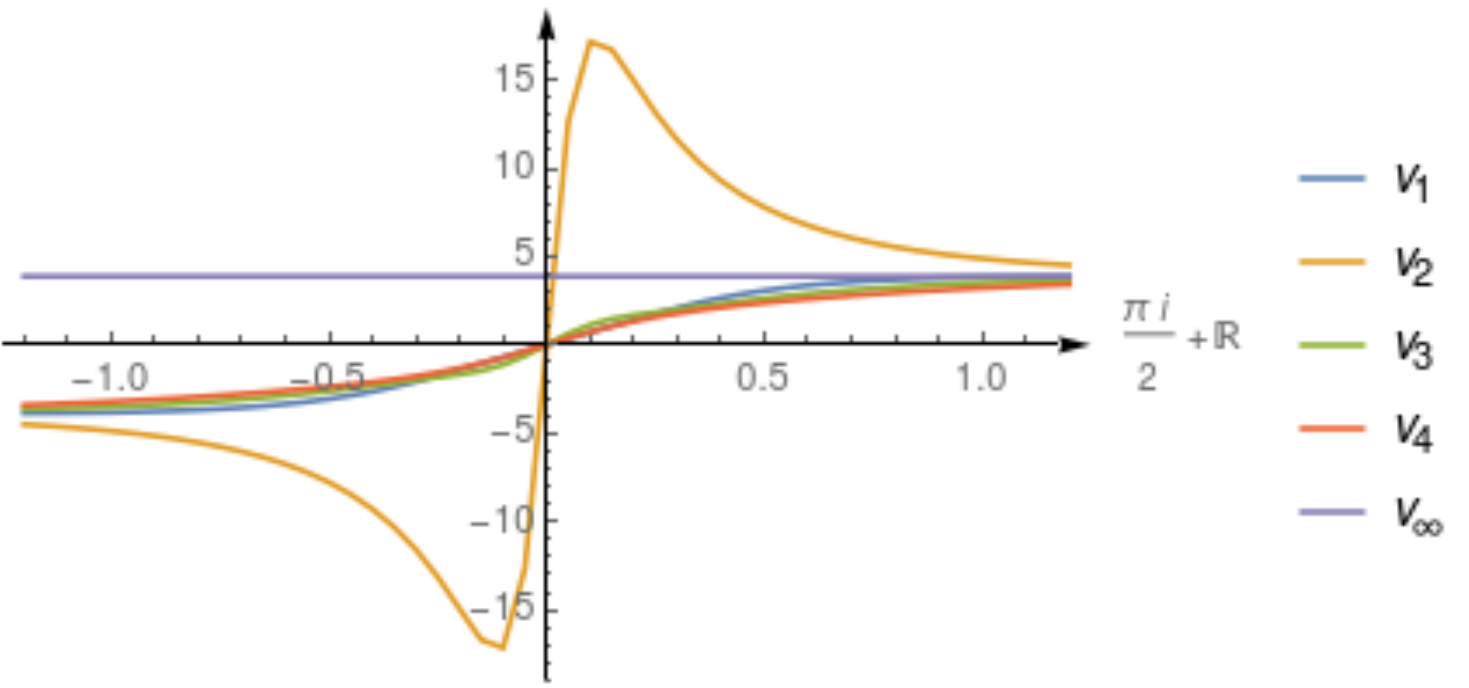}
& 
\includegraphics[width=.4\textwidth]{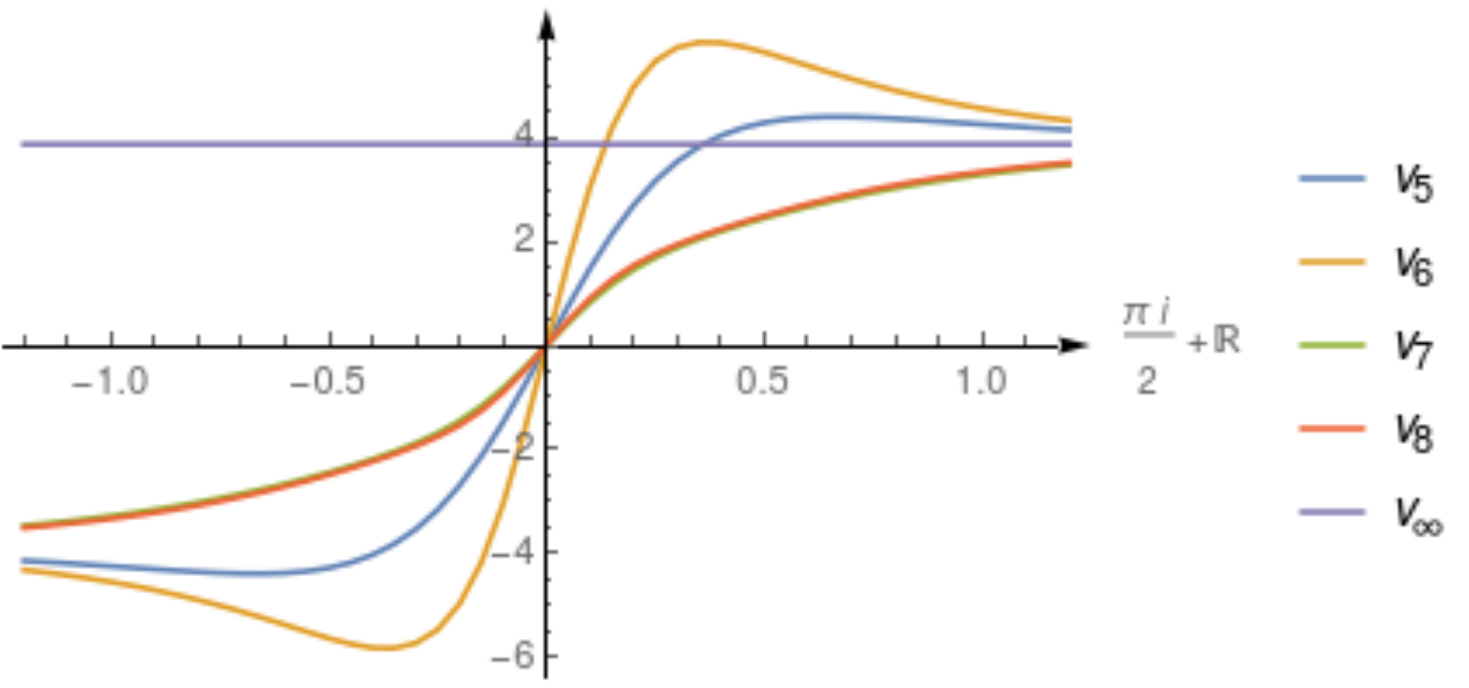} \\
\includegraphics[width=.4\textwidth]{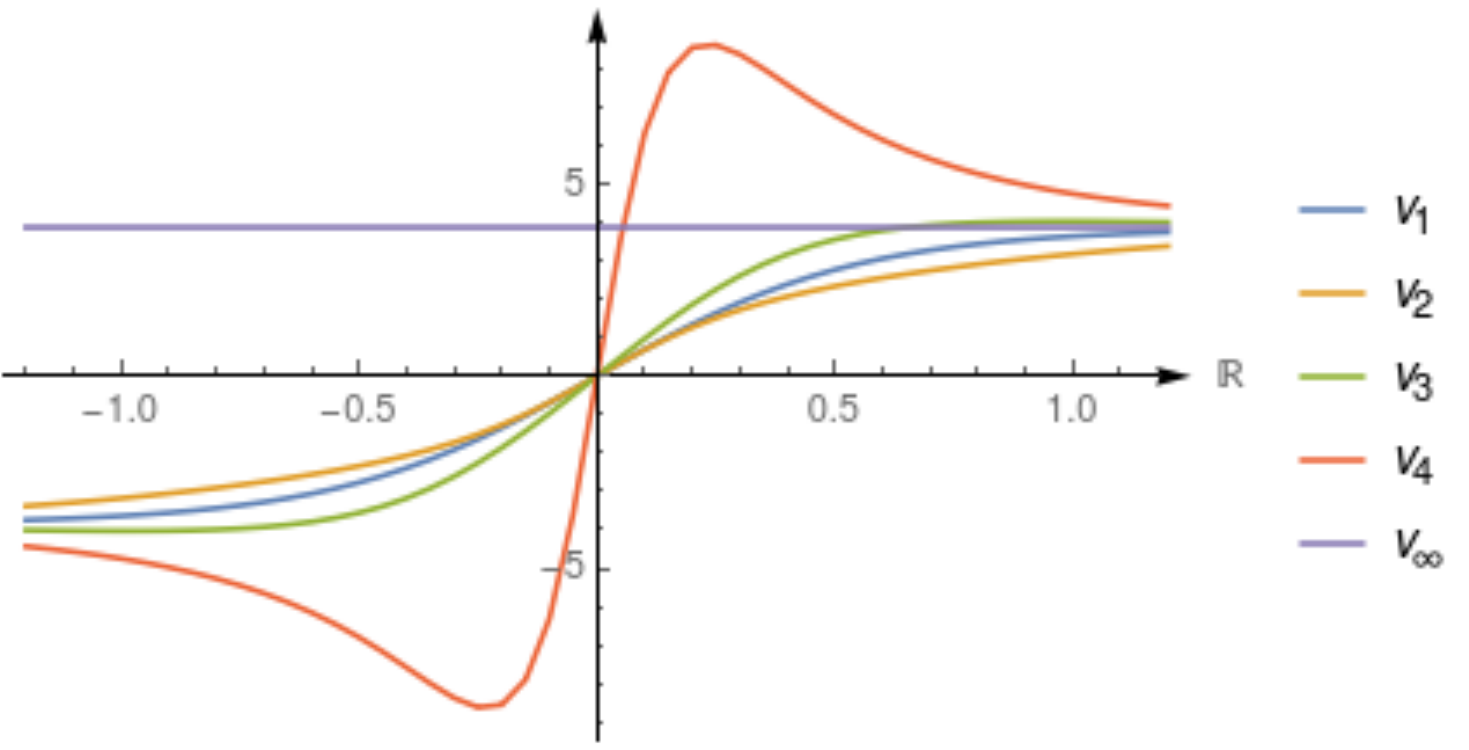}
& 
\includegraphics[width=.4\textwidth]{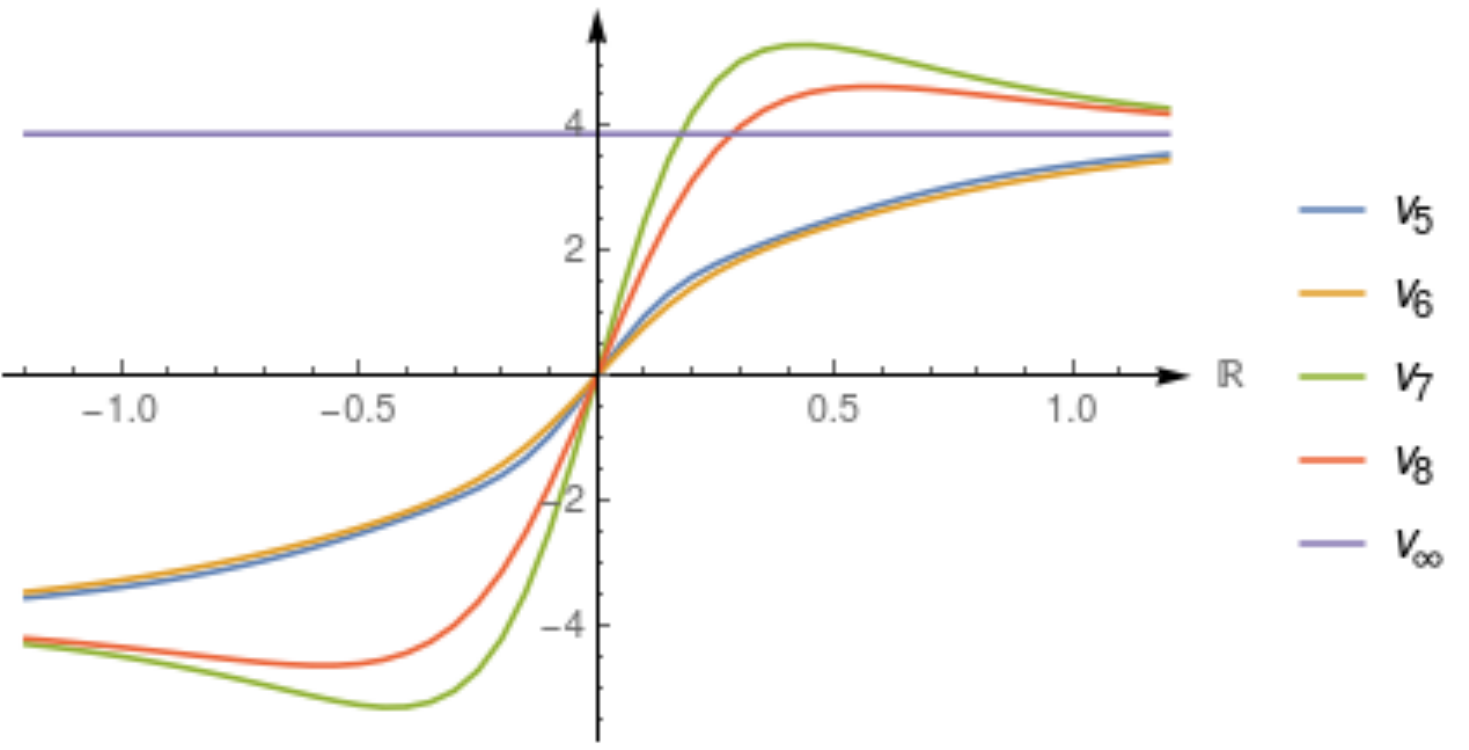}
\ea$
\caption{Velocities in the rapidity variable for the $r$-strings, $r=1,\dots, 4$ to the left and $r=5,\dots, 8$ to the right. The behaviour on the line $\R+\i\tf{\pi}{2}$ is depicted on top and 
the one on the line $\R$ is depicted on the bottom. Here, $\zeta=0.5365\pi$ and the endpoint of the Fermi zone is $q=0.2$ in units where $J=1$ leading to a per-site magnetisation $\mf{m}=1-2D$ with $D=0.1801$. The curves for $v_{2}$, $v_{5}$ and $v_{6}$ go beyond $\op{v}_{\infty}$
on $\R+\i\tf{\pi}{2}$. Likewise, the curves $v_{3}$, $v_{4}$, $v_{6}$ and $v_{7}$ go beyond $\op{v}_{\infty}$ on $\R$. However, at this value of zeta, $2,5$ and $6$ strings do not exist with parity $1$, while 
$3$, $4$, $6$ and $7$ strings do not exist with parity $0$. \label{Figure vitesse a zeta 05365}}
 
 \end{figure}

\begin{figure}[t]
 
$\ba{cc}
\includegraphics[width=.4\textwidth]{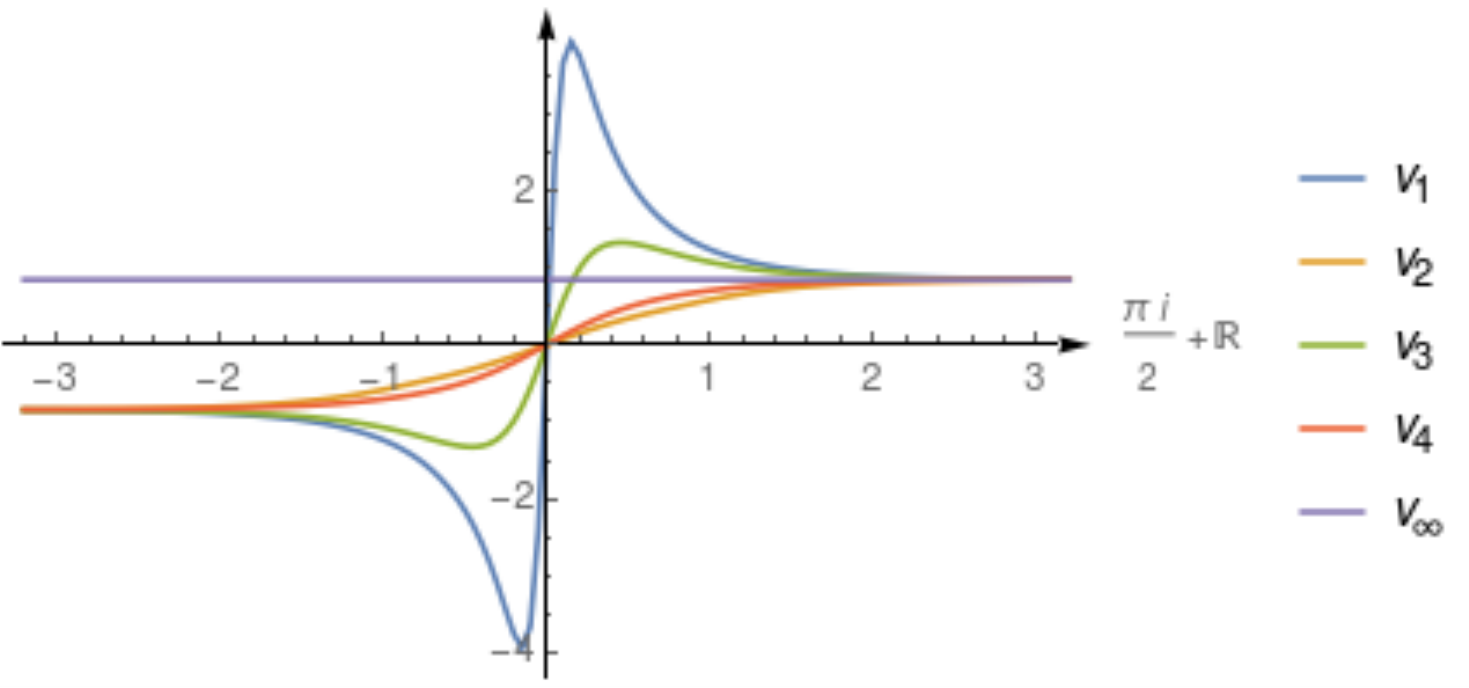}
& 
\includegraphics[width=.4\textwidth]{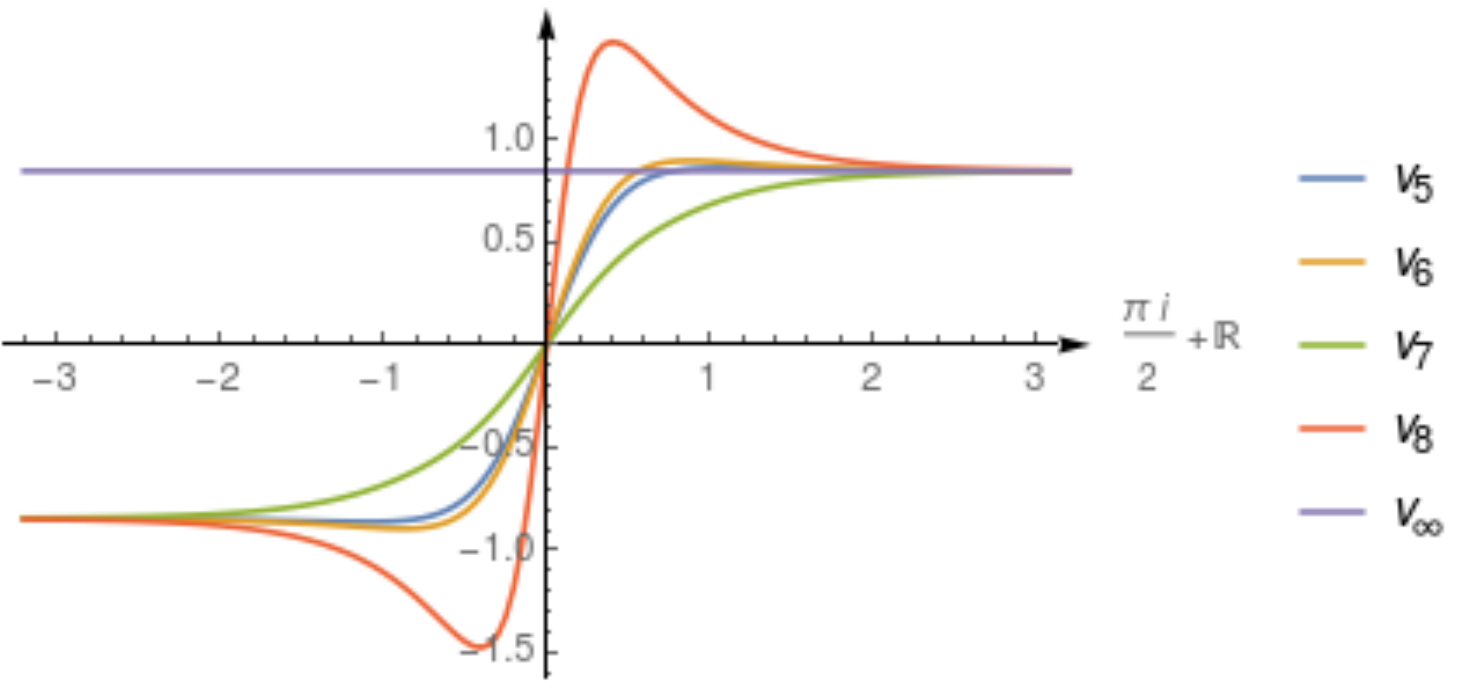} \\
\includegraphics[width=.4\textwidth]{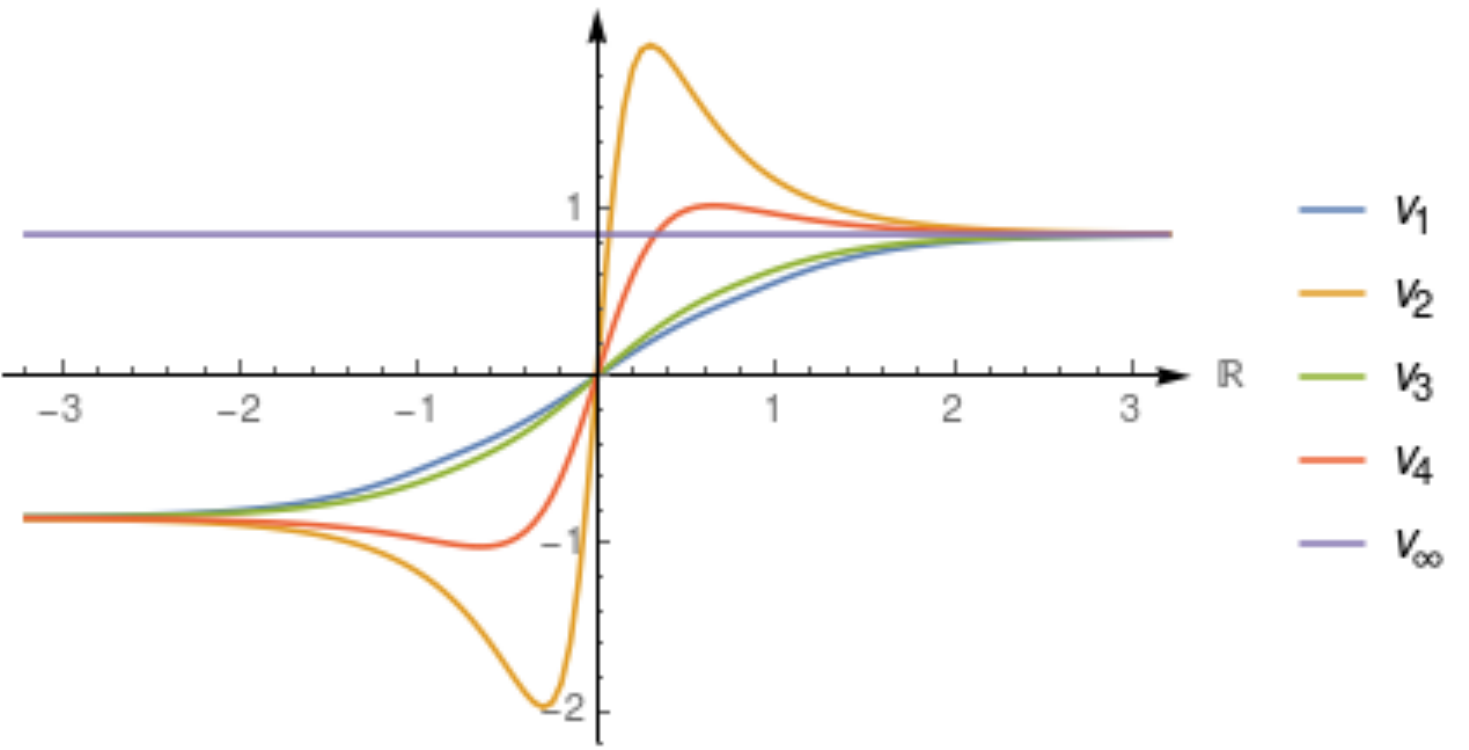}
& 
\includegraphics[width=.4\textwidth]{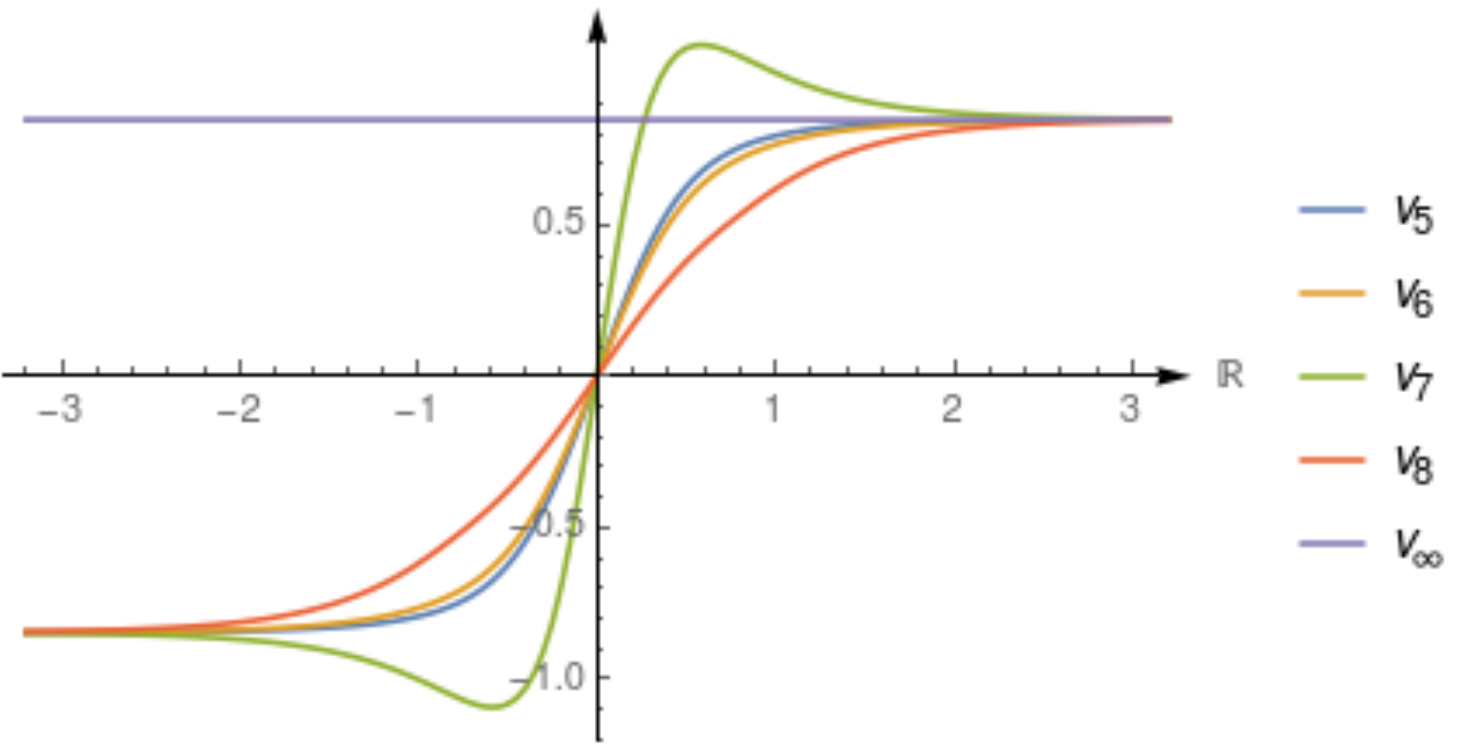}
\ea$
\caption{Velocities in the rapidity variable for the $r$-strings, $r=1,\dots, 4$ to the left and $r=5,\dots, 8$ to the right. The behaviour on the line $\R+\i\tf{\pi}{2}$ is depicted on top and 
the one on the line $\R$ is depicted on the bottom. Here, $\zeta=0.9065\pi$ and the endpoint of the Fermi zone is $q=0.8$ in units where $J=1$ leading to a per-site magnetisation $\mf{m}=1-2D$ with $D=0.1125$. The curves for $v_{1}$, $v_{3}$, $v_{5}$, $v_{6}$ and $v_{8}$ go beyond $\op{v}_{\infty}$
on $\R+\i\tf{\pi}{2}$. Likewise, the curves $v_{2}$, $v_{4}$, and $v_{7}$ go beyond $\op{v}_{\infty}$ on $\R$. At this value of zeta, $6$ and $8$ strings do not exist with parity $1$, while 
$7$ strings do not exist with parity $0$. However, $2$ and $4$ strings exist with parity $0$ while $3$ and $5$ strings exist with parity $1$. Furthermore, it appears that $v_1$ goes beyond $\op{v}_{\infty}$
on $\R+\i\tf{\pi}{2}$, thus there will be two solutions to $v_1(\la)=\op{v}$ on $\R+\i\tf{\pi}{2}$ with $\op{v}_{\infty}<|\op{v}|<\op{v}_1^{(M)}$. \label{Figure vitesse a zeta 09065}}
 \end{figure}

\begin{figure}[t]
 
$\ba{cc}
\includegraphics[width=.4\textwidth]{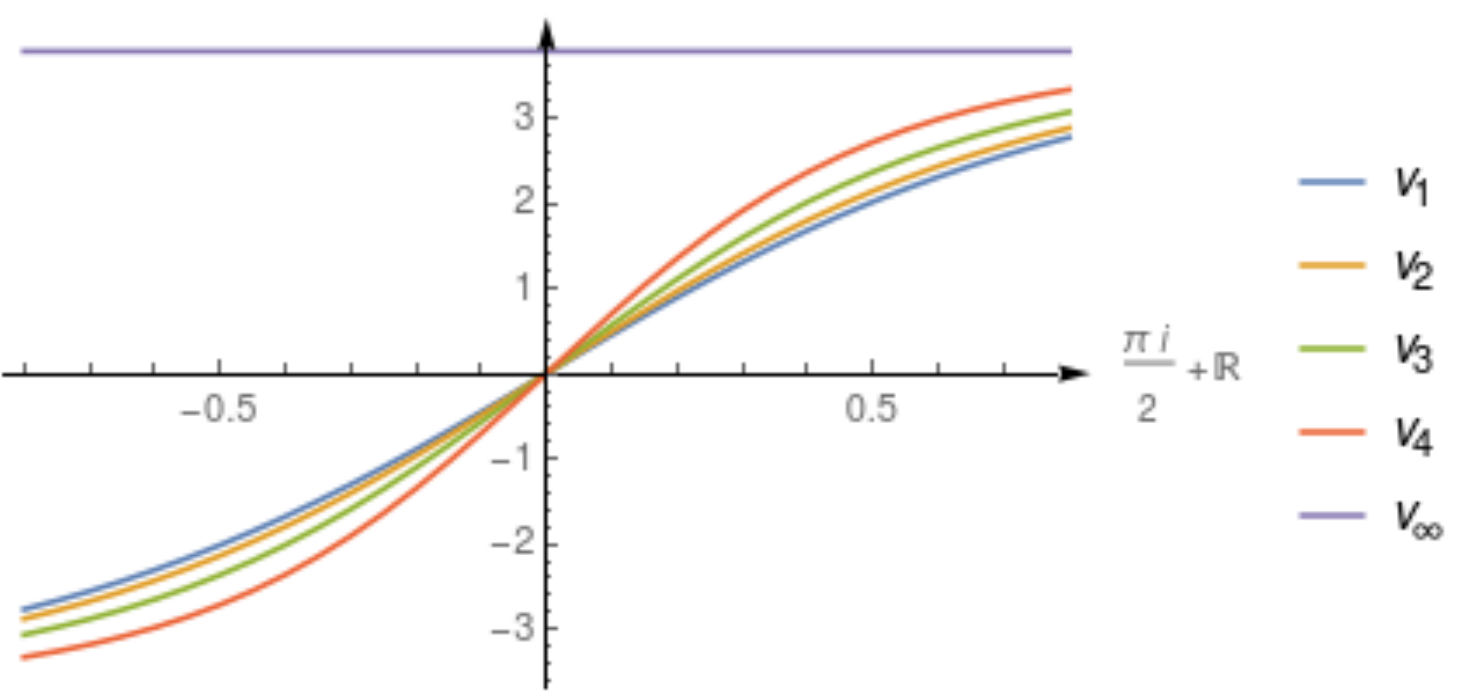}
& 
\includegraphics[width=.4\textwidth]{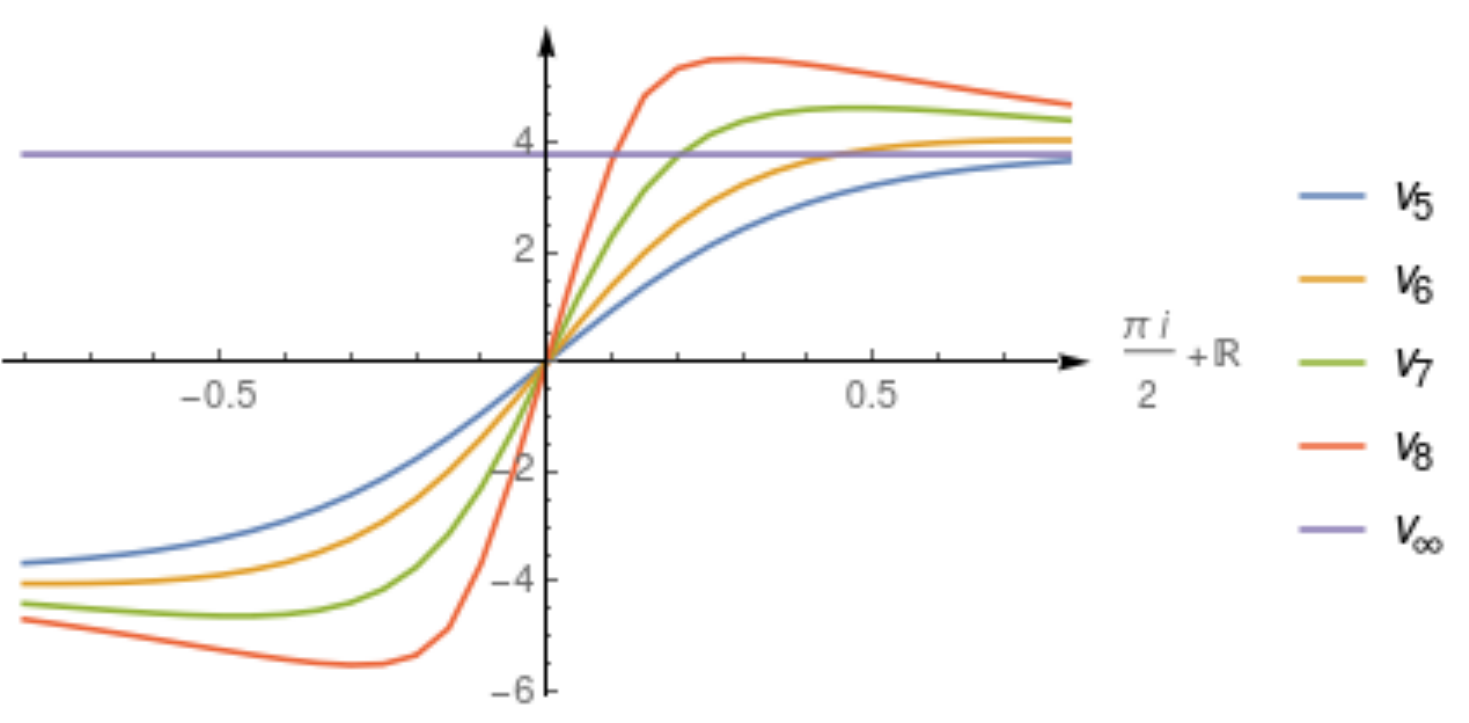} \\
\includegraphics[width=.4\textwidth]{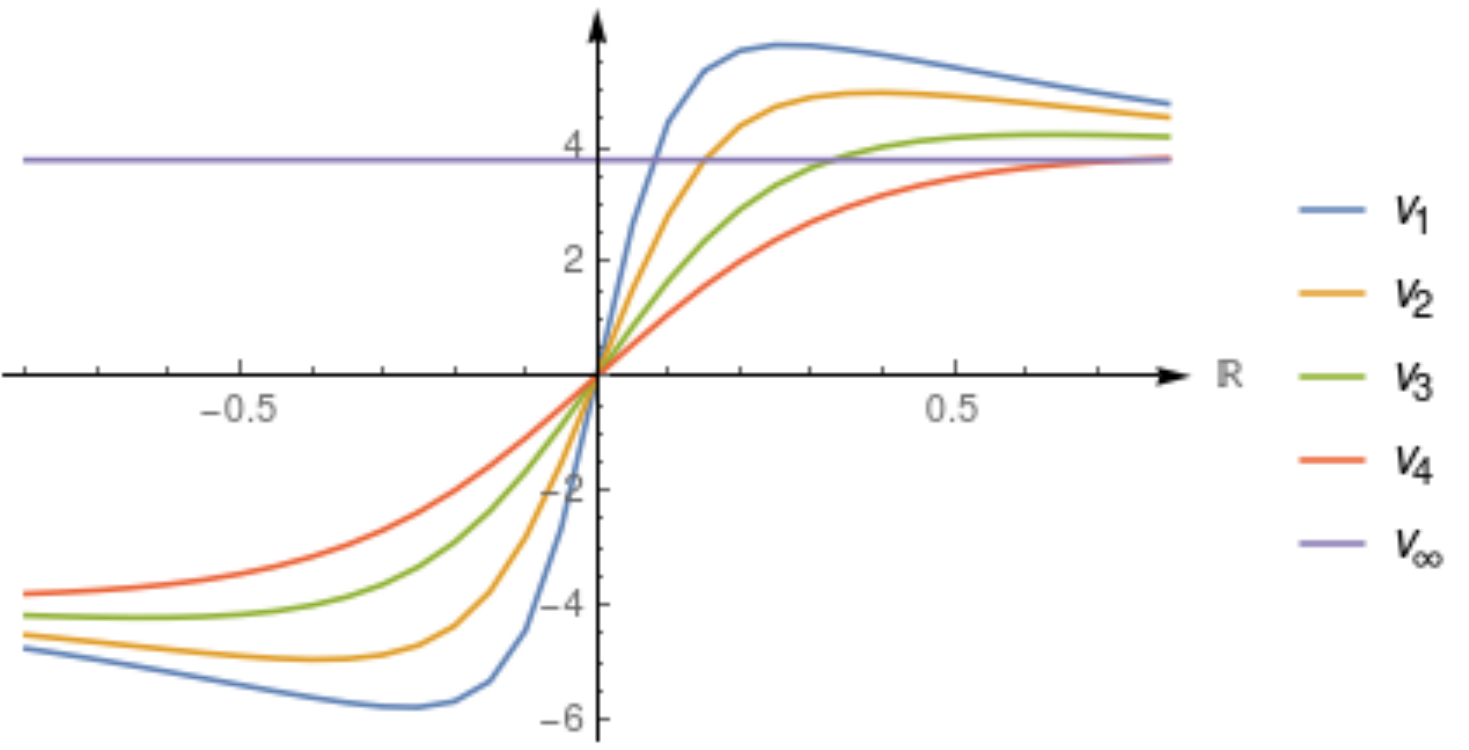}
& 
\includegraphics[width=.4\textwidth]{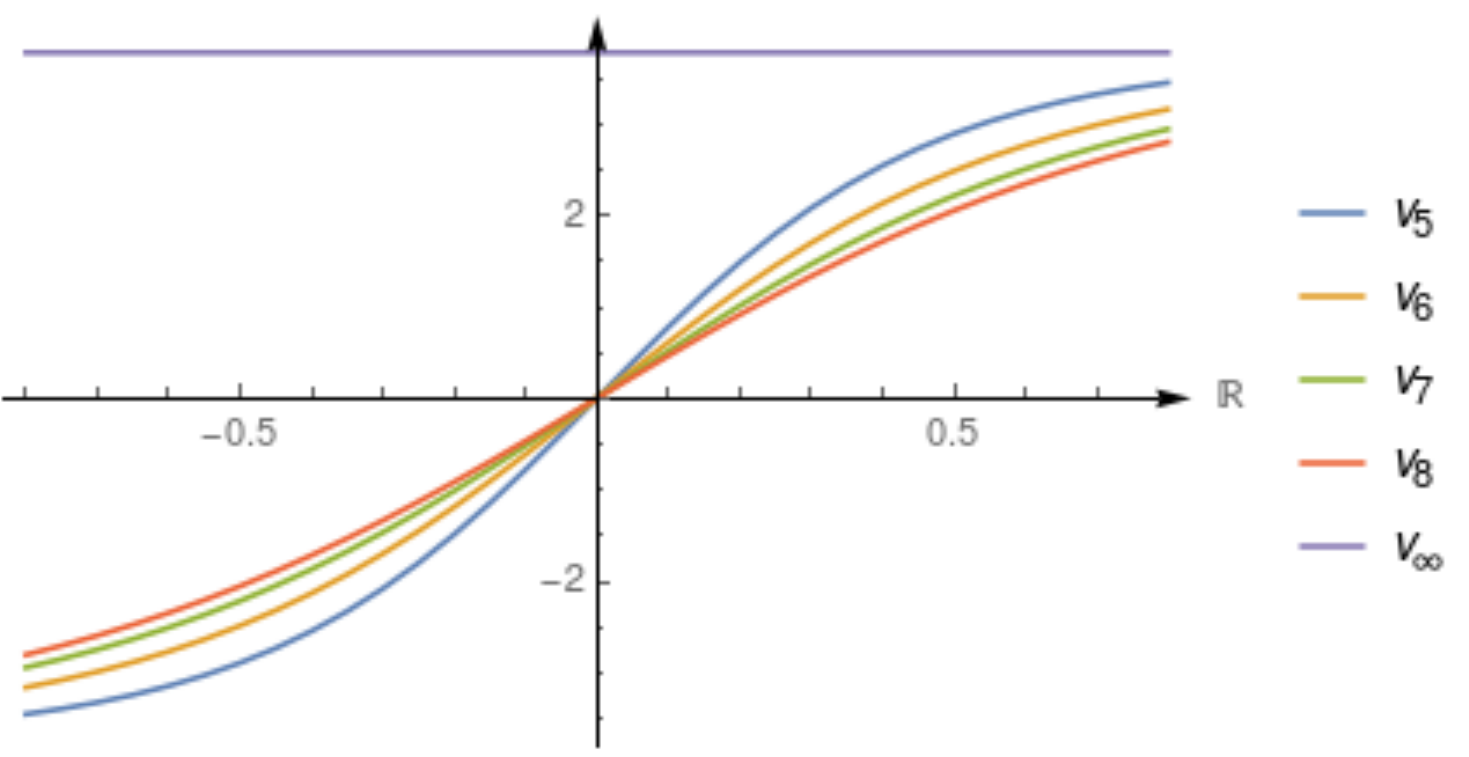}
\ea$
\caption{Velocities in the rapidity variable for the $r$-strings, $r=1,\dots, 4$ to the left and $r=5,\dots, 8$ to the right. The behaviour on the line $\R+\i\tf{\pi}{2}$ is depicted on top and 
the one on the line $\R$ is depicted on the bottom. Here, $\zeta=0.1065\pi$ and the endpoint of the Fermi zone is $q=0.2$ in units where $J=1$ leading to a per-site magnetisation $\mf{m}=1-2D$ with $D=0.4187$. 
The curves for  $v_{6}$, $v_{7}$ and $v_{8}$ go beyond $\op{v}_{\infty}$
on $\R+\i\tf{\pi}{2}$. Likewise, the curves $v_{1}$, $v_{2}$, and $v_{3}$ go beyond $\op{v}_{\infty}$ on $\R$. At this value of zeta, $6$, $7$ and $8$ strings do not exist with parity $1$.
However, $1$, $2$ and $3$ strings exist with parity $0$. Furthermore, it appears that $v_1$ goes beyond $\op{v}_{\infty}$
on $\R$, thus there will be two solutions to $v_1(\la)=\op{v}$ on $\R$ with $\op{v}_{\infty}<|\op{v}|<\op{v}_1^{(M)}$.\label{Figure vitesse a zeta 01065}}
 \end{figure}

 One may also study the shape of $\op{v}_r$ by solving the linear integral equation numerically. Some typical results are gathered in Figures \ref{Figure vitesse a zeta 05365}, \ref{Figure vitesse a zeta 09065} 
and \ref{Figure vitesse a zeta 01065}. Such a numerical analysis seems to indicate that $\op{v}_{r}^{(m)}=\op{v}_{\infty}$ independently of $r$. 
However, for some $r$-strings, depending on the value of $\zeta$, it may happen that  $\op{v}_{r}^{(M)}>\op{v}_{\infty}$
while for other $r$-strings one will have $\op{v}_{r}^{(M)}=\op{v}_{\infty}$. However, for $ \op{v}_{\infty} <|\op{v}|$ the number of zeroes of 
$u_{r}^{\prime}$ does not seem to exceed two. More precisely, 
\begin{itemize}
 
\item $u^{\prime}_r(\la,\op{v})$ \textit{does not vanish} on $\R+\i \sg_r \tfrac{\pi}{2}$ for $|\op{v}|>\op{v}_{r}^{(M)}$;

\item $u^{\prime}_r(\la,\op{v})$ \textit{vanishes twice}  on $\R+\i \sg_r \tfrac{\pi}{2}$ for $ \op{v}_{\infty} <|\op{v}|<\op{v}_{r}^{(M)}$;
 
\item  $u^{\prime}_r(\la,\op{v})$ only vanishes \textit{once} on $\R+\i \sg_r \tfrac{\pi}{2}$ when $-\op{v}_{\infty} < \op{v} <\op{v}_{\infty}$. 
\end{itemize}
Here, the second condition is to be omitted if $\op{v}_r^{(M)}=\op{v}_{\infty}$.

\end{document}